\documentclass[12pt]{article}

\usepackage{scicite}

\usepackage{times}

\usepackage{graphicx}

\usepackage{rotating}

\newenvironment{sciabstract}{%
\begin{quote} \bf}
{\end{quote}}

\title{Illuminating Gravitational Waves: \\
A Concordant Picture of Photons \\
from a Neutron Star Merger \\
}

\author{M. M. Kasliwal$^{1\ast}$, E. Nakar$^{2}$, L. P. Singer$^{3,4}$, D.~L.~Kaplan$^{5}$,\\ 
D.~O.~Cook$^{1}$, A. Van Sistine$^{5}$, R.~M.~Lau$^{1}$, C. Fremling$^{1}$,\\ 
O. Gottlieb$^{2}$, J.~E.~Jencson$^{1}$, S.M. Adams$^{1}$, U. Feindt$^{6}$, K.\\ 
Hotokezaka$^{7}$, S. Ghosh$^{5}$, D.~A.~Perley$^{8}$, P.-C. Yu$^{9}$, T. Piran$^{10}$,\\
J. R. Allison$^{11,12}$, G. C. Anupama$^{13}$, A.Balasubramanian$^{14}$,\\
K.~W Bannister$^{15}$, J. Bally$^{16}$, J. Barnes$^{17}$, S. Barway$^{18}$, E. \\
Bellm$^{19}$, V. Bhalerao$^{20}$, D. Bhattacharya$^{21}$, N. Blagorodnova$^{1}$, \\
J.~S.~Bloom$^{22,23}$, P.~R.~Brady$^{5}$, C. Cannella$^{1}$,  D. Chatterjee$^{5}$,\\ 
S. B. Cenko$^{3,4}$, B. E. Cobb$^{24}$, C. Copperwheat$^{8}$, A. Corsi$^{25}$,\\ 
K. De$^{1}$, D. Dobie$^{11,26,15}$, S.~W.~K.~ Emery$^{27}$, P. A. Evans$^{28}$,\\ 
O. D. Fox$^{29}$, D. A. Frail$^{30}$, C. Frohmaier$^{31,32}$, A. Goobar$^{6}$,\\ 
G.  Hallinan$^{1}$, F. Harrison$^{1}$, G. Helou$^{33}$, T. Hinderer$^{34}$,\\ 
A.~Y.~Q. Ho$^{1}$, A. Horesh$^{10}$, W.-H. Ip$^{9}$, R. Itoh$^{35}$, D. Kasen$^{22,36}$,\\ 
H. Kim$^{37}$, N.P.M. Kuin$^{27}$, T. Kupfer$^{1}$, C. Lynch$^{11,26}$,\\ 
K. Madsen$^{1}$, P.~A.~Mazzali$^{8,38}$, A. A. Miller$^{39,40}$, K. Mooley$^{41}$,\\ 
T. Murphy$^{11,26}$, C.-C. Ngeow$^{9}$, D. Nichols$^{34}$, S. Nissanke$^{34}$,\\ 
P. Nugent$^{22,23}$, E.~O.~Ofek$^{42}$, H. Qi$^{5}$, R.~M.~Quimby$^{43,44}$,\\ 
S. Rosswog$^{45}$, F. Rusu$^{46}$, E. M. Sadler$^{11,26}$, P. Schmidt$^{34}$,\\ 
J.  Sollerman$^{45}$, I. Steele$^{8}$, A.~R.~Williamson$^{34}$, Y. Xu$^{1}$, L. \\
Yan$^{1,33}$, Y. Yatsu$^{35}$, C. Zhang$^{5}$, W. Zhao$^{46}$}

\date{}

\usepackage[utf8]{inputenc}
\usepackage{graphicx}
\usepackage{scicite}
\usepackage{times}
\usepackage{hyperref}
\usepackage{amsmath}
\usepackage{ccaption}

\newcommand{\hst}{\textit{HST}}
\newcommand{\galex}{\textit{GALEX}}

\newcommand{\wise}{\textit{WISE}}
\newcommand{\ngc}{NGC~4993}
\newcommand{\ot}{EM170817}

\newcommand{\degr}{\ensuremath{^{\circ}}}
\newcommand{\A}{{\cal A}}
\newcommand{\phn}{\phantom{0}}

\newcommand{\pasp}{Publ. Astron. Soc. Pac.}
\newcommand{\apj}{Astrophys. J.}
\newcommand{\apjl}{Astrophys. J.}
\newcommand{\aap}{Astron. Astrophys.}
\newcommand{\aj}{Astron. J.}
\newcommand{\nat}{Nature}
\newcommand{\apjs}{Astrophys. J. Suppl. Ser.}
\newcommand{\mnras}{Mon. Not. R. Astron. Soc.}
\newcommand{\prd}{Phys. Rev. D}
\newcommand{\prl}{Phys. Rev. Letters}
\newcommand{\physrep}{Phys. Rep.}
\newcommand{\procspie}{Proc. SPIE}
\newcommand{\araa}{Annual Review of Astron and Astrophys}

\begin{document}


\baselineskip24pt


\maketitle 


\normalsize{$^{1}$Division of Physics, Math and Astronomy, California Institute of Technology, }\\
\normalsize{1200 East California Boulevard, Pasadena, CA 91125, USA}\\
\normalsize{$^{2}$The Raymond and Beverly Sackler School of Physics and Astronomy,}\\
\normalsize{Tel Aviv University, Tel Aviv 69978, Israel}\\
\normalsize{$^{3}$Astroparticle Physics Laboratory, NASA Goddard Space Flight Center, }\\
\normalsize{Mail Code 661, Greenbelt, MD 20771, USA}\\
\normalsize{$^{4}$Joint Space-Science Institute, University of Maryland, College Park, MD 20742, USA}\\
\normalsize{$^\ast$To whom correspondence should be addressed; E-mail:  mansi@astro.caltech.edu.}
\normalsize{$^{5}$Department of Physics, University of Wisconsin, Milwaukee, WI 53201, USA}\\
\normalsize{$^{6}$The Oskar Klein Centre, Department of Physics, Stockholm University,}\\
\normalsize{AlbaNova, SE-106 91 Stockholm, Sweden}\\
\normalsize{$^{7}$Center for Computational Astrophysics, Simons Foundation, }\\
\normalsize{Flatiron Institute, 162 5th Ave, New York, 10010, NY, USA}\\
\normalsize{$^{8}$Astrophysics Research Institute, Liverpool John Moores University, }\\
\normalsize{IC2, Liverpool Science Park, 146 Browlow Hill, Liverpool, L3 5RF, UK}\\
\normalsize{$^{9}$Graduate Institute of Astronomy, National Central University, }\\
\normalsize{No. 300, Zhongda Rd., Zhongli Dist., Taoyuan City 32001, Taiwan}\\
\normalsize{$^{10}$Racah Institute of Physics, The Hebrew University of Jerusalem,}\\
\normalsize{Jerusalem, 91904, Israel}\\
\normalsize{$^{11}$Sydney Institute for Astronomy, School of Physics A28,}\\
\normalsize{The University of Sydney, NSW 2006, Australia}\\
\normalsize{$^{12}$ARC Centre of Excellence for All-sky Astrophysics in 3 Dimensions (ASTRO 3D)}\\
\normalsize{$^{13}$Indian Institute of Astrophysics, II Block Koramangala, Bangalore-560034, India}\\
\normalsize{$^{14}$Indian Institute of Science Education and Research,}\\
\normalsize{Dr. Homi Bhabha Road, Pashan, Pune 411008, India}\\
\normalsize{$^{15}$Australia Telescope National Facility, Commonwealth Scientific and Industrial}\\
\normalsize{Research Organisation, Astronomy and Space Science,}\\
\normalsize{PO Box 76, Epping, NSW 1710, Australia}\\
\normalsize{$^{16}$University of Colorado, Boulder, USA }\\
\normalsize{$^{17}$Columbia Astrophysics Laboratory, Columbia University,}\\
\normalsize{New York, NY, 10027, USA}\\
\normalsize{$^{18}$South African Astronomical Observatory (SAAO),}\\
\normalsize{P.O. Box 9, Observatory, Cape Town 7935, South Africa}\\
\normalsize{$^{19}$Department of Astronomy, University of Washington, Seattle, WA 98195}\\
\normalsize{$^{20}$Department of Physics, Indian Institute of Technology Bombay,}\\
\normalsize{Mumbai 400076, India}\\
\normalsize{$^{21}$Inter-University Centre for Astronomy and Astrophysics,}\\
\normalsize{P. O. Bag 4, Ganeskhind, Pune 411007, India}\\
\normalsize{$^{22}$Department of Astronomy, University of California, Berkeley, CA 94720-3411, USA }\\
\normalsize{$^{23}$Lawrence Berkeley National Laboratory, 1 Cyclotron Road,}\\
\normalsize{MS 50B-4206, Berkeley, CA 94720, USA}\\
\normalsize{$^{24}$Department of Physics, George Washington University,}\\
\normalsize{Washington, DC 20052, USA }\\
\normalsize{$^{25}$Department of Physics and Astronomy, Texas Tech University,}\\
\normalsize{Box 41051, Lubbock, TX 79409-1051, USA}\\
\normalsize{$^{26}$Australian Research Council Centre of Excellence for All-sky Astrophysics (CAASTRO),}\\
\normalsize{$^{27}$University College London, Mullard Space Science Laboratory,}\\
\normalsize{Holmbury St. Mary, Dorking, RH5 6NT, U.K.}\\
\normalsize{$^{28}$University of Leicester, X-ray and Observational Astronomy Research Group}\\ 
\normalsize{Leicester Institute for Space and Earth Observation, Department of Physics\& Astronomy,}\\
\normalsize{University Road, Leicester, LE1 7RH, UK}
\normalsize{$^{29}$Space Telescope Science Institute, 3700 San Martin Dr, Baltimore, MD 21218}\\
\normalsize{$^{30}$National Radio Astronomy Observatory, Socorro, New Mexico, USA}\\
\normalsize{$^{31}$Department of Physics and Astronomy, University of Southampton,}\\
\normalsize{Southampton, Hampshire SO17 1BJ, UK}\\
\normalsize{$^{32}$Institute of Cosmology and Gravitation, Dennis Sciama Building,}\\
\normalsize{University of Portsmouth, Burnaby Road, Portsmouth PO1 3FX, UK}\\
\normalsize{$^{33}$Infrared Processing and Analysis Center, California Institute of Technology, Pasadena, CA 91125, USA}\\
\normalsize{$^{34}$Institute of Mathematics, Astrophysics and Particle Physics,}\\
\normalsize{Radboud University, Heyendaalseweg 135, 6525 AJ Nijmegen, The Netherlands}\\
\normalsize{$^{35}$Department of Physics, Tokyo Institute of Technology,}\\
\normalsize{2-12-1 Ookayama, Meguro-ku, Tokyo 152-8551, Japan}\\
\normalsize{$^{36}$Department of Physics, University of California, Berkeley,CA 94720, USA }\\
\normalsize{$^{37}$Gemini Observatory, Casilla 603, La Serena, Chile}\\
\normalsize{$^{38}$Max-Planck Institute for Astrophysics, Garching, Germany}\\
\normalsize{$^{39}$Center for Interdisciplinary Exploration and Research in Astrophysics (CIERA)}\\
\normalsize{and Department of Physics and Astronomy, Northwestern University, Evanston, IL 60208, USA}\\
\normalsize{$^{40}$The Adler Planetarium, Chicago, IL 60605, USA}\\
\normalsize{$^{41}$Astrophysics, Department of Physics, University of Oxford,}\\
\normalsize{Keble Road, Oxford OX1 3RH, UK}\\
\normalsize{$^{42}$Department of Particle Physics \& Astrophysics, Weizmann Institute of Science,}\\
\normalsize{Rehovot 7610001, Israel}\\
\normalsize{$^{43}$Department of Astronomy, San Diego State University, CA92182, USA }\\
\normalsize{$^{44}$Kavli Institute for the Physics and Mathematics of the Universe (WPI), The University of}\\
\normalsize{Tokyo Institutes for Advanced Study, The University of Tokyo,}\\
\normalsize{Kashiwa, Chiba 277-8583, Japan}\\
\normalsize{$^{45}$ The Oskar Klein Centre,  Department of Astronomy, Stockholm University,}\\
\normalsize{AlbaNova, SE-106 91 Stockholm, Sweden}\\
\normalsize{$^{46}$University of California Merced, USA}\\
\normalsize{$^{47}$Department of Astrophysical Sciences, Princeton University,}\\
\normalsize{Peyton Hall, Princeton, NJ 08544  USA}\\
\normalsize{$^\ast$To whom correspondence should be addressed; E-mail:  mansi@astro.caltech.edu.}

\pagebreak



\baselineskip24pt

\begin{sciabstract}

 Merging neutron stars offer an exquisite laboratory for simultaneously studying strong-field gravity and matter
 in extreme environments. We establish the physical association of an electromagnetic counterpart (\ot) to 
 gravitational waves (GW170817) detected from merging neutron stars. By synthesizing a panchromatic dataset,
 we demonstrate that merging neutron stars are a long-sought production site forging heavy elements by
 r-process nucleosynthesis. The weak gamma-rays seen in EM170817 are dissimilar to classical short gamma-ray bursts with
 ultra-relativistic jets. Instead, we suggest that breakout of a wide-angle, mildly-relativistic cocoon engulfing the jet elegantly explains the low-luminosity gamma-rays, 
the high-luminosity ultraviolet-optical-infrared and the delayed radio/X-ray emission. We posit that all merging neutron stars may 
lead to a wide-angle cocoon breakout; sometimes accompanied by a successful jet and sometimes a choked jet. 

\end{sciabstract}


%

\begin{figure*}[!hbt]
\centering
\includegraphics[width=\textwidth]{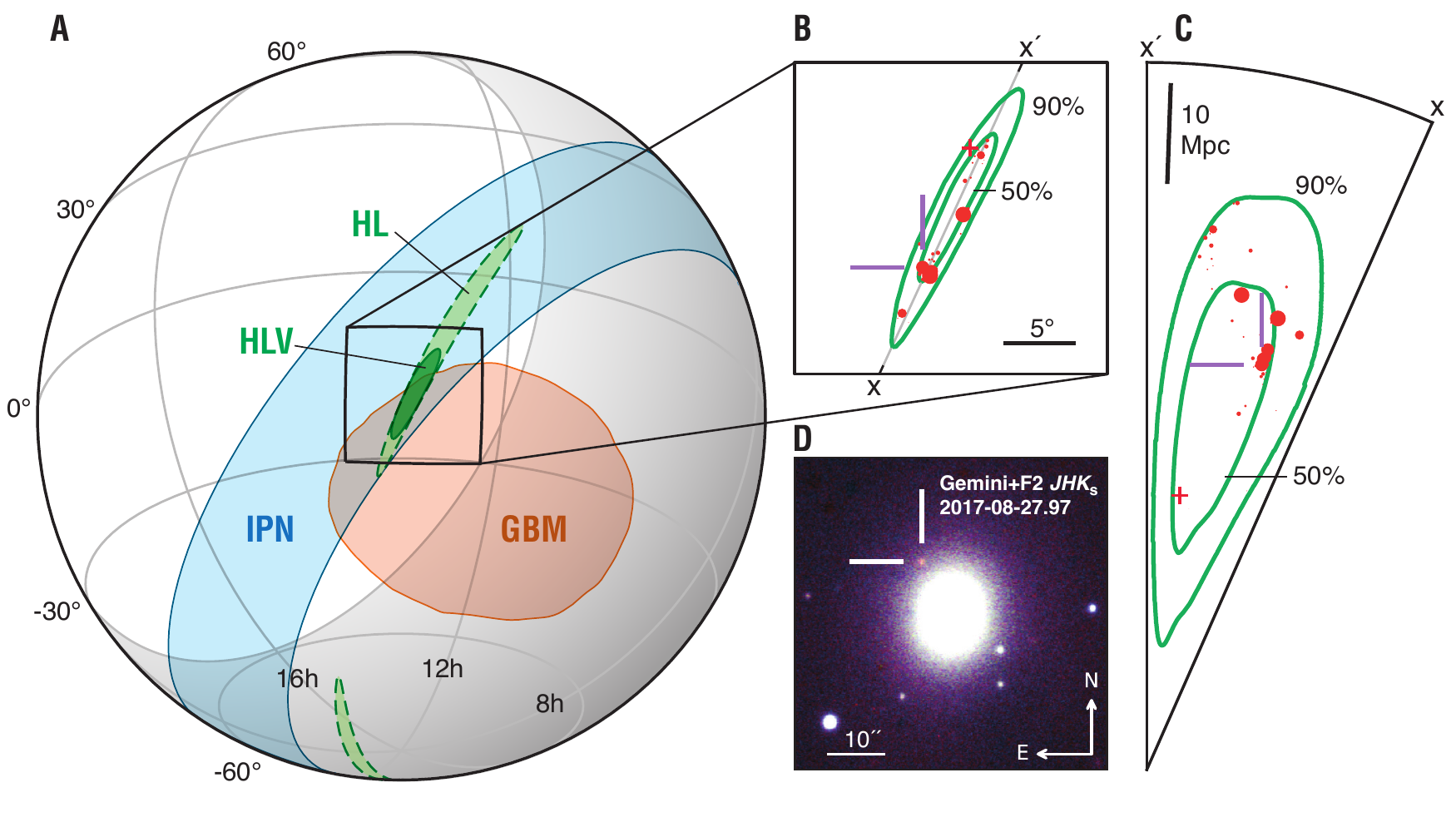}
\caption{
Localization of GW170817 and associated transient \ot. (A) Constraints at the 90\% confidence level on the sky position from gravitational-wave and $\gamma$-ray observations. The rapid LIGO localization is indicated by the green dashed contour, and the  LIGO/Virgo localization by solid green. {\it Fermi} GBM \cite{Goldstein2017} is shown in orange, and the Interplanetary Network triangulation from {\it Fermi} and {\it INTEGRAL} in blue \cite{LVCC21515}. The shaded region is the Earth limb as seen by {\it AstroSat} which is excluded by the non-detection by the Cadmium Zinc Telluride Imager instrument. (B) 49 galaxies from the Census of the Local Universe catalog (Table S3; red, with marker size proportional to the stellar mass of the galaxy) within the LIGO/Virgo three-dimensional 50\% and 90\% credible volumes (green).  One radio-selected  optically-dark galaxy whose stellar mass is unknown is marked with a +. (C) Cross-section along the X-X' plane from panel B, showing the luminosity distances of the galaxies in comparison to the LIGO/Virgo localization. (D) False-color near-infrared image of \ot\ and its host galaxy \ngc, assembled from near-infrared observations with the FLAMINGOS-2 instrument on Gemini-South \cite{sot}, with $J$, $H$, and $K_s$ shown as blue, green, and red, respectively.  Our $K_s$-band detections span  2017 Aug 18.06 to 2017 Sep 5.99 and we show 2017-08-27.97 above. 
\label{fig:localization}}
\end{figure*}

On 2017 August 17 at 12:41:04 UTC,  gravitational waves from the merger of two neutron stars (NS-NS) were detected by the Laser Interferometer Gravitational-Wave Observatory (LIGO) and dubbed GW170817 \cite{GW170817}. Two seconds later, the first temporally-coincident photons were detected as $\gamma$-rays by the {\it Fermi} satellite \cite{LVCC21528,GBM2017,Goldstein2017}. GW170817 was such a loud event that the joint on-sky localization from the LIGO and Virgo interferometers was only 31 square degrees (Figure~\ref{fig:localization}) with an initial distance estimate of 40$\pm$8\, Megaparsec \cite{LVCC21513}. To identify potential host galaxies \cite{Gehrels2016,Nissanke2013}, we cross-matched to our Census of the Local Universe (CLU; \cite{Cook17}) galaxy catalog and found only 49 galaxies in this volume \cite{LVCC21519,sot}. To prioritize follow-up, we ranked the galaxies by stellar mass (see Table~S1 and supplementary online text \cite{sot}). A multitude of telescopes promptly began multi-wavelength searches for an electromagnetic counterpart in and around these galaxies. Ground-based searches were systematically delayed (due to the Southern location) by half a day until sunset in Chile \cite{Kasliwal2014,Singer2014}. A bright optical transient was identified and announced by the Swope telescope team at Las Campanas Observatory  \cite{LVCC21529,Coulter2017} in the 3rd ranked galaxy in our list, named \ngc. This source, SSS\,17a, is located at right ascension 13$^{\rm h}$09$^{\rm m}$48.071$^{\rm s}$ and declination $-$23$^{\rm d}$22$^{\rm m}$53.37$^{\rm s}$ (J2000 equinox, \cite{sot}), with a projected offset from the nucleus of \ngc\ of 2.2\,kiloparsec and away from any sites of star formation (Figure~S1; \cite{sot}). We also detect this transient in the infrared and ultraviolet wavelengths (see companion paper; \cite{Evans17}). 
Nine days later, an X-ray counterpart was identified \cite{LVCC21765,Troja2017}. Fifteen days later, a radio counterpart was identified (see companion paper; \cite{Hallinan17}). 

Initially, the bright luminosity and the blue, featureless optical spectrum of SSS17a appeared to be consistent with a young supernova explosion that should brighten (see Figure~S2, Figure~S3).  However, on the second night, the source faded substantially in the optical and brightened in the infrared (Figure~\ref{fig:lightcurve_comparison}). Combining ultraviolet-optical-infrared (UVOIR) data from 24 telescopes on 7 continents, we constructed a bolometric light curve (Figure~\ref{fig:lc}; see \cite{sot} for details). The bolometric luminosity evolves from 10$^{42}$\,erg s$^{-1}$ at 0.5\,d to 3$\times$10$^{40}$\,erg s$^{-1}$ at 10\,d (Figure~\ref{fig:lc}). By estimating the black-body effective temperature evolution, we find that the source rapidly cools from $\approx\,$11000\,K to $\approx\,$5000\,K in a day to $\approx$1400\,K in ten days. The inferred photospheric expansion velocities span 0.3$c$ to 0.1$c$, where c is the speed of light \cite{sot}. Furthermore, infrared spectroscopy shows broad features that are unlike any other transient seen before (Figure~\ref{fig:spectra}, Figure~S4, Figure~S5). The combination of high velocities, fast optical decline, slow infrared evolution and broad peaks in the infrared spectra are unlike any other previously known transient and unlikely to be due to a chance coincidence of an unrelated source. We thus establish that the panchromatic photons, hereafter \ot, are spatially, temporally and physically associated with GW170817. With this firm connection, we now turn our attention to understanding the astrophysical origin of \ot.

\section*{Evidence for Nucleosynthesis of Heavy Elements}
It is well established that chemical elements up to iron in the periodic table are produced either in the Big Bang or in cores of stars or in supernova explosions. 
However, the origin of half the elements heavier than iron, including gold, platinum and uranium, has remained a mystery. These heavy elements are synthesized
by the rapid capture of neutrons (r-process nucleosynthesis). Some models have proposed that the decompression of neutron-rich matter in a NS-NS merger may
provide suitable conditions to robustly synthesize heavy r-process elements \cite{Lattimer1974,Freiburghaus1999}. Radioactive decay of freshly synthesized unstable isotopes should drive 
transient electromagnetic emission known as a ``kilonova" or ``macronova" (e.g, \cite{Li1998,Fernandez2016}). We test this hypothesis with the optical and infrared data of \ot.

First, we compare the spectra of \ot\ to a library of astronomical transients \cite{sot} and theoretical models for macronova spectra \cite{Barnes2013}. The optical spectra exhibit a featureless continuum (Figure~S2, Figure~S3). Infrared spectra (Figure~\ref{fig:spectra}) have two distinct, broad peaks in $J$-band (10620 $\pm$ 1900 Angstrom) and $H$-band (15500 $\pm$ 1430 Angstrom). Due to the high velocities in the ejecta material, each peak may be produced by a complex blend of elements instead of a single element. Although the $J$-band peak is reminiscent of either Helium or Hydrogen, the corresponding feature in the $H$-band seen in core-collapse supernovae is not present (Figure~S5).  If instead we compare to Type Ia supernovae, the $J$-band peak could be similar to iron group elements. However, once again, the second $H$-band peak is dissimilar to that seen in Type Ia supernovae (Figure~S4). By comparing predictions of spectra of macronovae \cite{Barnes2013}, based on the assumption that Neodymium (Nd) is representative of lanthanides synthesized via the r-process, we find a reasonable match to both the $J$-band and $H$-band features for ejecta mass (M$_{\rm ej}$)  of 0.05\,M$_{\odot}$ and velocity (v) of 0.1$c$ (Figure~\ref{fig:spectra}).
Recent updates to these models, incorporating line transitions from 14 elements and tuning the relative abundance ratios, indicate that Nd plays a crucial role in explaining these features \cite{Kasen17}. 
We conclude that a blend of elements substantially heavier than elements produced in supernovae is a viable explanation for the spectra of \ot.

Next, we compare our infrared light curves of \ot\ (Figure~\ref{fig:lightcurve_comparison}) to a suite of existing macronova models by various groups
\cite{Kasen2013,Tanaka2013,Rosswog2017,Wollaeger2017}. The slow, red photometric evolution seen in \ot\ is a generic
feature of all macronova models despite their differing treatments of matter dynamics, matter geometry, nuclear heating, opacities and radiation transfer.
The observed late-time emission ($>$3\,d) is fully consistent with radioactive decay of the dynamical ejecta containing elements from all three r-process abundance peaks (Figure~S10). 
The observed luminosity, temperature and temporal evolution roughly matches model predictions for an ejecta mass of $\sim0.05$ solar masses (M$_{\odot}$), an ejecta velocity 
of $\sim0.1$c and an opacity ($\kappa$) of $\sim10\,$cm$^{2}\,$g$^{-1}$. 

We examine this match further with simple analytics. Dividing the observed bolometric luminosity ($\approx\,6\times\,10^{41}$ erg s$^{-1}$ at 1 day) by the beta-decay heating 
rate of r-process elements ($\approx\,1.5\,\times\,10^{10}\,$erg s$^{-1}$ g$^{-1}$; \cite{Hotokezaka2016}) gives a lower limit on the r-process ejecta mass of $>$0.02 M$_{\odot}$. 
The decline rate of the bolometric luminosity also matches that expected from the beta-decay heating rate of r-process elements with the time-dependent 
thermalization efficiency of the decay products (Figure~\ref{fig:lc}).
The expansion velocity of the ejecta, $0.1$--$0.3c$, 
derived from the photospheric radius is consistent with 
 the results of merger simulations \cite{Hotokezaka2013,Bauswein2013,Rosswog2013}.  Ejecta mass estimates based on
 observed emission are necessarily lower limits as a significant amount of additional matter can be hidden at lower velocity.


Next, we focus on the early-time emission of \ot\ that is hotter, more luminous and faster-rising than predicted by the suite of macronova model predictions discussed above (Figure~\ref{fig:lightcurve_comparison}). 
Decay of free neutrons would give an unphysically large ratio of neutron mass to ejecta mass \cite{sot}. Ultraviolet flashes predicted by \cite{Aloy2005,Gottlieb2017} are on much shorter timescale 
than observed for \ot.  
Instead, we propose two possible explanations:
(i) If some fraction of the ejecta is boosted to mildly relativistic speeds, the relativistic expansion shortens the observed peak time and the Doppler effect results in 
bluer, brighter emission. The jet cocoon model (see below) can
accelerate enough material at higher latitudes. All material would
have $\kappa$$>\,\approx1\,{\rm cm}^2\,{\rm g}^{-1}$ in
this scenario. 
(ii) A disk-driven wind enriched with lighter r-process elements with
$\kappa\,\approx 0.5\,{\rm cm}^2\,{\rm g}^{-1}$ could also produce early, blue emission \cite{Evans17}. This
wind could be driven from a merger remnant that is a massive neutron star with an accretion torus. 
We could have distinguished between these two possibilities if data were available at even earlier times.

\section*{
A Synthesized Model Explaining the Panchromatic Photons}

We discuss three models in an effort to build a self-consistent picture that explains the $\gamma$-ray, X-ray, ultraviolet, optical, infrared and radio photons (Figure~\ref{fig:schematic}). 

\subsection*{A Classical, On-Axis Short Hard Gamma Ray Burst: Ruled Out}
A classical short hard gamma ray burst (sGRB) is produced by a jet in the line-of-sight of the observer (Model A in Figure~\ref{fig:schematic}) that is narrow (opening angle $\theta_{\rm jet}\,\sim$\,10$^{\circ}$) and ultra-relativistic (Lorentz factor $\Gamma\,>\sim$\,100). The progenitors of sGRBs have long been hypothesized to be NS-NS mergers \cite{Eichler1989}.
However, the observed $\gamma$-ray luminosity of \ot\ ($\sim$10$^{47}$ erg s$^{-1}$, \cite{GBM2017,Goldstein2017}) 
is lower than typical sGRBs by four orders of magnitude \cite{Nakar2007,Fong2015}. If EM170817 were simply an extremely weak sGRB,  
then the  successful breakout of a narrow, ultra-relativistic jet would require $<$3$\times$10$^{-6}$ M$_{\odot}$ of material that was previously ejected in the direction of the jet \cite{sot}. If the jet opening angle were wider,  it would require even less material to successfully break out \cite{sot}. Such a low ejecta mass is in contradiction with the observed bright UVOIR counterpart, which indicates $\approx$ 0.05 M$_{\odot}$ of ejecta. Furthermore, this scenario cannot account for the delayed onset of X-ray emission \cite{Evans17} and radio emission \cite{Hallinan17}. 

\subsection*{A Classical, Off-Axis Short Hard Gamma Ray Burst: Unlikely}

Next, we consider the possibility of a classical off-axis sGRB where the observer is not in the line-of-sight of a strong, ultra-relativistic jet (Model B in Figure~\ref{fig:schematic}). 
Given the sharp drop in observed $\gamma$-ray luminosity with observing angle,   
we find that the observer could only be off-axis by $<$8$^{\circ}$ \cite{sot}. 
Such a slightly off-axis orientation is unlikely as only a small fraction ($\approx$5\%) of observing angles are consistent with the observational constraints. Moreover, in this scenario, \ot\ is expected to exhibit a bright afterglow at all wavelengths roughly one day after the NS-NS merger, when the external shock decelerates to $\Gamma\,\sim\,$10. Initial non-detections in the radio \cite{Hallinan17} and X-ray \cite{Evans17} observations at this phase constrain the circummerger environment to an implausibly low density ($<$10$^{-6}$ cm$^{-3}$).
Another problem is that a hypothetical on-axis observer to such a sGRB would expect to see photons harder than we have thus far seen in sGRBs \cite{sot}. 
Thus, it is unlikely that the $\gamma$-rays are produced by a slightly off-axis sGRB. 

We conclude that \ot\ is not similar to the classical population of previously observed sGRBs. While the observed $\gamma$-rays are indicative of a relativistic outflow (with or without a jet), they must originate in a different physical mechanism \cite{sot}. We explore the possibility of a structured jet in sGRBs with a distribution of Lorentz factors and identify multiple challenges with this model \cite{sot}. 
Therefore, next, we propose a model with a wide-angle mildly relativistic outflow that propagates in our direction with a relatively small Lorentz factor.  






\subsection*{Cocoon Breakout: A concordant picture}
Based on our UVOIR observations, we estimate that
a few hundredths of a solar mass of ejecta are propelled
into the circummerger medium of a NS-NS merger with velocities spanning a few tenths the speed of light.
We consider a model where a relativistic jet is launched after a short delay, perhaps on account of a delayed collapse of the hyper-massive neutron star into a black hole. 
As the jet drills through the ejecta, the material enveloping the jet inflates to form a pressurized cocoon that expands outward at mildly relativistic speeds.
There are two possibilities: If the jet is wide-angle ($\approx$30$^{\circ}$), it will become choked and fail to drill out (Model C in Figure~\ref{fig:schematic}). 
If the jet is narrow ($\approx$10$^{\circ}$) and long-lived, it could penetrate the ejecta and look like a classical sGRB to an on-axis observer (Model D in Figure~\ref{fig:schematic}).  

Independent of the fate of the jet that created the cocoon, recent numerical simulations \cite{Gottlieb2017} 
show that the cocoon would expand at mildly relativistic velocities ($\Gamma \approx 2-3$ ) over a wide opening angle ($\approx 40^o$) with energy comparable to the jet. 
The cocoon has a wide enough angle and sufficient kinetic energy to easily explain the observed $\gamma$-rays.
However, it remains unclear how a cocoon would dissipate its energy internally at the radius where $\gamma$-rays are observed, 
given its ballistic and homologous expansion (unlike sGRB jets which are expected to be variable with irregular internal velocities and structure that can dissipate the jet energy by internal shocks or magnetic reconnection). A wide angle mildly relativistic cocoon, found by \cite{Gottlieb2017}, was recently proposed as a source of wide-angle $\gamma$-ray emission \cite{Lazzati2017}. However, this was based on an ad-hoc dissipation process that is somehow at work near the photosphere \cite{Lazzati2017}. Here, we suggest that the dissipation mechanism is the interaction of the cocoon with the ejecta and that the observed $\gamma$-rays result from the breakout of the mildly relativistic shock (driven by the cocoon) from the leading edge of the ejecta. We find that such a breakout can explain all properties of the observed low-luminosity $\gamma$-rays if its Lorentz factor is $\approx\,2-3$ and the breakout radius is $\sim 3 \times 10^{11}$cm \cite{sot}. 

We performed a relativistic hydrodynamical simulation in which a jet is injected into expanding ejecta to verify this picture for \ot\ \cite{sot}. We find that even if a minute amount of ejecta ($\approx\,3 \times 10^{-9}\,M_{\odot}$) moves at 0.8$c$, the breakout radius and velocity match those needed to produce the observed $\gamma$-rays for a wide range of ejecta and jet properties \cite{sot}. For example, in the simulation shown in Figure~\ref{fig:hydro}, a shock with $\Gamma \approx 2.5$ breaks out 10\,s after the merger at a radius of $2.4 \times 10^{11}$cm, generating $\gamma$-ray emission that would be observed with a delay of 2\,s with respect to merger time (consistent with the {\it Fermi} observations; \cite{GBM2017,Goldstein2017}). After the cocoon breaks out, the photons that were deposited by the shock diffuse outwards and produce cooling emission that fades on timescales of hours \cite{Gottlieb2017}. After a few hours, radioactive decay of r-process elements becomes the dominant source of the observed emission. The emission during the first day is dominated by fast cocoon material (v$\approx$0.4c), which is composed of high-latitude, low-opacity ($\kappa\sim$1 cm$^2$ g$^{-1}$)  ejecta that was accelerated by the jet to high velocities. After a few days, the slower, higher-opacity ($\kappa\sim$10 cm$^2$ g$^{-1}$) dynamical ejecta begins to dominate the emission. We find that the bolometric light curve evolution and the temperature evolution predicted by this simulation is consistent with our UVOIR observations (Figure~\ref{fig:lc}). 

The available radio and X-ray data are broadly consistent with both cocoon scenarios albeit with slightly different circummerger densities \cite{Evans17,Hallinan17}. If the jet is choked, the radio and X-ray data could be explained by the forward shock that the expanding cocoon drives into the circummerger medium. If the jet is successful,  the radio and X-ray data could be explained as 
a widely off-axis afterglow of the jet.  
If this emission is from the forward shock of a cocoon, we predict that the X-rays and radio will continue to rise. On the other hand, if this emission is from a widely off-axis afterglow of the jet, we predict that it will evolve slowly and eventually fade. In both scenarios, a cocoon would be needed to explain the $\gamma$-rays. We conclude that the cocoon model can self-consistently explain the multi-wavelength properties of \ot\ spanning $\gamma$-rays to radio.  

\section*{Implications}

Now we consider the question of whether \ot\ was an exceptional event or whether multi-messenger detections will soon become routine. 
The large ejecta masses and high velocities seen in \ot\ suggests that intrinsically luminous UVOIR macronova emission should accompany every NS-NS merger. 
If our proposed mildly relativistic cocoon model is correct, the wide opening angle of the cocoon implies that $\gamma$-rays would be emitted towards the observer in about 30\% of NS-NS mergers. If the jet is choked, we expect to see late onset of radio and X-ray emission from the cocoon forward shock.  If the jet producing the cocoon successfully breaks out, the source would appear either as a classical wide off-axis afterglow or a classical on-axis afterglow depending on the observer's line-of-sight. The launch of a successful on-axis cocoon jet may already have been seen in previous reports of possible late-time excess optical/infrared emission in sGRBs attributed to macronovae. In Figure~\ref{fig:lightcurve_comparison}, we find that the excess seen in GRB\,130603B \cite{Tanvir2013}, GRB\,160821B \cite{Kasliwal2017} and GRB\,050709 \cite{Jin2016} are roughly consistent with our observed light curve for \ot. Separately, a plateau in the distribution of durations of sGRB may indicate that a large fraction of sGRBs may have choked jets \cite{Moharana2017}. Joint gravitational wave and electromagnetic observations of NS-NS mergers will shed light on the relative fraction of cocoons with choked jets and cocoons with successful jets.


Now we consider whether NS-NS mergers could be the primary sites of
r-process nucleosynthesis. This depends on both the rate of NS-NS
mergers and the average amount of r-process material synthesized per
merger. Based on the macronova light curve, we estimated a lower limit
on the mass of the produced r-process elements in EM170817 to be
$M_{\rm ej}\approx 0.05M_{\odot}$. The solar abundance pattern shows
that the first of three r-process peaks accounts for $\approx$80\% of
the total r-process abundance (Figure~S10 and \cite{sot}). To account for the observed solar abundance in all three r-process peaks with NS-NS mergers, we would need a rate of $\sim500\,{\rm Gpc^{-3}\,yr^{-1}}\,(M_{\rm ej}/0.05M_{\odot})^{-1}$. To account for the observed abundance in the two heavier r-process peaks with NS-NS mergers, the rate would only need to be $\sim100\,{\rm Gpc^{-3}\,yr^{-1}}$. 
Based on the detection of GW170817, a NS-NS merger rate of 320--4740\,Gpc$^{-3}$\,yr$^{-1}$  was estimated at 90\% confidence \cite{GW170817}.
This is larger than the classical sGRB beaming-corrected rate
\cite{Wanderman2015,Jin2017} and larger than the predicted fraction of
NS-NS mergers based on the Galactic population \cite{Phinney1991}. 
Based on an archival search for transients like
\ot\ in the Palomar Transient Factory database, we find a 3-$\sigma$
upper limit on the rate of 800\,Gpc$^{-3}\,{\rm yr}^{-1}$ \cite{sot}. 
Therefore the large ejecta mass of \ot\ and the high rate estimates of GW170817/\ot\ are consistent with the scenario that NS-NS mergers are the main production sites of r-process elements of the Milky Way (as predicted by \cite{Lattimer1974}). 


The large rate, the wide angle for contemporaneous $\gamma$-rays, the bright UVOIR emission, the forward shock giving a late onset of X-rays and radio, the increase in sensitivity of GW interferometers, the increase in sensitivity of EM facilities (e.g. \cite{ztf,gattini,ultrasat,cutie}) --- all imply many more events like EM/GW170817.




\section*{Acknowledgments}
We thank Iva Kostadinova for seamlessly coordinating the GROWTH  (Global Relay of Observatories Watching Transients Happen) program and Britt Griswold for beautiful graphic arts. We thank Patricia Whitelock for facilitating IRSF observations.  We thank Scott Barthelmy for setting up an LVC GCN system that facilitated quick, citable communication between astronomers and maximized the science return. We thank Sterl Phinney, Shri Kulkarni and Lars Bildsten for valuable comments. We thank the staff of Gemini Observatory, in particular the director Laura Ferrarese for rapidly approving our Director's Discretionary Time request, and our program contact scientists Mischa Shirmer, Hwihyun Kim, Karleyne Silva, Morten Andersen, and Ricardo Salinas for supporting and executing observations. We especially grateful to Gemini for postponing scheduled maintenance on the FLAMINGOS-2 instrument in order to obtain as much data as possible on this extraordinary event.

This work was supported by the GROWTH (Global Relay of Observatories Watching Transients Happen) project funded by the National Science Foundation under PIRE Grant No 1545949. GROWTH is a collaborative project among California Institute of Technology (USA), University of Maryland College Park (USA), University of Wisconsin Milwaukee (USA), Texas Tech University (USA),  San Diego State University (USA), Los Alamos National Laboratory (USA), Tokyo Institute of Technology (Japan), National Central University (Taiwan), Indian Institute of Astrophysics (India), Inter-University Center for Astronomy and Astrophysics (India), Weizmann Institute of Science (Israel), The Oskar Klein Centre at Stockholm University (Sweden), Humboldt University (Germany), Liverpool John Moores University (UK). The data presented here is available in observatory archives and the PLUTO simulation input and output files are available online (URLs in supplementary online text). Additional acknowledgements are in the supplementary online text.

\section*{Supplementary materials}
www.sciencemag.org\\
Materials and Methods\\
Supplementary Text\\
Figs. S1, S2, S3, S4, S5, S6, S7, S8, S9, S10\\
Table S1, S2, S3\\
References (48-175)\\
Movie S1\\
%
%


\clearpage

\begin{figure*}[!hbt]
\centering
\includegraphics[width=\textwidth]{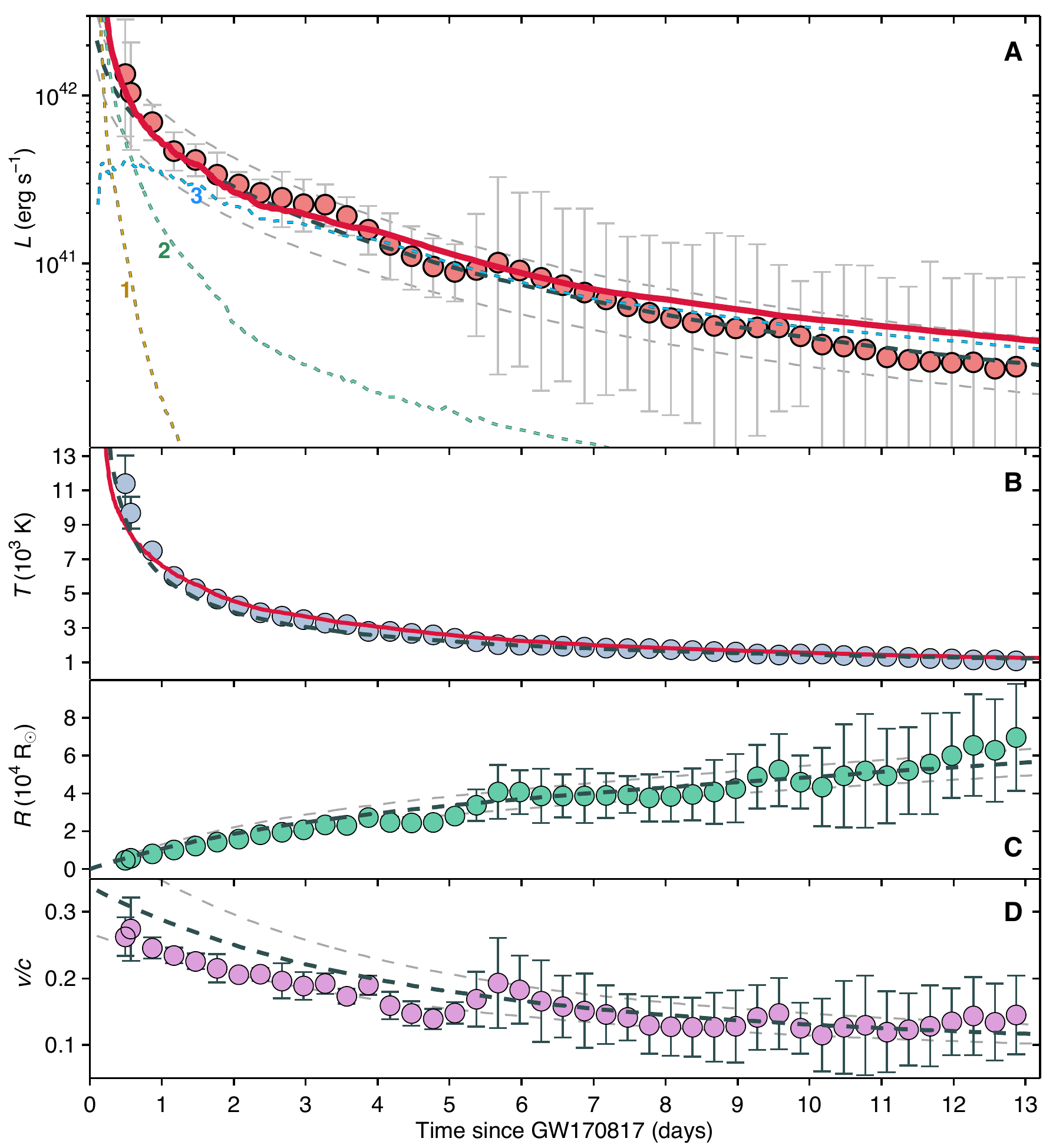}
\caption{
The evolution of \ot\ derived from the observed spectral energy distribution.  (A) Bolometric luminosity.  (B) Blackbody temperature.  (C) Photospheric radius.  (D) Inferred expansion velocity.  Individual points represent blackbody fits performed at discrete epochs to which the observed photometry has been interpolated using low-order polynomial fits. Dashed lines represent an independent Markov-Chain Monte Carlo fit without directly interpolating between data points (see \cite{sot} for methodology and best-fit parameter values). The solid red lines (in A and B) represent the results of a hydrodynamical simulation of the cocoon model where the UVOIR emission is composed of (in A) cocoon cooling (yellow dashed line labeled 1), fast macronova ($>$0.4$c$; green dashed line labeled 2), and slow macronova ($<$0.4$c$; blue dashed line labeled 3).
\label{fig:lc}}
\end{figure*}

\begin{figure*}[!hbt]
\centering
\includegraphics[width=\textwidth]{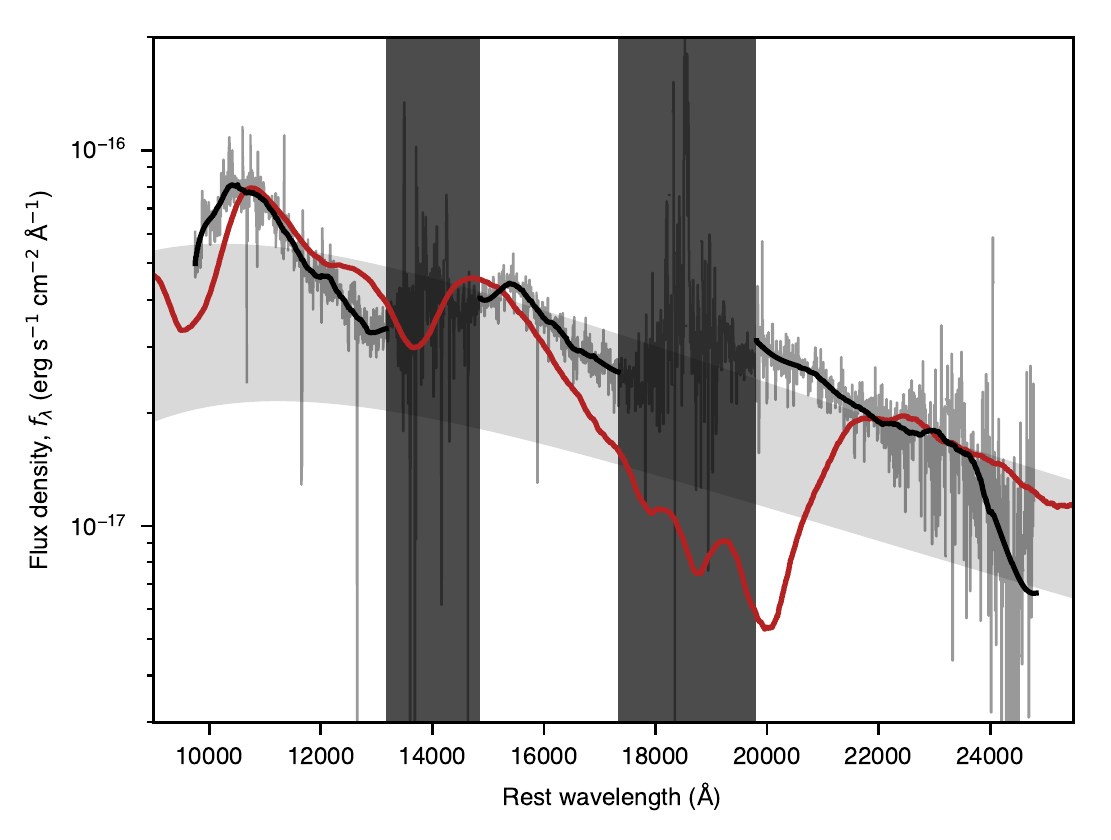}
\caption{
Near-infrared spectrum of \ot\ at 4.5 days after merger. For display purposes, the data have been smoothed using a Savitzky-Golay filter (solid black line), and the unfiltered data are shown in grey. 
A predicted model macronova spectrum \cite{Barnes2013} assuming an ejecta mass of $M_{\mathrm{ej}} = 0.05 M_{\odot}$ and a velocity of $v = 0.1c$ at a phase of 4.5 days post merger is shown in red. The spectra have been corrected for Milky Way extinction assuming reddening $E(B-V) = 0.1$ \cite{sot}. Regions of low signal-to-noise ratio from  strong telluric absorption by the Earth's atmosphere between the near-infrared $J$, $H$, and $K$ spectral windows are indicated by the vertical dark grey bars. The light grey shaded band is the blackbody which best fits the photometric measurements at 4.5 days \cite{sot}. 
\label{fig:spectra}}
\end{figure*}

\begin{figure*}[!hbt]
\centering
\includegraphics[width=\textwidth]{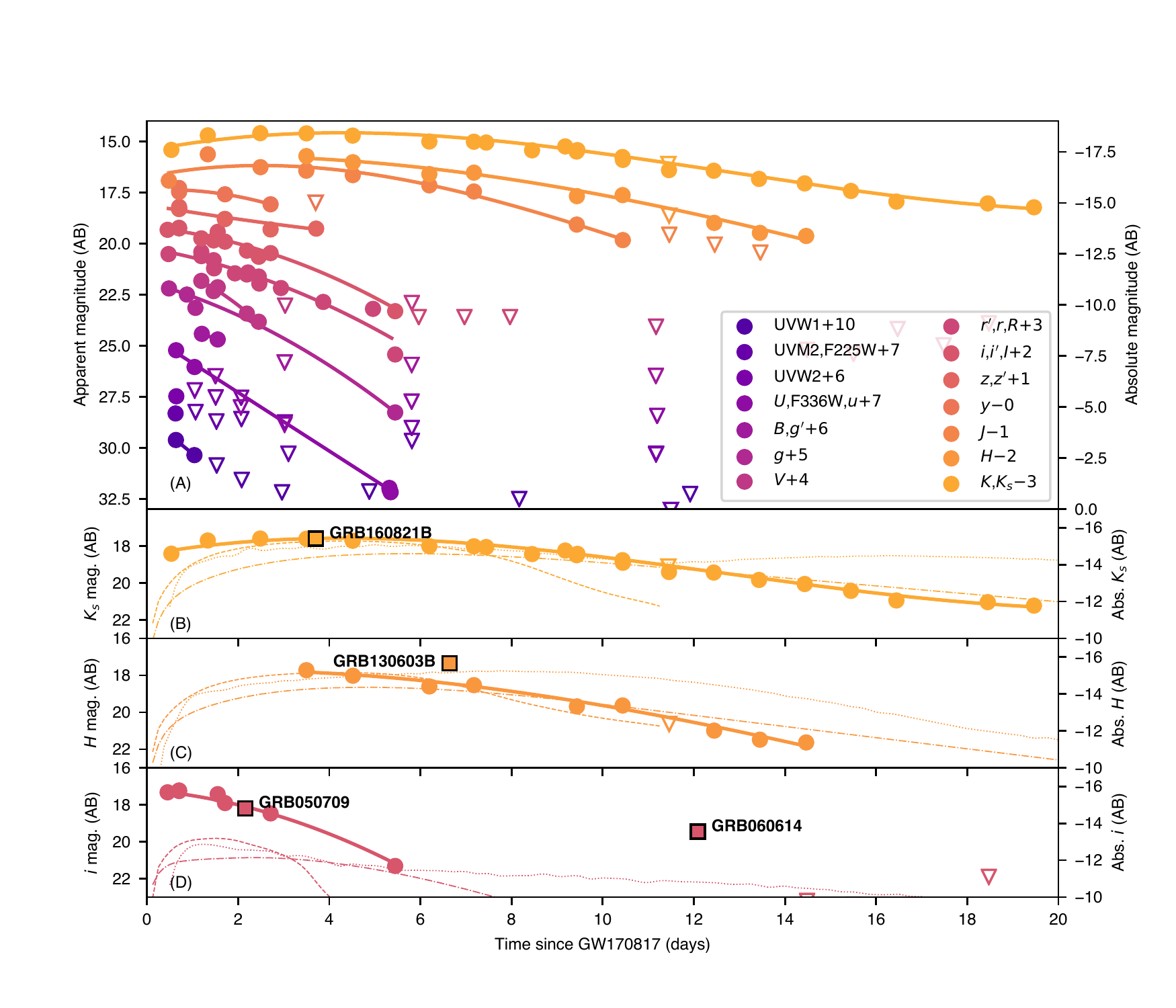}
\caption{
Lightcurves of \ot.  (A) Multi-wavelength lightcurve based on the ultraviolet/optical/near-infrared photometry of \ot\ (Table~\ref{tab:phot} and \cite{sot})
plotted as AB magnitude vs.\ time since merger, with open triangles
indicating 5$\sigma$ upper limits, colored by wavelength.  (B--D)
$K_s$, $H$, and $i$-band lightcurves of \ot\ with literature macronova
model lightcurves, which show a good match in the infrared but fail to
produce the observed blue emission.  For all lightcurves we plot both apparent magnitude and absolute magnitude assuming a distance of 40\,Mpc.  Detections are shown as circles, upper limits as triangles.  The models have been scaled
  to a distance of 40\,Mpc and reddened with $E(B-V)=0.1$ \cite{sot}. The model lightcurves are the following: $M_{\rm ej}=0.05 M_{\odot}, v_{\rm ej} = 0.1c$ from \cite{Barnes2016}, model N4 with the DZ31 mass formula from \cite{Rosswog2017} and $\gamma$A2 at a viewing angle of $30^{\circ}$ from \cite{Wollaeger2017}. Optical and near-infrared observations of previously observed short GRBs which appeared abnormally bright are shown as squares (scaled to 40\,Mpc and corrected for time dilation). GRB080503 \cite{Perley2009} would have had to be at a redshift of 0.22 to be consistent. GRB\,060614 \cite{Yang2015} is too luminous at late times. The excess emission noted in GRB\,160821B \cite{Kasliwal2017}, GRB\,130603B \cite{Tanvir2013} and GRB\,050709 \cite{Jin2016} appear to be similar to \ot. 
\label{fig:lightcurve_comparison}}
\end{figure*}

\begin{figure*}[!hbt]
\centering
\includegraphics[width=\textwidth]{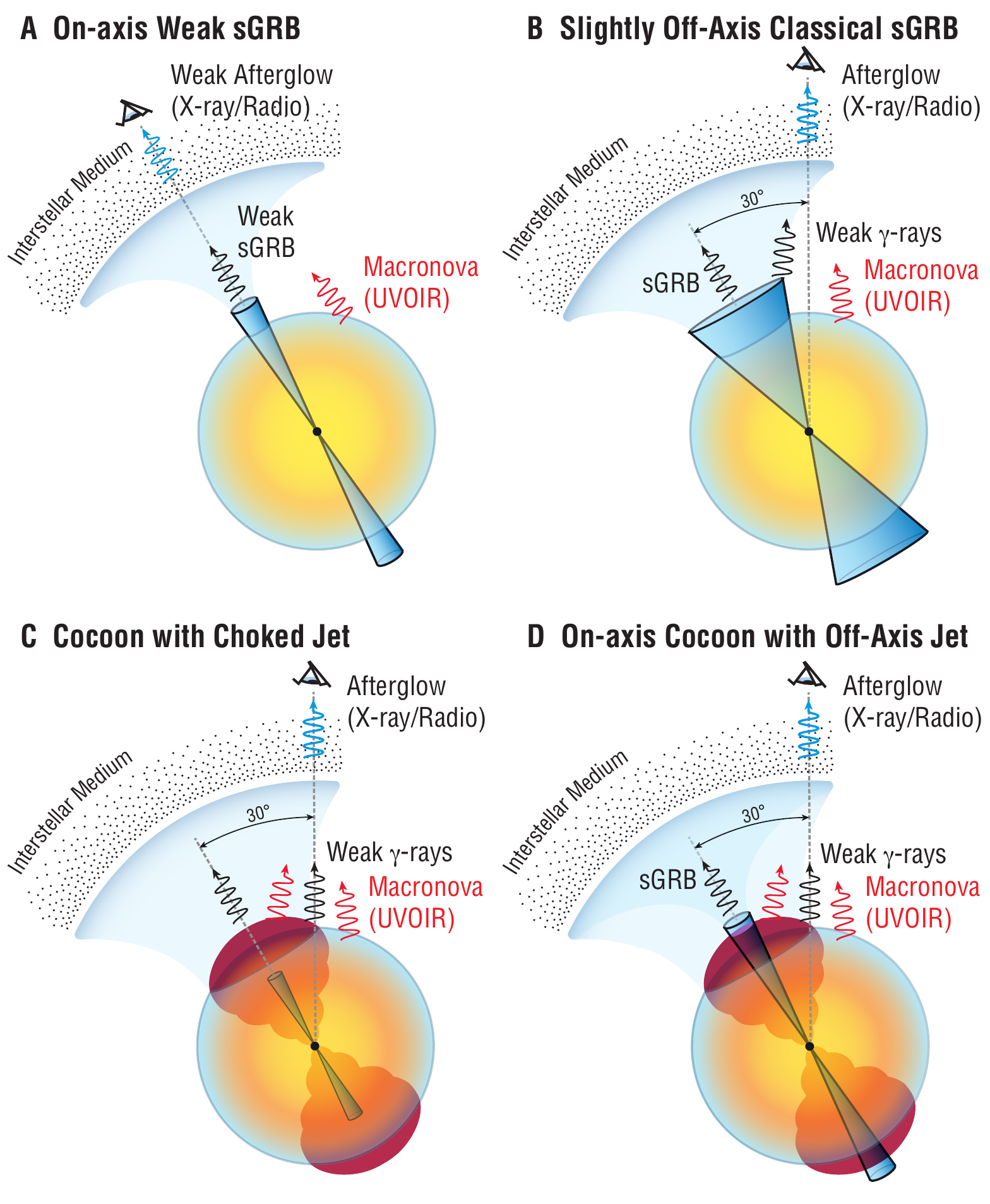}
\caption{  
Model schematics considered in this paper. In each panel, the eye indicates the line of sight to the observer.
(A) A classical, on-axis, ultra-relativistic, weak short gamma-ray burst (sGRB). 
(B) A classical, slightly off-axis, ultra-relativistic, strong sGRB.
(C) A wide-angle, mildly-relativistic, strong cocoon with a choked jet. 
(D) A wide-angle, mildly-relativistic, weak cocoon with a successful off-axis jet. 
\label{fig:schematic}}
\end{figure*}

\begin{figure*}[!hbt]
\centering
\includegraphics[width=0.9\textwidth]{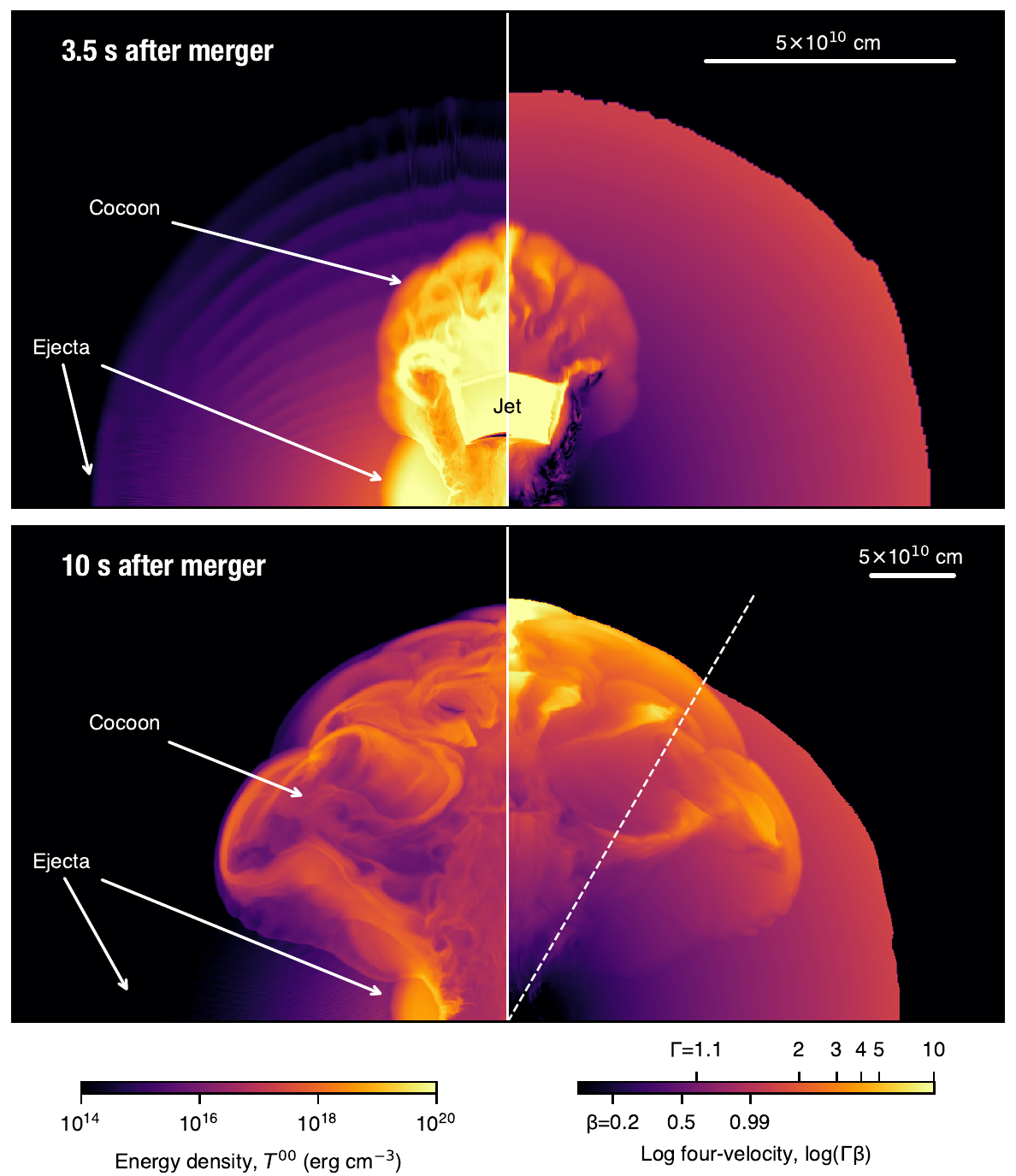} \\
\caption{
Snapshots from a hydrodynamic simulation of a cocoon generated by a choked jet with emission consistent with \ot\  (see \cite{sot} for details). The left half-plane is color-coded by logarithmic energy density (erg cm$^{-3}$) and depicts the energetics.  The right half-plane is color-coded by logarithmic four-velocity ($\Gamma\beta$) and depicts the kinematics. The observer is at an angle of 40$^{\circ}$, the ejecta mass is 0.1\,M$_{\odot}$ and the jet luminosity is 2.6$\times$10$^{51}$ erg s$^{-1}$. Based on this simulation, a bolometric light curve is calculated and shown in Figure~2.
(A) This snapshot is taken at 3.5\,s, shortly after the jet injection stops. The jet is fully choked by 4s. 
(B) This snapshot is taken at 10\,s when the cocoon breaks out.  The breakout radius is 2.4$\times$10$^{11}$\,cm which corresponds to 8 light-seconds. Thus, the delay
between the observed $\gamma$-ray photons and the NS-NS merger is the difference in these times, 2\,s. The Lorentz factor of the shock upon breakout is between 2 and 3.
\label{fig:hydro}}
\end{figure*}

\clearpage
\setcounter{figure}{0}    

\renewcommand{\thefigure}{S\arabic{figure}}
\renewcommand{\thetable}{S\arabic{table}}
\renewcommand{\theequation}{S\arabic{equation}}

\section*{Materials and Methods}
\subsection*{Data Reduction}
Below we describe the ultraviolet/optical/near-infrared photometric and spectroscopic observations, along with the relevant data reduction.  The photometric results from our observations as well as the literature are collected in Table~\ref{tab:phot}.  A log of our spectroscopic observations is provided in Table~\ref{tab:spec}. 



\subsubsection*{Gemini-S FLAMINGOS-2 near-infrared imaging}
We observed \ot\ fifteen times between 2017~August~18 and
2017~September~15 with the Florida Multi-Object Imaging Near-Infrared
Grism Observational Spectrometer (FLAMINGOS-2) imaging
spectrograph \cite{F2} on the 8.1-m Gemini-S Telescope on Cerro
Pach\'{o}n in Chile (PI Singer, Programs GS-2017B-DD-1,
GS-2017B-DD-6).  Near-infrared  $JHK_s$ imaging data were reduced
using standard procedures and calibrated relative to sources from the
Two Micron All-Sky Survey (2MASS)  \cite{2006AJ....131.1163S}. Flat
fields at each filter position were taken each night of FLAMINGOS-2
observations using the Gemini Facility Calibration Unit (GCAL). Dark
frames with identical exposure times to the flat fields and science
frames were also taken at least every other observing night. Median
sky images were produced from a series of dithered science frames and
then subtract to remove atmospheric  OH and thermal emission. On nights with poor observing conditions where the sky varied on timescales faster than a full dither series, a subset of the science frames was selected to create the sky frame. The sky-subtracted science frames were then aligned using the centroid of the bright core of the galaxy \ngc\ and then co-added. 

Three 2MASS stars (2MASS~J13094767$-$2321594,
2MASS~J13094842$-$2323499, and 2MASS~J13094767$-$2321594)
in the vicinity of \ot\ with the highest photometric quality flag (“A”) in all $JHK_s$ filters were used for flux calibration for all of the FLAMINGOS-2 observations. These three bright standard stars were consistently within the field of view of all the FLAMINGOS-2 observations and thus provide robust calibrators for measuring variability from \ot.  

Due to its proximity to the core of the host galaxy \ngc, photometric
measurements of \ot\ are contaminated by the extended galaxy emission
unless the galaxy is properly subtracted. In order to subtract the
extended host galaxy emission, we perform a median filter subtraction
of each image before measuring the flux from \ot.  Median filtering
removes small spatial scale features (i.e. point sources) and returns
the extended galaxy emission as well as large spatial scale background
structures. We then subtract the median-filtered image from the
original to remove the extended emission. To avoid
self-subtracting emission from \ot\  we set the dimensions of the median filter to $\sim5$ times the full-width at half maximum (FWHM) of the point-spread function (PSF) as measured from the 2MASS calibrator stars for each image. 

Depending on the signal-to-noise ratio of \ot, aperture or PSF-fitting photometry was
performed. PSF-fitting photometry is preferred due to possible
residual emission from the host galaxy \ngc\ overlapping with the
projected position of \ot\ even after the median filter
subtraction. However, as \ot\ faded the PSF fitting failed to yield
robust photometry (SNR $> 5$) and thus aperture photometry was
used. For aperture photometry, the inner radius of the aperture was
set to the FWHM of the PSF as measured from the bright 2MASS
calibrator stars. Emission from the sky and residual \ngc\ background at
the position of \ot\ was estimated from a circular annulus centered
on \ot\ with an inner and outer radius of 1 and 2 times the FWHM,
respectively. 



\subsubsection*{Palomar WIRC near-infrared imaging}
We observed  \ot\ on 2017~August~27 in the $K_s$-band  using the Wide
Field Infrared Camera (WIRC; \cite{2003SPIE.4841..451W}) on the
200-in.\ Hale Telescope at Palomar Observatory (P200). Due to the low
declination of \ot\ and proximity to the Sun at the time of these
observations, we observed the target before $12\degr$ twilight at
airmass $> 7$ (elevation $<10\degr$). To deal with the bright and rapidly varying sky background we used the minimum exposure time for the WIRC detector of $0.92$\,s with 8 coadds, allowing us to dither the telescope every $\sim 10$\,s. We obtained 4 well-dithered frames before reaching the elevation limit of the telescope, for a total integration time of 29.44\,s. Individual frames were dark-subtracted and flat-fielded. A median sky frame was constructed from the dithered sciences images, scaled to the sky level in each individual frame and subtracted to remove the bright background. The individual frames were then aligned and coadded using centroid measurements of 4 well-detected stars, and an astrometric solution was found by comparing to 2MASS. The photometric zero point was also determined by comparing aperture photometry of these stars to their 2MASS magnitudes. 

\subsubsection*{IRSF near-infrared imaging}
We observed \ot\ on 2017~August 23, 24,  26, and 28   with the
Simultaneous Infrared Imager for Unbiased Survey (SIRIUS) 
near-infrared ($JHK_s$)  camera \cite{2003SPIE.4841..459N} installed
on  the 1.4-m telescope InfraRed Survey Facility (IRSF)  telescope at the
South African Astronomical Observatory in Sutherland, South Africa (PI
Barway, Program Barway-2017-01-IRSF-57).  We obtained 10 dithered
exposures of 30\,s each with dithering radius of 60\,arcsec per
observing sequence. This was repeated  seven to eight
times to obtain good SNR. Twilight flat frames were obtained before
and after the observations. Dark frames were obtained at the end of
the nights. The data reduction steps which include dark frame
subtraction, flat-field correction, sky-subtraction, dither
combination and astrometric calibration were carried out using the
SIRIUS data reduction pipeline software \cite{irsf}. Similar to the FLAMINGOS-2 image processing, we perform a median filter subtraction on the IRSF images to remove the extended galaxy emission before measuring the flux from \ot.

\subsubsection*{ANDICAM near-infrared imaging}
We observed \ot\ on 2017~August~24--28 with A Novel Dual Imaging
CAMera (ANDICAM) instrument \cite{2003SPIE.4841..827D} mounted on the 1.3-m telescope at Cerro Tololo Inter-American Observatory (CTIO) in Chile (PI Cobb, Program NOAO 2017B-0160). 
On 2017~August~24, a total of 16 individual 45\,s $K$-band frames were
obtained by imaging at 4 different dither positions at each of 4
slightly offset telescope pointings. Each observation from 2017~August
25 to 28 consisted of a total of 20 individual 90\,s $K$-band frames
obtained by imaging at 5 different dither positions at each of 4
slightly offset telescope pointings. After $2\times 2$ binning and
flat-fielding of the individual frames, sky frames were produced at
each dither position by median-combining all images taken at the same
dither position (but with different telescope
pointings). Corresponding dither position sky frames were subtracted
from each image with rescaling to compensate for variability in
background brightness. All sky-subtracted images were aligned and
averaged to produce a single master $K$-band image for each
epoch. Finally, we perform a median filter subtraction  to remove the extended galaxy
emission before measuring the flux from \ot. 

\subsubsection*{Apache Point Observatory near-infrared imaging}
We observed \ot\ in the near-infrared ($K_s$-band) on 2017~August~27  using the Near-Infrared Camera \& Fabry-Perot Spectrometer (NICFPS; \cite{2003SPIE.4841..367V}) instrument on the Apache Point Observatory 3.5-m telescope (PI Chanover, Program 2017 Q3DD04).  Observations were acquired towards the end of evening twilight at high airmass (6--8) through highly variable, partially cloudy conditions.   Forty 6\,s exposures were acquired, alternating between the source and a position $5^\prime$ away using a dither pattern to move the target to different parts of the detector.  Because of rapidly varying clouds, only frames showing stars were used in the image processing.  Unregistered frames were combined to form a sky image which was subtracted from each on-source frame.  On-source frames were registered using the $K_s=9.75$\,magnitude star 2MASS~J13094158$-$2323149 $1.4^{\prime}$ away from \ngc\ and combined into a final image using a median filter after adjusting for the variable background level using the mode of each image.   Photometric calibration used the same star as that used for registration.  

\subsubsection*{VLT/VISIR mid-infrared imaging}
\ot\ was observed \cite{visir} using the Very Large Telescope (VLT) spectrometer and imager for the mid-infrared (VISIR) \cite{Lagage2004} at the Cassegrain focus of Unit Telescope 3 (UT3) on 2017 August~23, 2017 August~31, 2017 September~01, and 2017~September~06 with the J8.9 filter (central wavelength $8.72\,\mu$m). 
Total on-source integration times were 44.8, 17.5, 12.2, and 44.8 minutes, respectively. 
Chopping and nodding in perpendicular directions with $8^{\prime \prime}$ amplitudes were used to remove the sky and telescope thermal background.

Observations of \ot\ and standard stars observed on the same night were reduced following the imaging data reduction processes outline in the VISIR Pipeline User Manual: chop and nod frame subtraction, source detection, and shift and add. However, since we did not detect \ot\ or any other point-source  in the field of view, the chopping and nodding amplitudes and directions provided in the image headers were used to shift the images before coadding. As a test, we performed the same ``blind” shift and add procedure on the images of the standard stars and found that there were negligible differences in the image quality  from the properly coadded standard star images.
Calibration was performed based on mid-infrared standard stars
obtained within same night using the standard-star flux catalog for
VISIR imaging filter based on \cite{Cohen1999}.


 

\subsubsection*{Gemini-S GMOS optical imaging}
We observed \ot\ in the optical ($g$- or $i$-band) several times
between 2017~September~1 and 2017 September~5 with the Gemini
Multi-Object Spectrographs (GMOS; \cite{GMOS,GMOSccd}) at the Gemini-S
observatory.  The data were reduced and coadded using
the \textsc{GMOS}  and  \textsc{GEMTOOLS} modules
in \textsc{PyRAF}.  \ot\ was not detected in any of these data.
Limiting magnitudes (listed in Table~\ref{tab:phot}) were calculated
based on aperture photometry with photometric calibration done
relative to the Pan-STARRS PS1 catalogs \cite{ps1} after utilizing the
median subtraction technique described in the FLAMGINOS-2 imaging
subsection.

\subsubsection*{\textit{HST} ultraviolet imaging and astrometric referencing}
We observed \ot\ in the  ultraviolet on 2017~August~22 and 2017~August~29 with the \textit{Hubble Space Telescope} (\hst) Wide Field Camera 3 (WFC3) using the UVIS detectors (PI Kasliwal, Program HST-GO-15436).  The first epoch used the F225W and F336W filters ($2359\,$\AA\ and $3355\,$\AA\ central wavelengths, the latter similar to $U$-band) while the second used the F275W ($2704\,$\AA\ central wavelength) and F336W filters.
We calculated PSF magnitudes with the software package \textsc{DOLPHOT}\ (v.2.0) \cite{Dolphin00}. \ot\ was only detected in the 2017~August~22 F336W image (see figure~\ref{fig:hst}).  For the other images we calculated 5-sigma limiting magnitudes based on aperture photometry (see Table~\ref{tab:phot}).  

We registered the 2017~August~22 stack of F336W images with
the \textit{Gaia} astrometric catalog \cite{gaiaDR1} to derive a more
precise position of \ot.  With 22 matched \textit{Gaia} sources the
refined astrometric solution has root-mean-square residuals of
$0.05^{\prime \prime}$ and we measure the position of \ot\ to be
(J2000) Right Ascension $13^{\rm h}09^{\rm m}48.071^{\rm s}$,
Declination $-23\degr22^{\prime}53.37^{\prime \prime}$.



\subsubsection*{Gemini-S FLAMINGOS-2 near-infrared spectroscopy }
We observed \ot\ on 2017~August~22 and 2017~August~29 with the
FLAMINGOS-2 spectrograph \cite{F2} on the 8.1-m Gemini-S Telescope on
Cerro Pach\'{o}n in Chile. For the first epoch we used the 3-pixel
($0.54^{\prime \prime}$ wide) slit and obtained spectra with both the
JH and HK grism/filter sets to achieve a spectral resolving power of $600$ across the 1.0--2.4\,$\mu$m spectral range. Using an exposure time of 150\,s, we obtained a sequence of  6 frames in the JH setup and 8 frames in the HK setup. The target was nodded along the slit between frames in an ABBA pattern to allow for accurate subtraction of the sky background. Baseline calibrations were also obtained, including dark frames, spectral flat fields, Ar arc lamp spectra for wavelength calibration, and observations of the A0V star HIP 69718 immediately after the science observations at similar airmass.  For the second epoch we increased the total integration time to 16 frames of 150\,s each using the HK setup.

Dark current subtraction, flat-fielding, sky background subtraction,
coadding of the 2D spectra, wavelength calibrations, and 1D
extractions were performed using standard tasks in the
Gemini \textsc{IRAF} package. Telluric corrections and flux calibrations were performed using the \textsc{IDL}
tool \textsc{xtellcor} \cite{2003PASP..115..389V}. A model spectrum of Vega was used to match and remove the hydrogen lines of the Paschen and Brackett series from the A0V telluric standard and construct a telluric correction spectrum free of stellar absorption features. The resulting telluric correction spectrum was also used for the initial flux calibration. The absolute flux calibration was then found by comparing synthetic photometry derived from the spectra to the $H$-band photometric measurement of the transient at the same phase.
In the second epoch, despite the increased exposure time we
did not detect \ot. 


\subsubsection*{Gemini-S GMOS optical spectroscopy}
\label{sec:gmos_spec}
We observed \ot\ on 2017~August~20 and 2017~August~21 using the Gemini
Multi-Object Spectrographs (GMOS; \cite{GMOS,GMOSccd}) on the 8.1-m
Gemini-S Telescope on Cerro Pach\'{o}n in Chile with the R400 and B600
gratings. The spectra were reduced using the Gemini \textsc{IRAF}
package \cite{gemini}. The standard star EG274 was observed on
2017~August~20 and used to flux-calibrate spectra from both of the
observed epochs. Both spectra show largely featureless continua. The
observations taken on 2017~August~21 show a clear trace on both the
blue and red sides of the detectors. The trace on the blue side of the
observation on 2017~August~20 was too weak to extract.

\subsubsection*{Keck LRIS optical spectroscopy}
We observed \ot\  on 2017~August~25 using the Low-Resolution Imaging
Spectrometer (LRIS; \cite{1995PASP..107..375O}) on the 10-m Keck I
telescope.  The visibility of the target  was
poor and the observations were conducted under non-optimal conditions
at the end of evening twilight, with the airmass in the direction of
the target increasing from 3.8 at the start of the sequence to $\sim$6
at the end of the final exposure.  We used the D560 dichroic to split
the light between the red and blue sides, and used the 400/8500
grating on the red side and the 400/3400 grism on the blue side.  The
observations were processed using the analysis
pipeline \textsc{LPIPE} \cite{lpipe} and summed.  A weak trace is visible at the
transient location on the red-side camera.  Our low-signal-to-noise ratio extraction
of the spectrum (figures~\ref{fig:lrisspec} and \ref{fig:lrisspec2})
shows a featureless red continuum with no significant features (in
particular, no broad or narrow emission lines), although we note that
a clean extraction is complicated by the significant contamination
from the spatially-varying host galaxy continuum.

\subsection*{Bolometric Light Curve Construction}
\label{sec:lc}
We combined  available ultraviolet, optical, and near-infrared
photometric data (including our data along with data published in the LIGO/Virgo collaboration (LVC)
Gamma Ray Coordinates Network (GCN) circulars; Table~\ref{tab:phot})
on \ot\ to build a bolometric light curve using two independent approaches illustrated in figure~\ref{fig:lc}.   All analysis used a distance of 40\,Mpc \cite{2001ApJ...553...47F,2000ApJ...529..698S} and a reddening of $E(B-V)=0.1$\,mag \cite{schlafly11}.

Our first method uses low-order polynomials to enable interpolation of
the photometry in the individual filters. A blackbody is then fitted
to the interpolated photometry for a set of discrete epochs starting
from 0.49~d after GW170817 (0.04\,d after the first $i$-band
detection) up until 12.9\,d after GW170817. Second, we fit the data
with an evolving blackbody model by assuming a functional form for the
time-dependence of the blackbody temperature $T(t)$ and the radius
$R(t)$.  After experimenting with different models we found the best
functional forms to be $R(t)=R_0(1-e^{-\lambda_R t}) + vt$ (a model that
initially decelerates and then coasts) and $T(t)=T_0 t^{\alpha_T}$,
where $R_0$ is a reference radius, $v$ is an expansion speed,
$\lambda_R$ is a deceleration rate, $T_0$ is a reference temperature at
time $t=0$, and $\alpha_T$ is a temperature power-law exponent.
  The data were fit by computing the monochromatic flux density as a function of time and wavelength appropriate for each observation (corrected for extinction assuming $A_V=0.31$\,mag, and using the reddening law of \cite{1989ApJ...345..245C} when no instrument-specific reddening data was available) and comparing with the data; no interpolation or binning was used on the data themselves.  The parameter fitting was done using a Markov Chain Monte Carlo method \cite{2013PASP..125..306F}: we used 80 walkers, ran for 50 iterations to achieve ``burn-in'', and then ran for a further 1000 iterations, only using every 41st value to account for finite autocorrelation in the resulting chains. We then determined the best-fit parameter values and uncertainties from the posterior distributions.  We restricted the fit to times $\leq 12.5\,$d after GW170817, since after that time the multi-wavelength coverage was significantly diminished and we typically only had detections in a single band ($K_s$).  Our best-fit values and uncertainties (68\% confidence limits) are $R_0=24000 \pm 3000\,R_\odot$, $v=2500 \pm 300\,R_\odot\,{\rm d}^{-1} = 20000\pm3000\,{\rm km\,s}^{-1}$, $\lambda_R=0.42 \pm 0.05\,{\rm d}$,
$T_0= 6050\pm50 \,$K, and $\alpha_T=-0.62 \pm0.01 $, with $t$ in days.  A model with $\lambda_R$ finite (i.e., initially decelerating) is preferred over a model with constant velocity: $\chi^2$ is reduced by 600 with roughly 140 degrees-of-freedom.

Overall both approaches give us consistent results for effective
temperature, photospheric radius and kinematics (see main text and
figure~\ref{fig:lc}).  With neither approach do we compute a formal
goodness-of-fit, as our data-set is too inhomogeneous.  We are drawing
data from 24 different telescopes, with  many photometric systems,
filter transformations, extinction coefficients, and zero-points, not
all of which are fully documented.  There are also significant
variations in calibration methodology and host-galaxy subtraction, all
of which can affect the resulting photometry.  We have manually
removed the data where host subtraction was known to be problematic
but do not perform quantitative model evaluation (e.g., for deviations
from blackbodies).

\section*{Supplementary Text}

\subsection*{Census of the Local Universe Galaxy Catalog}
\label{sec:CLU}
We used a galaxy catalog to narrow down our
candidate transients within the gravitational wave trigger volume and provide basic and physical properties for each of the galaxies within that volume. We find that the galaxies in the 90\% volume span a large range of properties containing low-mass dwarfs as well as larger galaxies (spirals and ellipticals). 

The  Census of the Local Universe (CLU) galaxy catalog consists of two
parts: a compilation of known galaxies from many sources; and new
galaxies (i.e., those with no previous distance information) from a
four-filter, narrowband survey designed to find redshifted H$\alpha$
emission out to 200\,Mpc (based on conservative predictions for the
horizon of NS-NS detections with  LIGO/Virgo
\cite{2006ApJ...650..281N, 2016ApJ...832L..21A})  across $\approx
3\pi$ of the northern sky. In this case, the gravitational wave localization was confined to a  declination (near $-25\degr$) below the southern limit of the narrowband survey ($\delta\geq-20\degr$), so only the compiled galaxy catalog overlapped with the GW event and will be described here.

The compiled portion of the CLU galaxy catalog
\cite{2011PhDT........35K} was taken from existing galaxy databases:
NASA/IPAC Extragalactic Database (NED), Hyperleda \cite{hyperleda},
the Extragalactic Distance Database (EDD; \cite{EDD}), the Sloan
Digital Sky Survey data release 12 (SDSS DR12; \cite{sdss12}), and the
Arecibo Legacy Fast Arecibo L-band Feed Array survey (ALFALFA; \cite{haynes11}). The catalog contains $\sim$260,000 galaxies with  spectroscopic distances less than 200\,Mpc. The current version of the catalog contains updates from both NED and SDSS in early 2016. Distances based on Tully-Fisher methods were favored over kinematic (i.e., redshift) distances because of the large contribution of peculiar velocities in this local sample; however, the majority of the distances are based upon redshift information. 

In addition to distances, the catalog also contains compiled
multi-wavelength photometric information. We have cross-matched the
CLU catalog with \textit{Galaxy Evolution Explorer} (\galex) all sky
catalog \cite{galex}, the \textit{Wide-field Infrared Survey Explorer}
(\wise) all sky catalog \cite{wise}, and SDSS DR12 \cite{sdss12}  to
obtain fluxes from the ultraviolet (UV) to the infrared (IR), with
$\sim$104,000 matches for the \galex\ far-ultraviolet (FUV, 1500\,\AA)
band, $\sim$226,000 for the \wise\ 3.4 and 22\,$\mu$m bands, and
$\sim$114,000 for the SDSS $r$-band.

We  spatially cross-matched the CLU galaxy catalog to the 90\%
containment volume of the LIGO/Virgo trigger \cite{LVCC21513} reported
by the BAYESian TriAngulation and Rapid localization (BAYESTAR) probability sky map \cite{bayestar} with no assumption on orientation and found 49 galaxies. In addition, we have used the \galex\ FUV and \wise\ 3.4 and 22\,$\mu$m bands to measure the physical properties of these galaxies. The star formation rates (SFRs) are derived from \galex\ FUV Kron-magnitudes \cite{murphy11} after correcting for Milky Way extinction \cite{schlafly11} as tabulated by NED and internal dust extinction using a combination of observed \galex\ FUV and \wise\ 22\,$\mu$m fluxes \cite{hao11}. The stellar masses (M$_{\star}$) are derived from 3.4\,$\mu$m ALLWISE fluxes and a mass-to-light ratio of 0.5\,$M_\odot/L_\odot$ \cite{mcgaugh15}.

Table~\ref{tab:clumatch} lists the basic properties of the 49 galaxies (sorted by stellar mass) along with the UV and IR fluxes, SFRs, stellar masses, and the probabilities in the containment volume \cite{LVCC21535}. The galaxies span ranges of stellar mass ($10^7\,M_\odot$ to $10^{10.5}\,M_\odot$)  and SFR ($10^{-2.6}\,M_\odot\,{\rm yr}^{-1}$ to $10^{0.5}\,M_\odot\,{\rm yr}^{-1}$) suggesting that the list contains dwarfs as well as larger more massive galaxies.

%
%


\subsection*{Host Galaxy \ngc}
\subsubsection*{Basic Properties}
\begin{figure}
\includegraphics[width=\textwidth, angle=0]{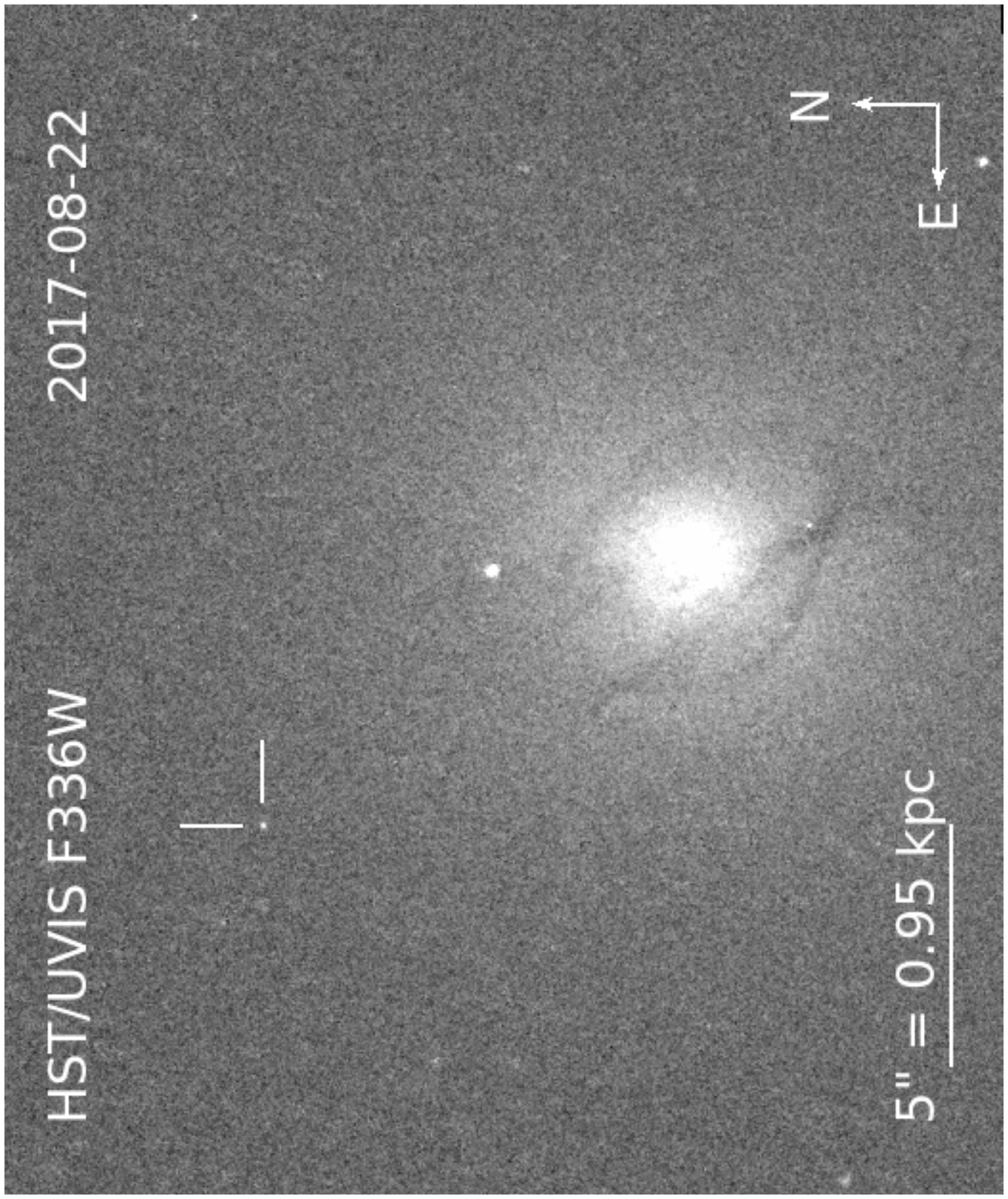}
\caption{\emph{Hubble Space Telescope} WFC3/F336W ultraviolet image of \ot\ and \ngc, taken 2017~August~22.  North is up, east is the to left, and a $5^{\prime \prime}$ scale-bar is indicated.  The position of the transient is shown with tick marks.  Dust lanes are visible toward the center of \ngc.}
\label{fig:hst}
\end{figure}

\begin{figure}
\includegraphics[width=1.0\textwidth, angle=0]{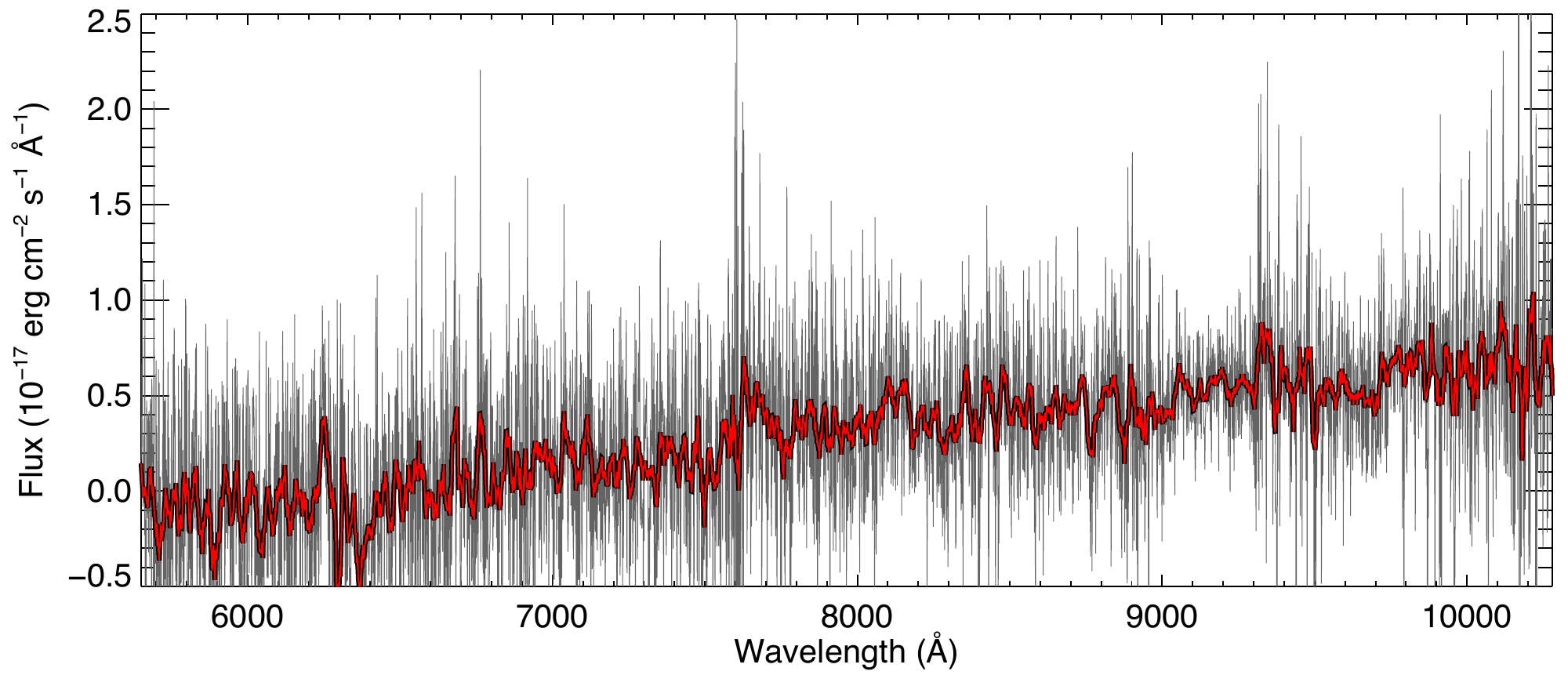}
\caption{Optical spectrum of \ot\ taken 2017~August~25 with the Low Resolution Imaging Spectrometer on Keck I.  The measured spectrum is shown in gray, and a version smoothed with a kernel of width 20\,\AA\ is shown in red.  \ot\ was already quite faint at this time and no unambiguous features are evident.}
\label{fig:lrisspec}
\end{figure}

\begin{figure}
\includegraphics[width=1.0\textwidth, angle=0]{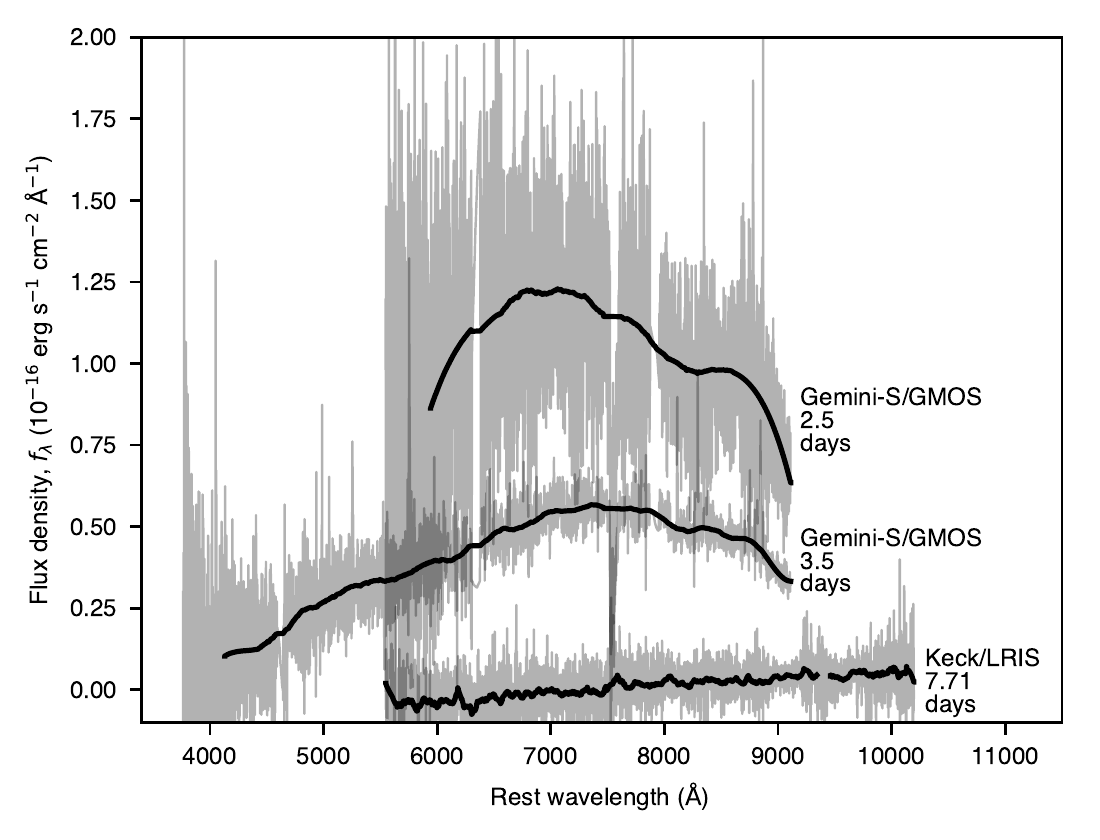}
\caption{Optical spectral sequence of \ot\ including the Gemini-S/GMOS spectra from 2.5 and 3.5 days and the Keck/LRIS spectrum from 7.7 days. The measured spectra are shown in gray, and versions smoothed with a Savitsky-Golay filter are shown in black.}
\label{fig:lrisspec2}
\end{figure}

The EM counterpart for GW170817, \ot, was reported near \ngc, the
third most massive of the galaxies cross-matched to the LIGO/Virgo
trigger by the Census of the Local Universe project \cite{LVCC21535} (\ngc\ was not present in the top 15 galaxies  in
the Galaxy List for the Advanced Detector Era catalog
\cite{2016yCat.7275....0D} and reported in \cite{LVCC21516}, although
it is present in their on-line catalog). \ngc\ has been classified as
an E-S0 galaxy with a morphological T-type of $-3$ \cite{capaccioli15} at a
distance of roughly 40\,Mpc, based on Tully-Fisher measurements of
other galaxies in its group
\cite{2001ApJ...553...47F,2000ApJ...529..698S}. Archival \hst/Advanced
Camera for Surveys (ACS) F606W data (Program ID: 14840; PI Bellini) show
 complicated dust lanes extending a few kpc from the nucleus \cite{LVCC21536,LVCC21669} superimposed on an elliptical galaxy, similar to 
many other early-type galaxies. The dust lanes may be the result of a minor merger that occurred as long as several Gyr ago. 
Note that the dust lanes do not extend across the position of \ot\ (figure~\ref{fig:hst}), suggesting \ot\ is not located in a heavily-obscured region. This is echoed by examination of spatially-resolved spectra, which show no emission-lines within $\pm 5^{\prime \prime}$ ($\pm 1\,$kpc) of the transient, and our \hst\ UV images which  show very little emission at the position of the \ot\ (Figure~\ref{fig:hst}).

Archival optical spectra \cite{2008AJ....135.2424O,2009MNRAS.399..683J} show a
continuum dominated by old stars with a pronounced 4000\,\AA\ break indicative of little if any active star formation; [N{\sc II}], [S{\sc II}], and weak [O{\sc III}] emission lines are present in the nucleus with likely some H$\alpha$ emission filling in the Balmer absorption from the stellar continuum, but the relatively high [N{\sc II}]/H$\alpha$ ratio is suggestive of a
low-luminosity active galactic nucleus (LLAGN) rather than star
formation.

We estimated the Eddington ratio for the central black hole by
computing the bolometric luminosity $L_{\rm bol}$ from the X-ray luminosity
(5.6$\times$10$^{39}$\,erg\,s$^{-1}$; \cite{LVCC21612}) assuming a
ratio of bolometric to X-ray flux of 16 for LLAGNs \cite{2008ARA&A..46..475H}, while the Eddington luminosity $L_{\rm Edd}$
is determined from  black hole mass obtained from the central velocity
dispersion  (163\,km\,s$^{-1}$; \cite{2008AJ....135.2424O}); these
imply $L_{\rm bol}/L_{\rm Edd} = 1.4\times 10^{-5}$, which is similar to ratios of other LLAGNs \cite{2008ARA&A..46..475H}.


\subsubsection*{Constraints on the NS-NS Merger Timescale}
To further constrain the timescale of NS-NS mergers, we investigated star formation histories of \ngc\ by fitting the spectral energy distribution (SED) with the package \textsc{MAGPHYS}.  \textsc{MAGPHYS}\ uses  stellar population syntheses code \cite{2003MNRAS.344.1000B,2007IAUS..241..125B} to provide spectral evolution at
wavelengths from 912\,\AA\ to 1\,mm and at ages between $1\times 10^{5}$ and $2\times10^{10}$\,yr.  We collected photometric data
from the \galex\ (NUV and FUV; \cite{galex}), Pan-STARRS1 ($grizy$; \cite{ps1}), 2MASS
($JHK_{s}$; \cite{2006AJ....131.1163S}), \wise\ (W1, W2, W3, W4; \cite{wise}) and  \textit{IRAS} (60\,$\mu$m; \cite{1984ApJ...278L...1N}) surveys, where we used upper limits for the \galex/FUV and IRAS/60\,$\mu$m bands (Figure~\ref{fig:host-SED}).  The best-fit model gives the stellar
mass of the galaxy $M_{\star}\sim 3\times 10^{10}\,M_\odot$  and the star
formation rate (SFR)  $\sim 3 \times 10^{-3}\,M_{\odot}\,{\rm
  yr}^{-1}$.  There appears to be an offset between the data and the
model in the region near the near-infrared (2MASS $JHK_s$) and
mid-infrared (\textit{WISE}) observations.  This may reflect different
methods of measuring the entire extent of \ngc\ used for the different catalogs.  

For comparison, the SFR derived from \galex/FUV is $\sim 4 \times 10^{-2}\,M_{\odot}\,{\rm yr}^{-1}$, the SFR from \galex/NUV for stars more massive than $5\,M_\odot$ is $\sim 3 \times 10^{-3}\,M_{\odot}\,{\rm yr}^{-1}$ \cite{1998PhDT........40H,1997ApJ...481L...9C}, and the SFR from the \textit{IRAS}/60\,$\mu$m upper limit is $<0.08\,M_\odot\,{\rm yr}^{-1}$ \cite{1992ARA&A..30..575C,LVCC21645}.
Given the varying systematics and uncertainties in both the data and the methods, we consider these to be largely consistent and indicative of small levels of ongoing star-formation, with SFR of 10$^{-3}$ to 10$^{-2}\,M_{\odot}\,{\rm yr}^{-1}$.
This is consistent with estimates based on the non-detection of neutral
hydrogen \cite{2004MNRAS.350.1195M}.  The estimate of the SFR is an upper limit because of the possible contamination of the central LLAGN.

The time since the last burst of star formation ended  is
$\sim$2\,Gyr, implying a relatively long timescale for the merger of
the NS-NS binary system since progenitors of neutron stars are short-lived.
Even a more conservative limit based on the absence of early-type stars in the spectrum puts the last episodes of star-formation more than a Gyr ago.
This is not consistent with previous predictions that the delay time
of NS-NS mergers is short (1--100\,Myr;
\cite{1963PhRv..131..435P,2002ApJ...572..407B}), although there are
Galactic NS-NS binaries with merger times 100\,Myr--10\,Gyr (e.g.,
\cite{2008LRR....11....8L}).  In addition, population synthesis of compact object mergers finds significant numbers of sources produced after Gyr delays, with the progenitor systems formed at high redshift during the peak epochs of star formation (e.g., \cite{2010ApJ...716..615O}).



\subsection*{Models}
In the sections below, we discuss several details of models to explain
the overall electromagnetic emission from \ot.  We estimate the
maximum ejecta mass possible to have a weak, on-axis,
ultra-relativistic jet break out.  We then consider a model where the
observed $\gamma$-rays are produced by an off-axis short $\gamma$-ray
burst (sGRB). We consider a structured jet and a strictly Newtonian
source.  This motivates us to describe the analytical and numerical
details of our preferred model: a shock breaking through a cocoon of
material.  Finally, we consider alternative sources of an
engine-driven wind and free-neutron decay to power the
early-time emission.

\subsubsection*{A Weak On-Axis Jet}
\label{sec:breakout}
The observed $\gamma$-rays carried an isotropic equivalent energy of $\sim 3 \times 10^{46}$\,erg over a duration of $\sim 2$\,s, corresponding to an average luminosity of $\sim 1.5 \times 10^{46}\,{\rm erg\,s}^{-1}$  \cite{GBM2017,Goldstein2017}. sGRB jets produce $\gamma$-rays very efficiently (e.g., \cite{Fong2015}), so if we observe a regular on-axis GRB the total jet isotropic equivalent energy is at most a few times larger than that of the observed $\gamma$-rays.
Therefore, to evaluate if the observed $\gamma$-rays could have been
produced by a weak on-axis jet we estimate the maximum ejecta mass
that a jet with an opening angle $\theta_{\rm j}$ and isotropic
equivalent luminosity $10^{47}\,{\rm erg\,s}^{-1}$ can cross within
2\,s. We approximate the ejecta as being static and spherical, with a
density profile $\rho \propto r^{-2}$, where $\rho$ is the density and
$r$ is the radius. We assume a typical ejecta velocity of $0.2$c, a
breakout time of 2\,s and $r=10^{10}$\,cm. We use a jet propagation
model \cite{bromberg11} which was calibrated numerically
\cite{harrison17} to estimate the breakout time \cite{applet}. We find that for $\theta_{\rm j}=10\degr$ a breakout of 2\,s is achieved for mass $<3 \times 10^{-6}\,M_{\odot}$. For  $\theta_{\rm j}=30\degr$ the upper limit on the mass is lower by about an order of magnitude.

To verify this calculation we carried out a numerical simulation where
a jet is launched into  expanding ejecta. We use an identical setup to
the one described in the main text with a few minor changes as
follows: we choose a simple radial density profile without an angular
component with a power-law index of 3.5 rather than 2, and omit the
extended ejecta. The jet has the same properties as in the main text but with a lower isotropic luminosity of $10^{-47}\,\rm{erg\,s^{-1}}$ and shorter delay time of $ 0.1\,\rm{s} $ between the merger and its launch (a short delay decreases the breakout time). The grid setup is similar, but extends only up to the breakout radius.
We varied the ejecta's total mass until the breakout took place at 2\,s, which happened near a mass of $ 2.5 \times 10^{-6}\,M_\odot$, very close to the analytic prediction.


Observational and theoretical considerations indicate that the amount
of mass that was ejected at high latitudes along the jet path is
higher than $10^{-5}\,M_\odot$ by orders of magnitude. First, the
bright optical emission during the first day implies that about
$0.02~M_\odot$ with a relatively low optical depth (assuming $\kappa \approx 1\,{\rm cm}^{2}\,{\rm g}^{-1}$) were ejected. This type of material is expected to be synthesized after the merger in the high latitude wind that is exposed to a high neutrino flux (e.g., \cite{perego2014}). Second, optical/IR emission indicates that the total ejected mass is at least $0.05 ~M_\odot$, and while this mass is most likely not distributed isotropically, numerical simulations show that all mass ejection processes throw a non-negligible fraction of mass at high latitudes (e.g., \cite{Hotokezaka2013,perego2014,siegel2014}). Moreover, all these simulations find the that the high latitude ejecta mass is larger than about $10^{-3}~M_\odot$. We therefore conclude that the observed $\gamma$-rays are highly unlikely to be produced by an on-axis low-luminosity jet.

\subsubsection*{An Off-Axis Jet}
\label{sec:offaxis}
A potential explanation for the extremely low luminosity of the observed $\gamma$-ray emission of \ot\ is that we observe a regular luminous sGRB, but our line-of-sight is outside of the GRB jet and the low luminosity is due to the lower Doppler boost compared to an on-axis observer. Below we examine the implications of such a configuration.  
	
Consider a jet with an opening angle $\theta_{\rm j}$ and a Lorentz
factor $\Gamma \gg 1$ that radiates $\gamma$-rays. An on-axis observer
sees $\gamma$-ray emission with a total energy $E$, a typical photon
energy $E_p$ and a total duration $\Delta t$.  We are interested  in what  an off-axis observer at a viewing angle $\theta_{\rm obs} >\theta_{\rm j}$ will see. We define the quantity $q=(\theta_{\rm obs}-\theta_{\rm j})\Gamma$ since for an off axis observer $q \gg 1$ and the Doppler boost ratio to an on-axis observer is $\propto q^{-2}$. We denote all the off-axis observables with prime. 

The effect of the Lorentz boost on the photons' energies implies:
\begin{equation}
	\frac{E_p}{E_p'}=q^2
\end{equation}
The observed total isotropic equivalent energy (fluence) ratio has
three regimes, depending on how far the observer is from the edge of the jet and how wide  the jet is compared with $1/\Gamma$:
\begin{equation}\label{eq:A}
  \A\equiv \frac{E}{E'}=\left\{ 	\begin{array}{ll} 
 q^4  ~~~&;~~~ \theta_{\rm obs}-\theta_{\rm j} \ll \theta_{\rm j}~~~(i)\\
&\\
 q^6(\theta_j\Gamma)^{-2}  ~~~&;~~~ \theta_{\rm obs}-\theta_{\rm j} \gg \theta_{\rm j} > 1/\Gamma ~~~(ii)\\
&\\
  q^6  ~~~&;~~~  \theta_{\rm j} < 1/\Gamma ~~~(iii)
\end{array} \right.
\end{equation}
where we define an amplification parameter $\A \equiv E/E'$. To understand this equation it is most convenient to consider first the last case, $\theta_{\rm j} < 1/\Gamma $, since then the jet can be regarded as a point source. If $\theta_{\rm obs}-\theta_{\rm j} \gg \theta_{\rm j} > 1/\Gamma$ then the whole solid angle of the jet ($\theta_{\rm j}^2$) contributes roughly equally to the observed fluence for an off-axis observer, compared to a solid angle $1/\Gamma^2$ that dominates the emission for an on-axis observer, thereby reducing the fluence ratio by a factor of $(\theta_{\rm j}\Gamma)^2$. Finally, when $\theta_{\rm obs}-\theta_{\rm j} \ll \theta_{\rm j}$ the whole jet does not contribute equally to an off-axis observer and the emission is dominated roughly by a solid angle of $(\theta_{\rm obs}-\theta_{\rm j})^2$. 

The observed duration depends on the details. If we assume that each observed pulse has a duration $\delta t$ and that it is generated by an episode of emission that takes place at a radius $r$ over some radii range $\Delta r \sim r$ then $\delta t \approx r/(2c\Gamma^2)$ and
\begin{equation}
	\frac{ \delta t}{\delta t'} = q^{-2}
\end{equation}
If the total on-axis duration is determined not by a single episode, but by the radial length of the jet (as in the case of internal shocks for example) then 
\begin{equation}
	 \Delta t' = \max\{q^2 \delta t,\Delta t\}
\end{equation}

For \ot\ $E' \approx 3 \times 10^{46}\,{\rm erg}$
\cite{GBM2017,Goldstein2017}. If the on-axis observer sees a regular
sGRB then ${\cal A}=10^3-10^6$ with a typical value of $\A=10^4$. A value $\A=10^3$ may be too low due to the requirement that the jet breaks out of the ejecta within 2\,s, as a jet with a luminosity of $10^{50}\,{\rm erg\,s}^{-1}$ can break out on time only if the ejecta mass along its path is $< 10^{-3}M_\odot$ (isotropic equivalent). This is much lower than the total mass we observed assuming an opacity $\kappa \approx 1 {\rm cm^2\,g}^{-1}$  of $\sim 0.02\,M_\odot$ (see main text), which is ejected presumably at high latitudes.

Classical sGRBs show non-thermal spectra and therefore the observed photons are expected to be generated above or near the photosphere. This expectation was one of the main indications that GRB jets are relativistic \cite{guilbert1983,lithwick2001} and it still provides the most robust lower limit on the Lorentz factor of GRB jets.
The optical depth of a relativistic jet $\tau_\Gamma$ that radiates a spectrum with
a power-law and  exponential cutoff, $N_\nu \propto \nu^\alpha
\exp[-h\nu/E_0]$, such as the one seen in \ot, is given by \cite{Nakar2007}:
\begin{equation}\label{eq:tauGamma}
	\tau_\Gamma \approx 10^{13} L_{51} \delta t_{-2}^{-1} \frac{m_e c^2}{E_0} \Gamma^{-(4-\alpha)}\exp\left[ -\frac{\Gamma m_e c^2}{E_0} \right] 
\end{equation}
where $L_{x}$ is the burst luminosity in units of $10^{x}\,{\rm
  erg\,s}^{-1}$, $\delta t_x$ is the single-pulse duration in units of
$10^x$\,s, and  $m_e c^2$ is the electron rest-mass energy. This expression can be written using the off-axis observables and $\A$:
\begin{equation}
	\begin{array}{ll}
	\tau_\Gamma \approx  10^{7} \A L_{47}^\prime \delta t_{0}^{\prime-1} \frac{m_e c^2}{E_0^\prime} \Gamma^{-(4-\alpha)}\exp\left[ -\frac{\Gamma m_e c^2}{E_0} \right]
\end{array} 
\end{equation}
There is still a dependence on the on-axis observed $E_0$ in the exponent of this equation, for which the transformation to the off-axis frame depends on $\A$ differently for each of the three regimes in equation~\ref{eq:A}. Therefore the limit on $\Gamma$ that we derive below is different for each regime.

The requirement that the $\gamma$-ray source is optically thin, i.e.,
$\tau_{\Gamma} < 1$, provides a lower limit on the jet Lorentz factor
$\Gamma(\A)$, which in turn determines the maximal distance the
observer can be from the edge of jet (i.e., $\theta_{\rm obs}-\theta_{\rm j}$) for each value of $\A$. Figure~\ref{fig:the_obs} shows that distance for cases (i) and (iii) which are independent of $\theta_{\rm j}$. The results for case (ii) are similar for all realistic values of $\theta_{\rm j}$. This suggests that for $\theta_{\rm obs} \approx 0.5$\,rad we can never be in the far  regimes, i.e., (ii) or (iii) since  $\theta_{obs}-\theta_{\rm j} \ll \theta_j$ is always true. Figure~\ref{fig:Gmin} shows the minimal Lorentz factor of the jet only for case (i) (near miss). This implies that if the observed $\gamma$-rays are generated by an off-axis jet then it must have $\Gamma > 100$ and the angle between us and the jet's edge cannot exceed $0.1$\,rad.

A luminous jet that the observer missed by only 0.1\,rad is expected
to produce a very bright afterglow roughly a day after the burst. The
blast wave driven into the circum-merger medium, which produces the
afterglow, decelerates to a Lorentz factor $< 10$ by that time and
thus its cone of emission enters the observer's line of sight. Hence,
after about 1\,d the observer sees a regular sGRB afterglow. Limits on
the X-ray and radio rule out this option
\cite{Evans17,Hallinan17}. The only way to avoid a bright afterglow is
if the merger took place in  an environment with a very low
circum-burst density. In fact, taking the minimal Lorentz factor
allowed by compactness (i.e., the maximal distance to the jet edge),
and a typical sGRB energy with $\A=10^4$, the circum-burst number
density should be as low as $\sim 10^{-6}\,{\rm cm}^{-3}$ for the
afterglow to be consistent with the observed limits and detections in
the X-ray and the radio. This value is lower by orders of magnitudes
than the density inferred by sGRB afterglows \cite{berger14} and  for
this specific event it is unexpected given the modest offset of the
merger location from the center of \ngc\ (although we do not know the
offset along the line-of-sight). It would be more typical of the
intergalactic medium \cite{2009RvMP...81.1405M}. Moreover, based on neutral hydrogen mass functions, a density of 10$^{-6}\,{\rm cm}^{-3}$ is ten times less likely than 10$^{-3}\,{\rm cm}^{-3}$ in early type galaxies \cite{Serra2012}. In addition, if \ot\ were an off-axis sGRB, then an on-axis observer would have  seen a burst with a typical photon energy of $E_p \approx 10$\,MeV, much higher than the values observed in sGRBs \cite{ghirlanda2004,mazets2004}. Finally, the probability to have such a near miss of the jet is only a few percent. In fact the chance to see a regular on-axis GRB is significantly higher in that case, since the jet is relatively wide ($\approx 0.4\,$rad).

To conclude, in order for the \ot\ to be an off-axis sGRB, in addition
to fortuitous alignment all of  the parameters need to be at  their limits. The Lorentz factor would be the lowest one allowed by the compactness limit and the surrounding density would be extremely low. In addition, the on-axis sGRB should have an $E_p$ that is much higher than that observed in sGRBs. This implies that it is 
unlikely that \ot\ is generated by an off-axis observation of a typical sGRB.

\subsubsection*{A Structured Jet}
We consider the possibility that sGRB jets have structure -- a core with high-luminosity and high Lorentz factor;  wings with lower luminosity and/or lower Lorentz factor. This structure may be induced by the jet launching mechanism or by the interaction of the jet with the ejecta. The question then is whether the observed $\gamma$-rays could have been generated at the jet's wings, either on-axis or off-axis. Below we consider both options.

If we observe the wings on-axis, then these are wide-angle low-luminosity wings. It is unclear how such wings would be generated. If a jet is launched with a high luminosity narrow core and low-luminosity wide angle wings, then the situation is similar to the on-axis wide and weak jet discussed above. The propagation of the narrow core does not facilitate the propagation of the wide low-luminosity wings, which will be choked by the ejecta. The interaction with the ejecta is also unlikely to produce low-luminosity wide wings because a cocoon, which has a comparable energy to the jet itself, is expected to dominate the wide angle outflow.

Off-axis emission from moderate Lorentz factor (say $\Gamma \sim 10$) material at the jet wings is more plausible. In this scenario the lower Lorentz factor of the jet wings allows the $\gamma$-rays to be observed although the angle to the jet edge is large enough that the afterglow emission will not violate the X-ray and radio observations.  However, this configuration is also in tension with the observations. First, the compactness criterion that we used in order to constrain the off-axis emission (see above) shows that material with low Lorentz factor that is observed off-axis  cannot have a high amplification factor $\A$. Namely it must have low luminosity. For example a source with $\Gamma \approx 10$ is limited to $\A\approx 20$, or a total isotropic equivalent energy of $\sim 6 \times 10^{47}$ erg as seen by an on-axis observer. Such low energy material is again expected to be suppressed by the cocoon. Moreover, observations of sGRBs do not support a structured jet with a luminosity that gradually drops at the wings. The reason is that with such a structure we would expect to detect at least some sGRBs where the $\gamma$-rays are generated by the low-luminosity wings that point towards us. The afterglows from these GRBs would have shown after a day or so the signature of an off-axis jet that carries significantly more energy than the one observed in $\gamma$-rays, once the the emission becomes dominated by the high energy core. Such sGRBs are not observed (see for example a compilation of sGRB prompt and afterglow properties by \cite{Fong2015}). 

We conclude that while in principle the observed $\gamma$-rays may have been generated by low-luminosity low-Lorentz factor wings of a jet seen off-axis, current observations and theory disfavor this possibility.

\subsubsection*{A Newtonian $\gamma$-ray source}
Assume that the $\gamma$-ray source is Newtonian. The optical depth of
the source, due to produced pairs, can be estimated using equation
\ref{eq:tauGamma} with $L_{51}=10^{-5}-10^{-4}$
\cite{GBM2017,Goldstein2017}and $\delta t_{-2} \approx 100$. Thus, if $\Gamma=1$ the source is extremely opaque. If we assume the maximal possible radius given the burst duration $\sim 3 \times 10^{10}$ cm,  the lower limit on the optical depth at the source is $\approx 10^{5}-10^{6}$. Such a large optical depth is unrealistic for many reasons. For example, the observed spectrum below $E_p$ is much shallower (softer) than a blackbody, while the spectrum of such a source is expected to be a blackbody or a Wein \cite{GBM2017,Goldstein2017}. Another problem is the implied diffusion time. The burst duration must be shorter than the diffusion time, which with this optical depth implies that the width of the emitting region cannot exceed $\sim 10^5$cm during the entire emission. Such a narrow non-expanding emitting region is not expected in ejecta that expands at sub relativistic velocities at a radius of $\sim 3 \times 10^{10}$ cm. 

Thus, we conclude that the $\gamma$-ray source must have a low optical depth.  Equation \ref{eq:tauGamma} shows that $\Gamma=2-3$ (mildly relativistic) is sufficient for an optically thin source. 
 
\subsubsection*{A Mildly Relativistic Shock Breakout of a Cocoon}
\label{sec:cocoon}
When a fast shock propagates in a high optical depth medium, it is dominated by radiation. The
shock breaks out once the optical depth drops such that the radiation is not confined within the shock layer and escapes to infinity. The theory of relativistic shock breakout was developed in the context of a shock that propagates in a star \cite{Colgate1968,nakar2012}  where the unshocked material is static and the stellar structure dictates its propagation. In the scenario of a breakout from the ejecta of a binary neutron star merger the shock breaks out of expanding material,  which has a different density profile than a star. Nevertheless, since the pulse of radiation that is emitted upon the shock breakout is dominated by the radiation generated within a very thin layer with an optical depth of unity, its properties depend almost entirely on two physical parameters: the shock Lorentz factor, $\Gamma_{\rm bo}$, and the breakout radius, $R_{\rm bo}$. Thus, we can use the derivation of \cite{nakar2012} to test whether a shock breakout can produce the observed signal, and if it does then to estimate its properties.

Not every flare of $\gamma$-rays can be generated by a relativistic shock breakout. First, shock breakouts do not produce flares with a highly variable temporal structure. This by itself implies that almost no GRBs (short or long) could be generated by shock breakouts. Second, the three main observables of the flare: energy, $E_{\rm bo}$, duration 
$t_{\rm bo}$ and  temperature, $T_{\rm bo}$ (note that the spectrum is not expected to be a blackbody, and $T_{\rm bo}$ is just the 
typical photon energy), depend only on two physical parameters, $\Gamma_{\rm bo}$ and $R_{\rm bo}$. Therefore they must  roughly satisfy a closure relation \cite{nakar2012}:
\begin{equation}
	t_{\rm bo} \sim 1\,{\rm~s} \left(\frac{E_{\rm bo}}{10^{46}\,{\rm erg}}\right)^{1/2}	\left(\frac{T_{\rm bo}}{150\,{\rm keV}}\right)^{-\frac{9+\sqrt{3}}{4}}
\label{eq:BOrel}
\end{equation}
The flare of $\gamma$-rays that followed \ot \cite{GBM2017,Goldstein2017}, which released $E_{\rm bo} \approx 4 \times 
10^{46}$\,erg over a duration of $t_{\rm bo} \sim 1-2$\,s at a typical photon energy of $T_{\rm bo} \sim 100-150$\,keV, satisfies the relation. Again, almost all the regular GRBs (short and long) do not satisfy this relation. They are too energetic and soft for their duration. The only type of GRB that satisfies equation \ref{eq:BOrel} and shows a non-variable light curves are low-luminosity GRBs, for which we have strong evidence that the $\gamma$-rays are generated by a mildly relativistic shock breakout \cite{nakar2015}.

Using any two of the three equations 14, 16 17 from \cite{nakar2012} we can find that if a shock breakout is the source of the $\gamma$-rays that followed GW170817 then $R_{\rm bo} \sim 3 \times 10^{11}$\,cm and $\Gamma_{\rm bo} \approx 2-3$. Equation \ref{eq:BOrel} and the estimates of the breakout parameters both assume that the shock breakout takes place over a relatively wide angle ($>20\degr$) that includes the observer's line-of-sight. This implies that in the case of \ot\ the cocoon breakout must have taken place over a wide angle of at least 0.5--1\,rad if it was the source of $\gamma$-ray emission.  

\subsubsection*{Hydrodynamical Simulation  of A Cocoon Breakout}
\label{sec_simulation}
To verify that a model of a cocoon driven by a choked jet can explain
the full range of electro-magnetic observations, we carry out 2D
relativistic hydrodynamic simulations, followed by a post-processing
calculation of the UV/Optical/IR emission during the expansion of the
cocoon and the ejecta. We search for a model in which the delay
between the merger and the observed photons from the breakout is 2\,s
\cite{GBM2017,Goldstein2017} and that its breakout radius and velocity
match those that we calculate above. We do not directly calculate  the
$\gamma$-ray signal since the breakout takes place over a scale that
is much smaller than the scales that our simulation can resolve.
However, the model is required  to match the  UV/Optical/IR observations.  

For the 2D relativistic hydrodynamic simulations we use the public code \textsc{PLUTO}\ \cite{2007ApJS..170..228M}. The initial configuration at $t=0$ (defined as the merger time) is cold ejecta that expands radially, which is present from the base of the grid at $ r_{\rm esc} = 4\times 10^8$\,cm up to $ r_{\rm max} = 2 \times 10^9$\,cm. The ejecta have an angular profile, where most of the mass ($ 75\% $) is near the equator at $ \theta > 1.0\,$rad, where $\theta$ is the polar angle. The ejecta are also divided in the radial direction into two regions (with similar angular profiles) -- a massive slow core that extends at $t=0$ up to $r_{\rm c}=1.3 \times 10^9$\,cm and low-mass fast material that extends at $t=0$ between $r_{\rm c}$ and $2 \times 10^9$\,cm. The density profile of the  dense core is:
\begin{equation}
\rho_{\rm{core}}(r,\theta) = \rho_0r^{-2}\left(\frac{1}{4}+\sin^3\theta\right)~,
\end{equation}
where $ \rho_0 $ is the normalization which is chosen to fix the total ejecta mass. The velocity profile of the core is
\begin{equation}
v_{\rm{core}}(r) = v_{\rm c,max}\frac{r}{r_c}~,
\end{equation}
where $ v_{\rm c,max} = 0.2c $ is the maximal velocity of the ejecta's
inner part. The extended ejecta density profile is chosen as a very
steep power-law in $v$ between $v_{\rm c,max}$ and $0.8$c so its total
mass is $1\%$ of the core mass and the mass carried by material at
$v>0.7$c is about $10^{-5}$ of the total ejecta mass. The jet is
injected into the ejecta with a delay of $1\,$s for a total working
time of $ 2\,$s and a total luminosity of $L_{\rm j}$. The jet is
injected with a specific enthalpy of 20 (in units of its rest-mass energy) at an opening angle of $0.7\,$rad from a nozzle at the base of the grid with a size of $10^8\,$cm. We search for a model that fits the evolution of the
bolometric luminosity and  temperature fitted to the UV/optical/IR
data by varying the total ejecta mass, namely $\rho_0$, and the jet's
luminosity, $L_{\rm j}$. We find reasonable fits with ejecta masses
larger than $0.05\,M_\odot$ and jet luminosities that release an
energy that is comparable or slightly larger than the total ejecta kinetic energy. The specific model presented in the paper (figs 2 and 6) and which is discussed below has a total ejecta mass of $0.1\,M_\odot$ and jet luminosity $L_{\rm j}=2.6 \times 10^{51}\,{\rm erg\,s}^{-1}$. 

Given its large opening angle the jet is not collimated. Instead it
works its way through a large amount of ejecta dissipating its energy
in the process, forming a cocoon. The jet injection stops at 3\,s and
by $t \approx 4$\,s the jet is fully choked, leaving a hot cocoon that
continues to propagate driving a mildly relativistic ($\Gamma \approx
2-3$) shock into the ejecta. About 10\,s after the merger the shock
catches up with the leading edge of the ejecta and breaks out at a
radius of $R_{\rm bo}=2.4\times 10^{11}\,{\rm cm}=8\,$light-seconds.  Thus, photons released during the cocoon shock breakout are only 2\,s behind the gravitational waves released at the merger, thereby consistent with the delay between the merger and $\gamma$-rays observed by Fermi \cite{GBM2017,Goldstein2017}. The breakout also takes place at a radius and velocity that is expected to produce a signal  consistent with \ot.

We end the simulation at $t=15\,$s when all the expansion is homologous and all the material moves ballistically. The emission after the breakout is calculated as a post-processing of the final snapshot of the hydrodynamical simulation, following a similar procedure to that described in \cite{Gottlieb2017}. It contains two components: (i) diffusion of photons that where deposited by the shock that crossed the ejecta, which we call ``cooling emission", and (ii) radioactive decay of the elements that where synthesized in the ejecta, termed ``macronova". In short, for each time step in the observer's frame we calculate first the radial optical depth from every
radius to infinity along a radial path. We do that for different angles and we determine the trapping radius, $r_t(\theta)$, where the optical depth $\tau(r_t(\theta)) = c/v$. Above this radius, photons diffuse freely to the observer at infinity while below this radius they are trapped. This is an approximation since the outflow is not spherically symmetric. In a similar manner, we calculate for each angle the photospheric radius $r_{\rm ph}(\theta)$ for which $\tau(r_{\rm ph}(\theta)) = 1$.
To obtain an approximated calculation that includes correction due to the mildly relativistic motion, we first calculate the luminosity in the comoving frame. The cooling emission luminosity in a  comoving frame is determined by the diffusion of the rest frame energy flux at the trapping radius of the radiation that was carried by the outflow from
the last hydrodynamical snapshot. The macronova emission arises from
the radioactive heating $\dot \epsilon$ generated by all the material above the trapping radius, $m(r>r_t)$, and is estimated at the rest frame of the photosphere as
$\dot{\epsilon} =m(r>r_t) \epsilon_0 10^{10} \left(\frac{t}{1\,\rm{d}}\right)^{-1.3} ~~\rm{erg\,s^{-1}}$, 
where $\epsilon_0$ is of order unity  \cite{korobkin2012}.  We take
$\epsilon_0=2$ during the first two days and reduce it to
$\epsilon_0=1.5$ afterwards  \cite{perego2014}.

Next, we estimate the rest frame temperature at $r_{\rm ph}(\theta)$ by taking the rest frame luminosity (the sum of cooling and macronova emission) in each direction and finding the radiation energy density at the photosphere assuming that the radiation is in thermal equilibrium and that the local radiation spectrum is a blackbody at this point. We assume that this blackbody radiation is emitted isotropically at the matter's rest frame of the photosphere. Having the rest frame luminosity and spectrum along each angle at every time in the explosion frame, we integrate the contribution from material at all angles for observers at different viewing angles at different observer times, by properly accounting for the Lorentz boost and the light travel time.

Finally, in order to find the ejecta optical depth at each time and place we need to estimate its opacity. For that we use tracers which mark material that is ejected initially (at $t=0$) at different angles. By looking at the velocity distribution of the tracers at the end of the simulation we find that  material that was ejected initially near the equator is associated with the slowest velocities ($ v/c < 0.1 $) while material that was ejected initially at high latitudes is associated with faster material. This is not surprising given that the energy of the jet was mostly deposited in high latitude material. Following \cite{perego2014}  we adopt an opacity $ \kappa = 10\,\rm{cm^2\,g^{-1}} $ for material slower than $ 0.1c $, which is presumably dominated by neutron rich dynamical ejecta that contains Lanthanides, and $ \kappa = 1.0\,\rm{cm^2\,g^{-1}}$ for material that is faster than $0.1$c which was presumably ejected from the disk and/or the massive neutron star that were formed after the merger and does not contain r-process elements from the second and third peaks. 


Throughout the simulations we applied an ideal gas equation of state
with an adiabatic index of $ 4/3$ as appropriate for a
radiation-dominated gas, and neglect gravity given that kinetic energy
dominates the ejecta in the simulation domain. For the integration an
Harten-Lax-van Leer (HLL) Riemann solver and a third order Runge Kutta time-stepping have been used. We employed a cylindrical grid with $1620\times 1600$ cells. The grid is divided in three patches on the $x$-axis and two on the $z$-axis (parallel to the jet axis). The innermost patch on the $x$-axis is stretching from the origin to $ x = 2 \times 10^8\,$cm with 20 cells uniformly distributed, this patch makes sure the jet's nozzle contains enough cells. Most of the cocoon forms and experiences the mixing up to $ x = 2\times 10^{10}\,$cm and  $ z = 2\times 10^{10}\,$cm. In this region we employ a resolution of 800 cells on each axis. Outside of these coordinates, the grid contains a uniform distribution of $ 800 $ additional cells in each axis up to $ x = 4\times 10^{11}\,$cm and $ z = 4\times 10^{11}\,$cm.

\subsubsection*{Early-time emission powered by an engine-driven wind}
\label{sec:engine}
We proposed two options to explain bright and blue emission seen on the first day.  Both options rely on radioactive heating (one involved mildly relativistic material and the other material with relatively low opacity) to power the emission. Here we suggest a third possibility: the release  of internal energy that was deposited in the ejecta by an  engine-driven wide-angle wind that lasted for $\sim 20$s.  

In the cocoon model that we presented above, the cooling emission fades within hours and does not contribute to the observed emission. The energy is deposited by the jet over a short duration when the ejecta radius is small. Then adiabatic losses cool  the radiation  by the time that the radiation is emitted. This can be shown using the following simple arguments. Assume that energy  is deposited at radius $R_0$ in a mass $M$. This energy is deposited by a strong shock so about half of it is deposited  in the form of radiation while the other half accelerates the mass to a velocity $v=\beta c$.
The radiation is released at the time $t_{\rm rad}$ given in equation \ref{eq:trad} at a radius $R_{\rm rad} \approx v t_{\rm rad}$ after adiabatic losses have reduced the internal energy to  $M v R_0/t_{\rm rad }$ and the luminosity is $ L \sim M v R_0/t_{\rm rad}^2$. Then using equation \ref{eq:trad} for $t_{\rm rad}$ we find:
\begin{equation}
	L \approx 6 \times 10^{40}\,{\rm erg\,s}^{-1} \kappa^{-1} \frac{R_0}{10^{10}\,{\rm cm}} \left(\frac{\beta}{0.2}\right)^2
\end{equation}
independent of the mass. We observed a luminosity of $10^{42}$\,erg implying that in order to explain the early emission as shock cooling, the shock should hit the ejecta at a radius of $\sim 10^{11}$\,cm. Since the ejecta propagates at a velocity that does not exceed $\sim 0.2$c it should be shocked $\sim 20$\,s after the merger. Such an option is plausible given that we know that in some sGRBs there is engine activity that releases energy comparable to that seen in the burst over a duration of $\sim 100$\,s \cite{Nakar2007}.

To verify our simple estimates we carried out a 1D numerical
simulation with spherical symmetry (using the same code as in the main
text, where the $ z $-axis is now the spherical $ r $-axis). We inject
an engine-driven wind that lasts for 20\,s and contains a total energy
of $ 2 \times 10^{51}\,\rm{erg} $ into $ 0.05\,M_\odot $ of cold
ejecta which maintains a power-law density profile with index $-1$,
and expands homologously with an initial velocity of 0.1--0.2$c$. We
found that this can produce the observed emission during the first day
for opacity $ \kappa \approx 3\,\rm{cm^2\,g^{-1}} $. We conclude that long-lasting engine-driven wind is another possible explanation for the bright, blue emission of \ot\ seen during the first day.


\subsubsection*{Early-time emission powered by free neutron decay}
\label{sec:neutron}
The observed emission during the first day is brighter, bluer and
rises faster than predicted for the
standard models of radioactive decay of r-process elements \cite{Kasen2013,Tanaka2013,Rosswog2017,Wollaeger2017}. It was suggested \cite{2015MNRAS.446.1115M}  that the decay of free neutrons may give rise to an early blue emission. We therefore examine if decay of free neutrons may have made a substantial contribution to the UV/optical emission observed during the first day.

Consider an outflow with a mass $m_{-2}$ (in units of $10^{-2}\,M_\odot$), a typical
velocity $\beta_{0.5}$ (in units of $0.5c$) and an initial free neutron fraction
$X_n$. The evolution of internal energy confined to the flow, $E$, can be estimated as:
\begin{equation}\label{eq:de_dt}
	\frac{dE}{dt}=\frac{E}{t}+L_{\rm h}-L_{\rm rad}
\end{equation}
The first term takes into account adiabatic losses, the second heating term is  due to neutron decay, $L_{\rm h}=6 \times  10^{45} m_{-2} X_n  e^{(-t/900\,{\rm s})}\,{\rm erg\,s}^{-1}$, and the cooling term $L_{\rm rad}$ accounts for radiative losses. The trapped radiation is released roughly at 
\begin{equation}\label{eq:trad}
	t_{\rm rad} \approx \left( \frac{\kappa m}{4 \pi c v}\right)^{1/2} = 0.7 m_{-2}^{0.5} \kappa^{0.5} \beta_{0.5}^{-0.5}\,{\rm day},
\end{equation}
where opacity $\kappa$ is in cm$^{2}$ g$^{-1}$.  Up to $t \ll t_{\rm rad}$ equation \ref{eq:de_dt} can be integrated neglecting radiative losses. For  $900\,{\rm s} \ll t \ll t_{\rm rad}$ this integration yields $E = 6 \times 10^{46} m_{-2} X_{n} {t_d}^{-1}$\,erg, where $t_{\rm d}$ is time in days. The trapped radiation is radiated at $t_{\rm rad}$ with a  peak luminosity that is roughly $L_{\rm peak} \approx E/t_{\rm rad}$. \begin{equation}
	L_{\rm peak} \approx \frac{E(t_{\rm rad})}{t_{\rm rad}} \approx 2 \times 10^{42} X_n \kappa^{-1} \beta_{0.5}\,\rm{erg\,s}^{-1}
\end{equation}
This luminosity is independent of $m_{-2}$. The observed luminosity during the first day implies that for $\kappa=1\,{\rm cm}^{2}\,{\rm g}^{-1}$, X$_{n}$ is 0.5 and the ejecta must be at least $\sim 0.01M_\odot$ of almost pure free neutrons. For larger $\kappa$ neutrons cannot produce the observed signal while lower $\kappa$ reduces the neutron fraction, while increasing the total mass  by the same factor (to match $t_{\rm rad}$), so an unrealistically high mass of  $\sim 0.01M_\odot$ of free neutrons is always needed to explain the observations from the  first day.

\subsection*{Rates}
\subsubsection*{\ot-like events in the Palomar Transient Factory database}
The Palomar Transient Factory (PTF; \cite{2009PASP..121.1395L}) and
intermediate Palomar Transient Factory (iPTF;
\cite{Cao2016,Masci2017}) databases keep track of every detection with
significance greater than 5-$\sigma$ on every image subtraction
performed during 2009 to 2017. This facilitates our ability to
calculate the rates of a variety of astrophysical events in
hindsight. We determined the efficiency of every image subtraction for
a transient of a given magnitude superimposed on a galaxy of a given surface brightness for the PTF survey over the time frame of 2010~January  through 2012~December  \cite{2017ApJS..230....4F}.  By coupling these efficiencies with a Monte Carlo simulation of a particular transient event, we can calculate a rate based on comparison to the observed number of such transients found during the survey or, if no such events were found, we can provide an upper limit on this rate.

We used these efficiencies, coupled with the lightcurve for \ot\ to calculate the rate in the following manner. First we performed a search in $R$-band for all transients resembling \ot. We only performed the search within the search radius of the CLU (Census of the Local Universe; \cite{Cook17}) galaxy catalog within 200 Mpc as defined in \cite{2011PhDT........35K}. We required 3 detections above our machine-learning threshold for a real source \cite{2012PASP..124.1175B}  found within a 10-day time frame with at least one non-detection both the week prior to the event and the week after the event. No other significant detections in the database were allowed at any time during the survey.

The result of this search yielded 2 potential candidate events, one of
which is almost certainly a heretofore unreported nova in NGC 253. The
other event, similar in absolute brightness and decline rate as \ot,
may be a similar event but may be a false positive. For now we will assume there are no positive detections of such events in PTF and proceed to calculate an upper limit for this rate.

From here we simulated, through a Monte Carlo, the lightcurves of
\ot\ as seen in our survey and if they would meet the required
detection thresholds as defined above.

From here we simulated the lightcurves of \ot\ as seen in our survey
via a Monte Carlo, asking the question at each point in our past
observing schedule whether or not the transient would be seen above
our real-bogus threshold based upon the sky conditions, background
host galaxy light, seeing, and other factors as found through our
efficiency studies. Then we evaluated if enough data points met the
required 3 detections above our real-bogus threshold, found within a
10-day time-frame with at least one non-detection both the week prior
and post the event. If so, it was counted as a positive detection
towards the relative rate.  The result of this simulation is an event
rate of 320\,Gpc$^{-3}$\,yr$^{-1}$ and a 3-$\sigma$ upper limit of
800\,Gpc$^{-3}$\,yr$^{-1}$. Relating this back to the optical
counterparts of NS-NS mergers, the major uncertainty is the luminosity
function in the $R$-band of such events. If the typical NS-NS merger
is 50\% fainter, the upper limit on such events nearly doubles to over
1600\,Gpc$^{-3}$\,yr$^{-1}$.

A similar search was carried out in the iPTF database yielding no
corresponding detections using the techniques reported in
\cite{zhao2016automatic}, which means that the rate described above is
conservative.


\subsubsection*{Implications for rates based on r-process abundances}
Here we discuss the relation of GW/\ot\ to the origin of r-process elements
in the Universe. There are three peaks in atomic mass seen in
r-process abundances (figure~\ref{fig:abundance}), near atomic mass
number $A$ of 80, 130, and 195.
The mass of the ejected r-process elements is estimated to be at least $\sim 0.05M_{\odot}$ 
in order to explain the observed light curves and spectra.
The bright emission at early times suggests that
the Lanthanide-free material  with a mass of $> 0.02M_{\odot}$ is ejected
(with $A< 140$; see figure~\ref{fig:abundance}).
Given the ejecta mass and merger rate deduced above,  we can test  
the hypothesis that mergers produced all the r-process elements heavier than a
minimal atomic mass number, $A_{\rm min}> 70$, in the Milky Way. 
The value of $A_{\rm min}$ for the merger ejecta
depends on the stiffness of the neutron star equation of state \cite{Sekiguchi2015,Radice2016} and the lifetime of the remnant massive neutron star (\cite{Metzger2014}, see also recent studies on
the nucleosynthesis in the merger ejecta \cite{Wanajo:2014,Wu:2016}.
Roughly $80\%$ of r-process elements are around the first r-process
peak at $A$ of 80 (figure~\ref{fig:abundance}). Therefore
the rate estimate from the Galactic r-process abundance is 
quite sensitive to the choice of $A_{\rm min}$. 
Assuming that
the solar r-process abundance  is the typical of the Milky Way stars,
we estimate the volumetric rate \cite{Hotokezaka:2015}: 
\begin{eqnarray}
R \sim \begin{cases}
500 \left(\frac{M_{\rm ej}}{0.05M_{\odot}}\right)^{-1}\,{\rm Gpc^{-3}\,yr^{-1}}~~~~~(A_{\rm min}=70),\\
100 \left(\frac{M_{\rm ej}}{0.05M_{\odot}}\right)^{-1}\,{\rm Gpc^{-3}\,yr^{-1}}~~~~~(A_{\rm min}=90),\label{rmass}
\end{cases}
\end{eqnarray}
where we use the number density of galaxies of $\approx 0.01\,{\rm Mpc^{-3}}$. 
The former corresponds to  all the r-process elements being produced by mergers
and the latter corresponds to only the heavier r-process elements
being produced without the first peak. The large event rate and ejecta mass 
inferred from GW/\ot\
suggest that the majority of r-process elements are   produced by mergers. 
However, the abundance patterns of the first-peak elements of extremely 
metal poor stars do not agree with each other \cite{Sneden:2008}. This suggests that more than 
one type of astrophysical phenomenon may produce these elements or there is variation
in the composition and amount of the r-process ejecta of neutron star mergers.

\section*{Additional Acknowledgements}
Gemini observatory data were obtained primarily under programs GS-2017B-DD-1 and GS-2017B-DD-6 (PI L. P. Singer) and is available at https://archive.gemini.edu. CTIO1.3m data obtained under program NOAO 2017B-0160 (PI B. Cobb) is available at http://archive.noao.edu/. Hubble Space Telescope data obtained under program HST-GO-15436 (PI M. M. Kasliwal) is available at https://archive.stsci.edu/. Keck Observatory data (PI G. Hallinan) is available at https://koa.ipac.caltech.edu/. VLT data is available at http://archive.eso.org/.  

The data of GALEX presented in this paper were obtained from the
Mikulski Archive for Space Telescopes (MAST). STScI is operated by the
Association of Universities for Research in Astronomy, Inc., under
NASA contract NAS5-26555. Support for MAST for non-HST data is
provided by the NASA Office of Space Science via grant NNX09AF08G and
by other grants and contracts. PS1: The Pan-STARRS1 Surveys (PS1) have
been made possible through contributions of the Institute for
Astronomy, the University of Hawaii, the Pan-STARRS Project Office,
the Max-Planck Society and its participating institutes, the Max
Planck Institute for Astronomy, Heidelberg and the Max Planck
Institute for Extraterrestrial Physics, Garching, The Johns Hopkins
University, Durham University, the University of Edinburgh, Queen's
University Belfast, the Harvard-Smithsonian Center for Astrophysics,
the Las Cumbres Observatory Global Telescope Network Incorporated, the
National Central University of Taiwan, the Space Telescope Science
Institute, the National Aeronautics and Space Administration under
Grant No. NNX08AR22G issued through the Planetary Science Division of
the NASA Science Mission Directorate, the National Science Foundation
under Grant No. AST-1238877, the University of Maryland, and Eotvos
Lorand University (ELTE). 2MASS: This publication makes use of data
products from the Two Micron All Sky Survey, which is a joint project
of the University of Massachusetts and the Infrared Processing and
Analysis Center/California Institute of Technology, funded by the
National Aeronautics and Space Administration and the National Science
Foundation. WISE: This publication makes use of data products from the
Wide-field Infrared Survey Explorer, which is a joint project of the
University of California, Los Angeles, and the Jet Propulsion
Laboratory/California Institute of Technology, funded by the National
Aeronautics and Space Administration. This research has made use of
the NASA/IPAC Extragalactic Database (NED) which is operated by the
Jet Propulsion Laboratory, California Institute of Technology, under
contract with the National Aeronautics and Space Administration.
\textsc{IRAF} is distributed by the National Optical Astronomy
Observatory, which is operated by the Association of Universities
for Research in Astronomy (AURA) under a cooperative agreement with
the National Science Foundation. \textsc{PyRAF} is a product of the Space Telescope Science Institute, which is operated by AURA for NASA. Some observations were obtained at the Gemini Observatory, which is operated by the Association of Universities for Research in Astronomy, Inc., under a cooperative agreement with the NSF on behalf of the Gemini partnership: the National Science Foundation (United States), the National Research Council (Canada), CONICYT (Chile), Ministerio de Ciencia, Tecnolog\'{i}a e Innovaci\'{o}n Productiva (Argentina), and Minist\'{e}rio da Ci\^{e}ncia, Tecnologia e Inova\c{c}\~{a}o (Brazil).

A.H.\ acknowledges support by the I-Core Program of the Planning and Budgeting Committee and the Israel Science Foundation. 
T.M.\ acknowledges the support of the Australian Research Council through grant FT150100099.
 Parts of this research were conducted by the Australian Research Council Centre of Excellence for All-sky Astrophysics (CAASTRO), through project number CE110001020.
The Australia Telescope Compact Array is part of the Australia Telescope National Facility which is funded by the Australian Government for operation as a National Facility managed by CSIRO.
D.L.K.\ is additionally supported by NSF grant AST-1412421.
A.A.M.\ is funded by the Large Synoptic Survey Telescope Corporation in support of the Data Science Fellowship Program.
P.C.Y., C.C.N.\ and W.H.I.\ thank the support from grant MOST104-2923-M-008-004-MY5 and MOST106-2112-M-008-007.
A.C.\ acknowledges support from the National Science Foundation CAREER award 1455090, ``CAREER: Radio and gravitational-wave emission from the largest explosions since the Big Bang''.
T.P.\ acknowledges the support of  Advanced ERC grant TReX and the Templeton Foundation. 
B.E.C,\ thanks  SMARTS 1.3-m Queue Manager Bryndis Cruz for prompt scheduling of the SMARTS observations.
Basic research in radio astronomy at the Naval Research Laboratory (NRL) is funded by 6.1 Base funding. Construction and installation of VLITE was supported by NRL Sustainment Restoration and Maintenance funding.
K.P.M's research is supported by the Oxford Centre for Astrophysical Surveys which is funded through the Hintze Family Charitable Foundation.
J.S. and A.G. are grateful for support from the Knut and Alice
Wallenberg Foundation. Together with S.R they are also supported by
the research environment grant ``Gravitational
Radiation and Electromagnetic Astrophysical Transients (GREAT)"
funded by the Swedish Research Council (VR) under Dnr 2016-06012.
S.R. has also been supported by VR
(grant number 2016-036573), and by the Swedish National Space Board (107/16).
E.O.O. is grateful for the support by grants from the Israel Science Foundation, Minerva, Israeli ministry of Science, the US-Israel Binational Science Foundation, and the I-CORE Program of the Planning and Budgeting Committee and The Israel Science Foundation. 
We thank the staff of the GMRT that made these observations possible.  The GMRT is run by the National Centre for Radio Astrophysics of the Tata Institute of Fundamental Research.
A.Y.Q.H. was supported by a National Science Foundation Graduate Research Fellowship under Grant No. DGE1144469.
J.E.J.\ acknowledges supported by the National Science Foundation Graduate Research Fellowship under grant no. DGE-1144469.
S.G.\ acknowledges support from NSF Award PHY-1607585.
G.C.A.\ and V.B.\ acknowledge the support of the Science and Engineering Research Board, Department of Science and Technology, India and the Indo-US Science and Technology Foundation for the GROWTH-India project. 
J.B.\ acknowledges support from the Einstein Fellowship grant  PF7-180162.
N.K., R.I.\ and Y.Y.\ acknowledge the support of Japan Society for the Promotion of Science for the GROWTH project.
CZTI is built by a TIFR-led consortium of institutes across India, including VSSC, ISAC, IUCAA, SAC and PRL. The Indian Space Research Organisation funded, managed and facilitated the project.
Parts of this research were conducted by the Australian Research Council Centre of Excellence for All-sky Astrophysics in 3D (ASTRO 3D) through project number CE170100013.
F.R.\ acknowledges support from the U.S. Department of Energy Early Career Award DE-SC0012160 ``Scalable and Energy-Efficient Methods for Interactive Exploration of Scientific Data.''
E.N.\ and O.G.\ acknowledge the support of an ERC starting grant (GRB/SN) and an ISF grant (1277/13).
P.E.N.\ acknowledges support from the DOE through DE-FOA-0001088,
``Analytical Modeling for Extreme-Scale Computing Environments.'' This research used resources of the National Energy Research Scientific Computing Center, a DOE Office of Science User Facility supported by the Office of Science of the U.S. Department of Energy under Contract No. DE-AC02-05CH11231.


\begin{thebibliography}{100}
\bibitem{visir}
These observations were taken under proposal 60.A-9292(B), which the European
  Southern Observatory kindly made public to the LV-EM community.
\end{thebibliography}


\begin{figure}
\centerline{\includegraphics[width=0.75\textwidth]{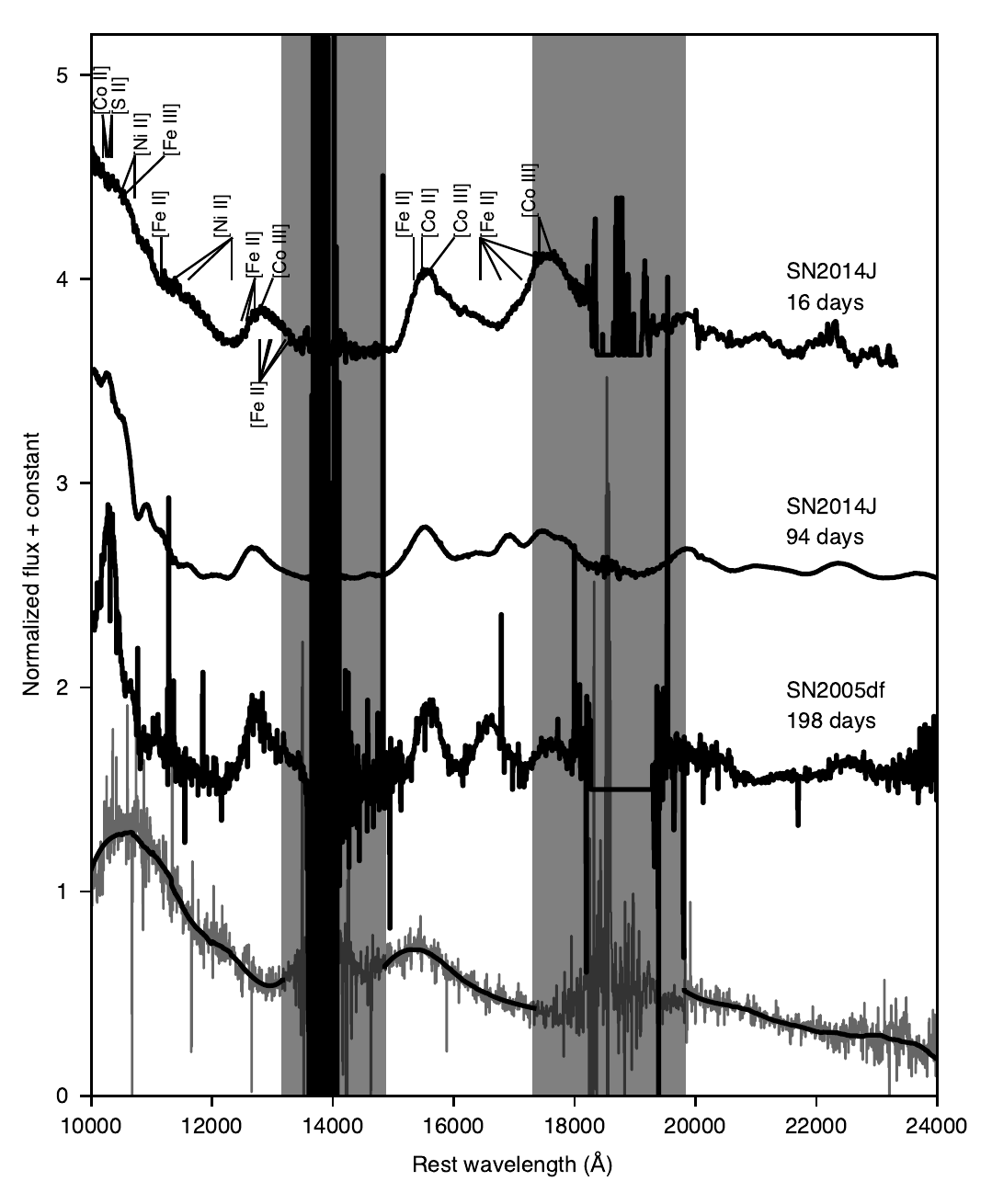}}
\caption{Near-infrared spectrum of \ot\ at $t = 4.5$~days post merger is shown at the bottom, along with the spectra of the type Ia SN 2014J at 16 and 94~days post maximum\cite{2016ApJ...822L..16S}, and the type Ia SN 2005df at 198~days post maximum \cite{2015ApJ...806..107D}. Each spectrum is normalized to the flux between $10000$--$10500$~\AA\ and shifted up from the one below for clarity. The spectrum of \ot\ was corrected for Milky Way reddening assuming $E(B-V) = 0.1$ and a standard $R_V = 3.1$ extinction law \cite{1999PASP..111...63F}, and smoothed using a Savitzky–Golay filter to clearly show the prominent, broad spectral features at $\sim 10600$ and $15000$~\AA. The unfiltered data are shown in gray. Regions of low S/N due to the strong telluric absorption features between the $J$, $H$, and $K$ spectral windows are indicated by the vertical, gray bars. We label several transitions of Fe-peak elements on the top spectrum of SN~2014J. While qualitatively similar, the broad Fe-peak features characteristic of SNe~Ia are inconsistent with the features observed in the spectrum of \ot.}
\label{fig:spec_Ia}
\end{figure}

\begin{figure}
\centerline{\includegraphics[width=0.75\textwidth]{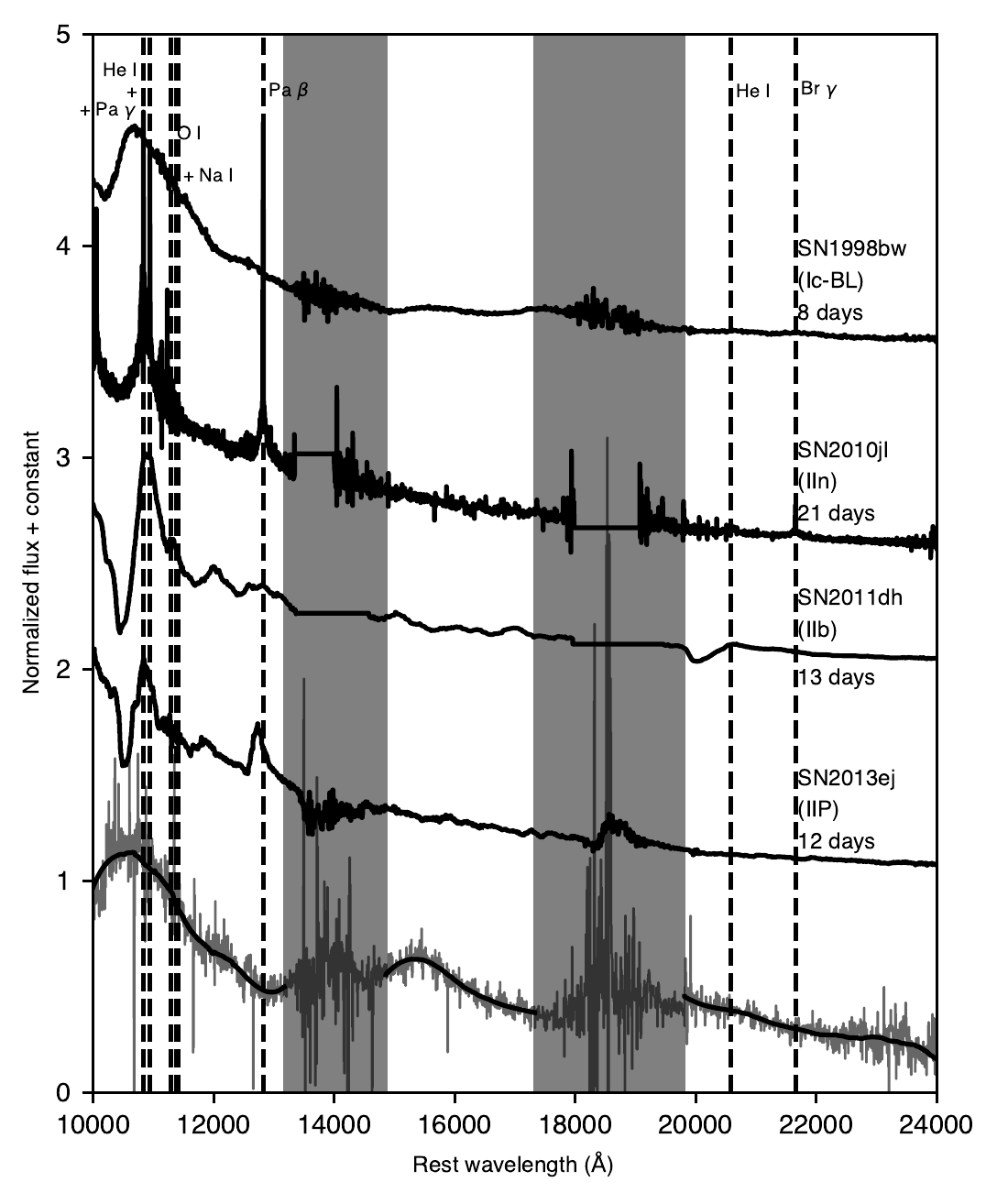}}
\caption{Near-infrared spectrum of \ot\ at $t = 4.5$~days post merger
  is shown at the bottom, along with the spectra of the type IIP SN
  2013ej at 12 days \cite{2016MNRAS.461.2003Y}, the type IIb SN 2011dh
  at 13~days \cite{2014A&A...562A..17E}, the type IIn SN~2010jl at 21
  days \cite{2015ApJ...801....7B}, and the broad-lined type Ic (Ic-BL)
  SN~1998bw at 8 days post maximum \cite{2001ApJ...555..900P}.  The
  spectra of \ot\ are presented as in figure~\ref{fig:spec_Ia}.
  We label several transitions commonly identified in the spectra of core-collapse SNe with vertical, dashed lines.}
\label{fig:spec_CCSNe}
\end{figure}

\begin{figure}
\includegraphics[width=\textwidth]{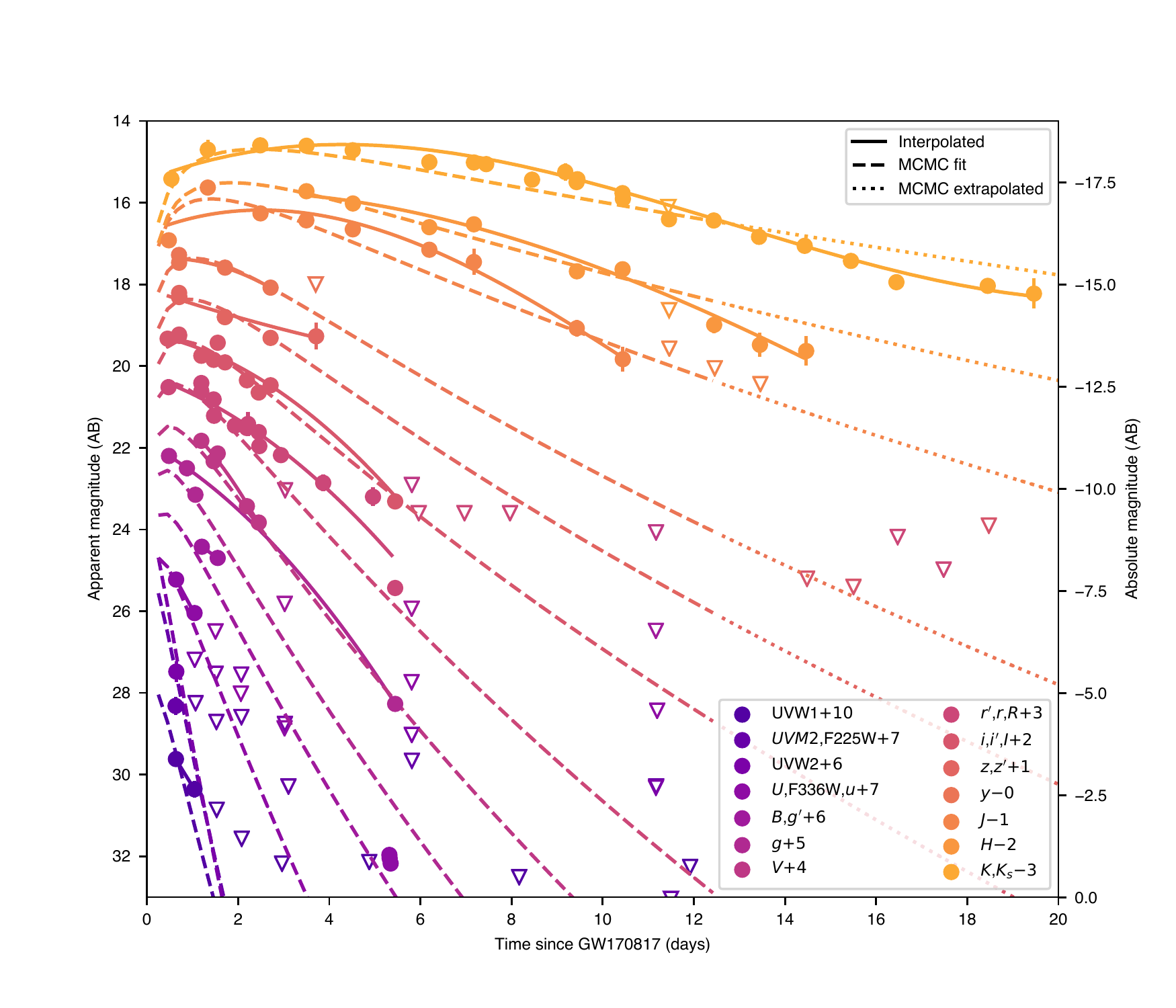}
\caption{Multiwavelength lightcurve based on the
  ultraviolet/optical/infrared photometry of \ot.  The data are our
  assembled photometry (Table~\ref{tab:phot}) plotted as AB magnitude
  vs.\ time since GW170817, with open triangles indicating 5$\sigma$
  upper limits, colored by wavelength.  We plot both apparent
  magnitude and absolute magnitudes assuming distance of 40\,Mpc.  We
  illustrate both methods of blackbody fitting used here: the solid
  curves are the low-order polynomial functions used at each wavelength to fit a separate absorbed blackbody at every epoch, while the dashed curves are the evolving absorbed blackbodies fit simultaneously with a Markov-Chain Monte Carlo (MCMC) algorithm (the dotted portions are extrapolations).  Filters with similar wavelengths (roughly within 300\,\AA) have been grouped together for clarity, and some filters have been omitted.}
\label{fig:lci}
\end{figure}


\begin{figure}
\includegraphics[width=\textwidth]{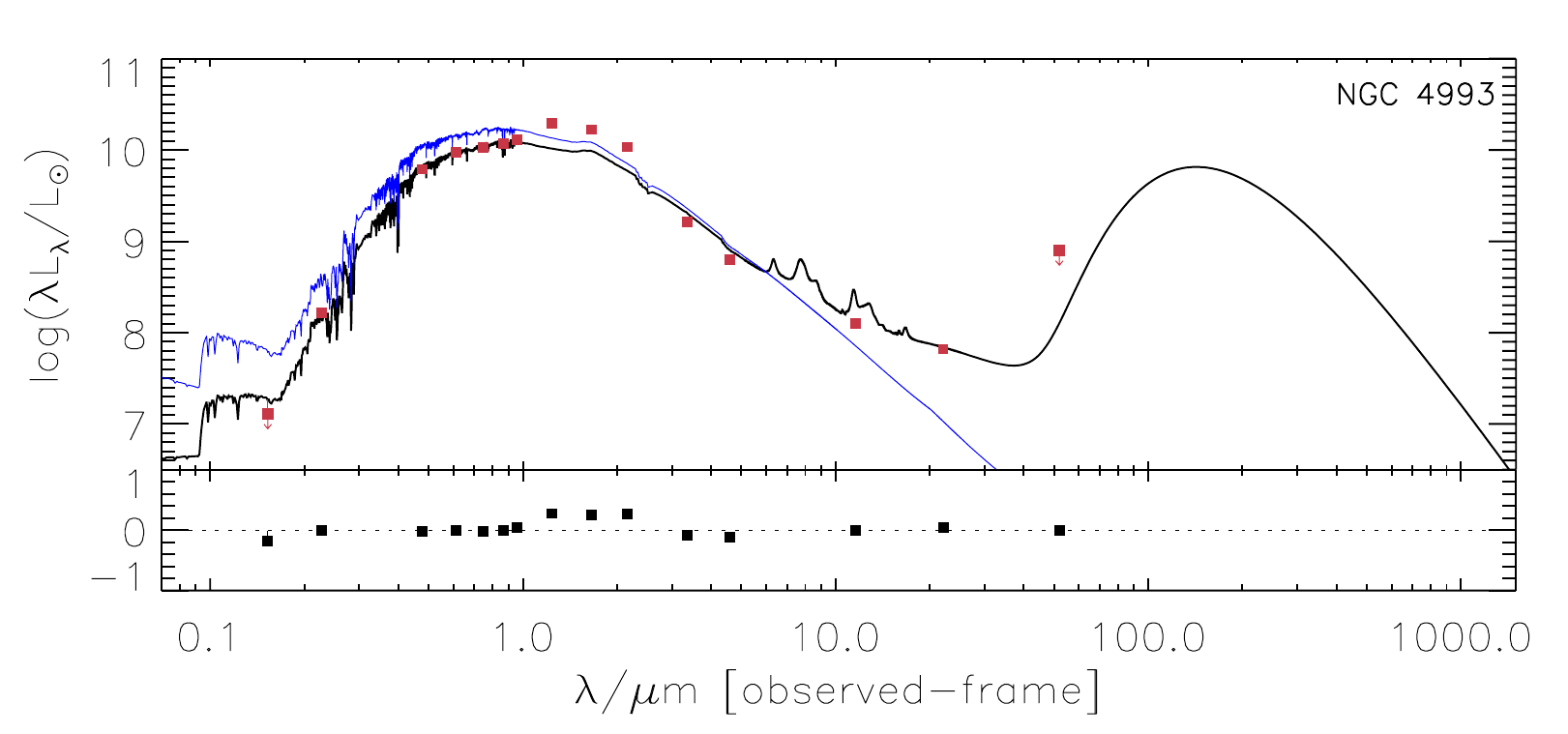}
\caption{Spectral energy distribution obtained by combining photometric data of \textit{GALEX}, Pan-STARRS, 2MASS, \textit{WISE}, and \textit{IRAS} surveys (red squares). The flux density was corrected for Galactic extinction before the fitting assuming $R_V= 3.1$ and $E(B-V) = 0.1$. We used upper limits for \galex/FUV and \textit{IRAS}/60\,$\mu$m bands. The upper panel shows the unattenuated stellar spectrum (blue line) and the sum of attenuated stellar spectrum and the infrared emission (black line). The lower panel shows the residuals $(L^{\rm obs}_{\lambda}-L^{\rm model}_{\lambda})/L^{\rm obs}_{\lambda}$ (black squares), where $L^{\rm obs}_{\lambda}$ and $L^{\rm model}_{\lambda}$ represent observed and predicted broad-band luminosity densities respectively.}
\label{fig:host-SED}
\end{figure}

\begin{figure}
	\includegraphics[width=\textwidth]{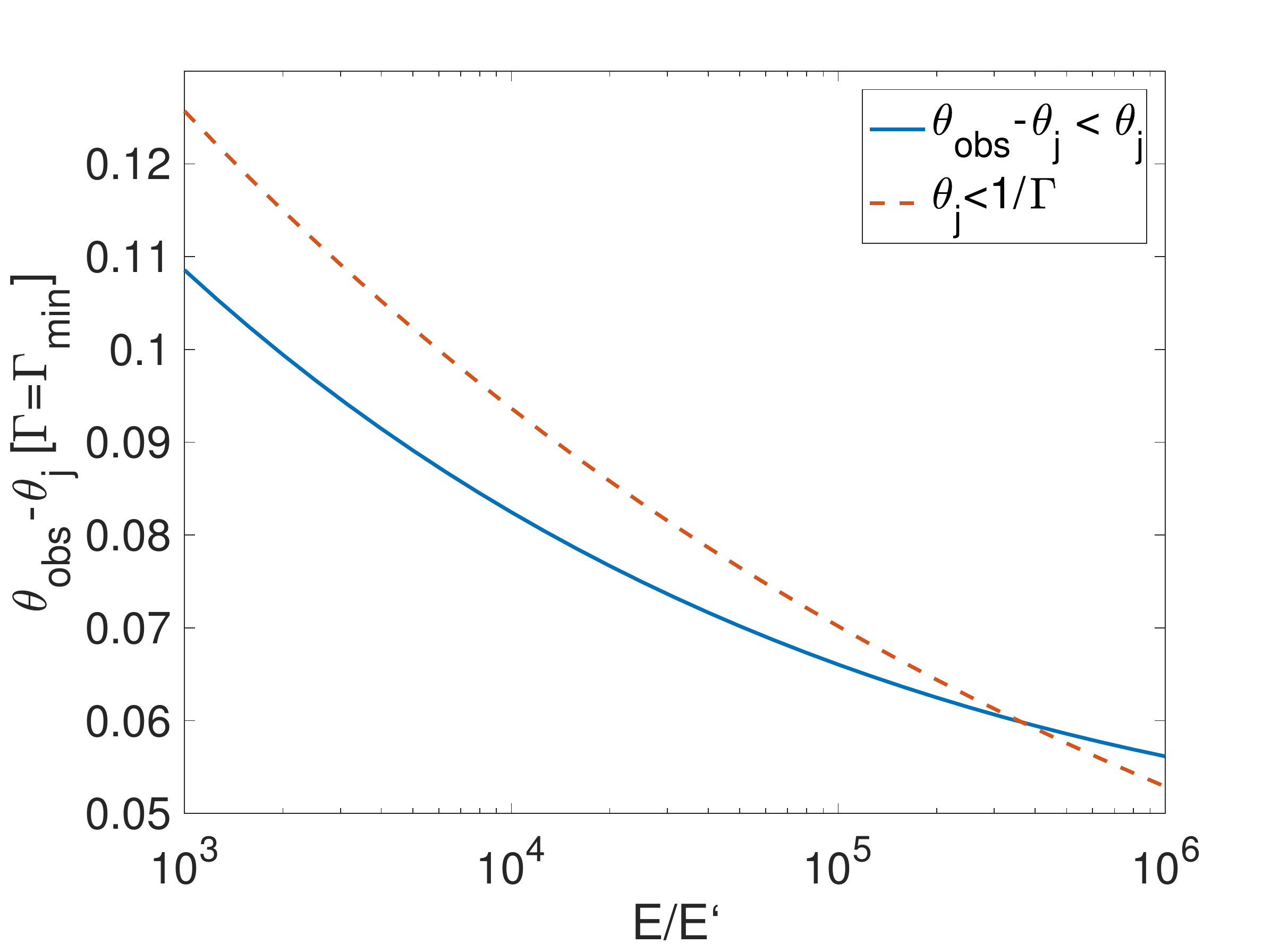}
	\caption{The maximal value of $\theta_{\rm obs}-\theta_{\rm j}$ as a function of the amplification parameter $\A\equiv E/E^\prime$ for observers  far and close to the jet axes (cases (i) and (iii) in equation~\ref{eq:A}). The observed values of \ot\ that we used are photon energy $E_p'=185$ keV and $\gamma$-ray spectral index $\alpha=-0.6$ \cite{GBM2017,Goldstein2017}.}
	\label{fig:the_obs}
\end{figure}

\begin{figure}
	\includegraphics[width=\textwidth]{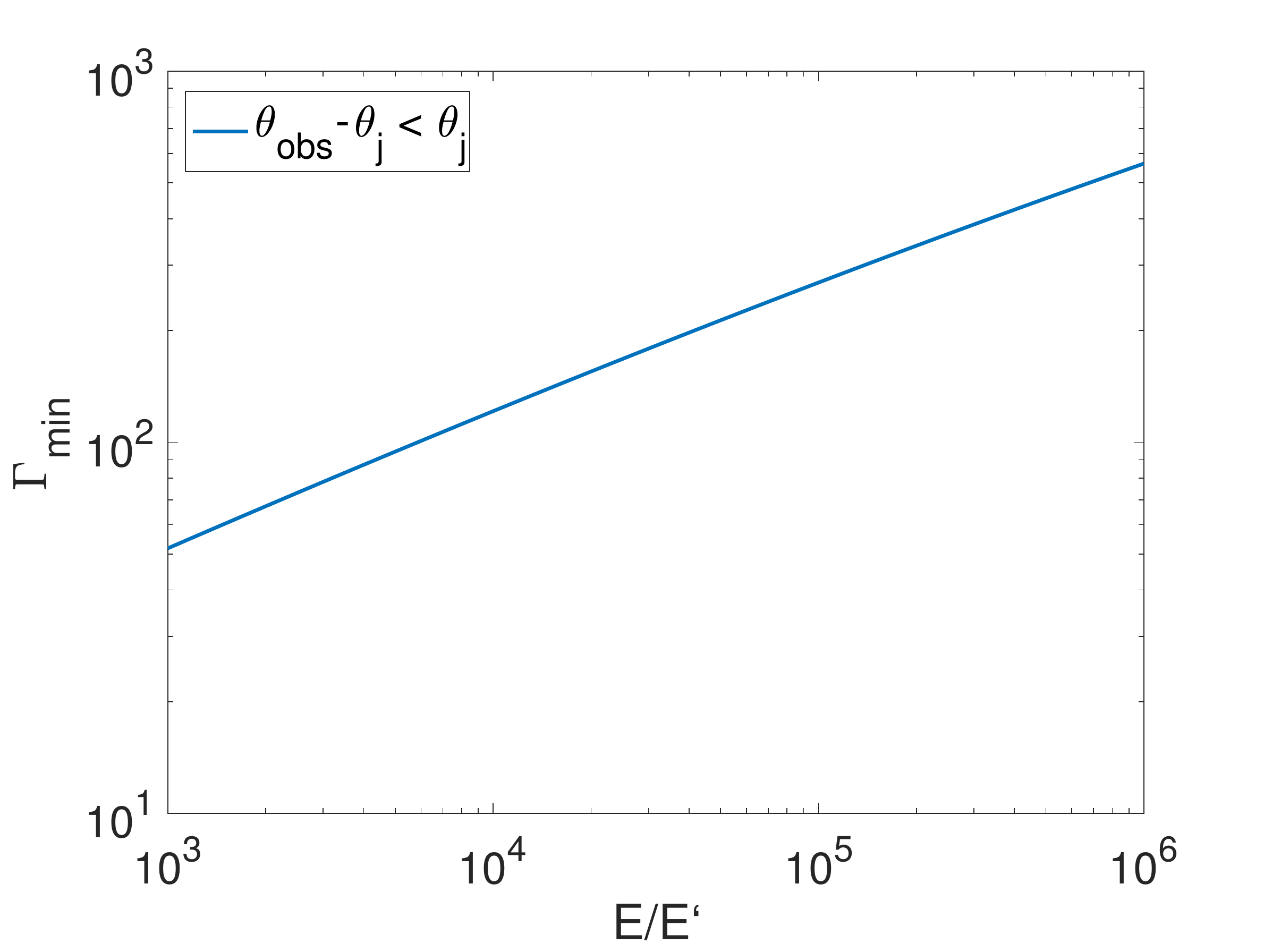}
	\caption{
		The minimal Lorentz factor of the jet, $\Gamma_{\rm min}$, required for the jet to be optically thin as a function of the amplification parameter $\A\equiv E/E^\prime$ for far off-axis observers, $\theta_{\rm obs}-\theta_{\rm j} \ll \theta_{\rm j}$ (case (i) in equation~\ref{eq:A}).  The parameters assumed here were the same as those assumed in figure~\ref{fig:the_obs}: photon energy $E_p'=185$ keV and $\gamma$-ray spectral index $\alpha=-0.6$ \cite{GBM2017,Goldstein2017}.
	}
	\label{fig:Gmin}
\end{figure}

\begin{figure}[h]
\centerline{\includegraphics[width=0.8\textwidth]{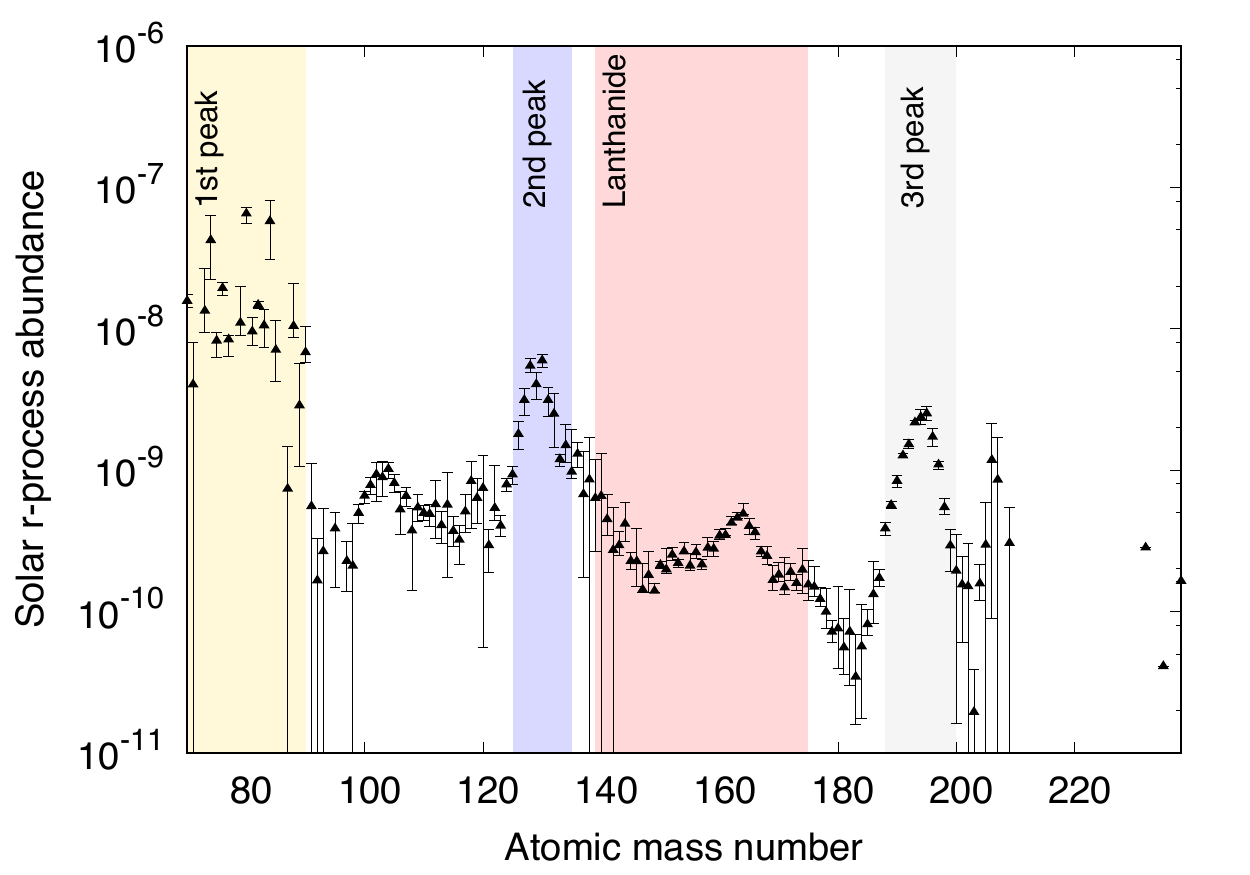}}
\centerline{\includegraphics[width=0.8\textwidth]{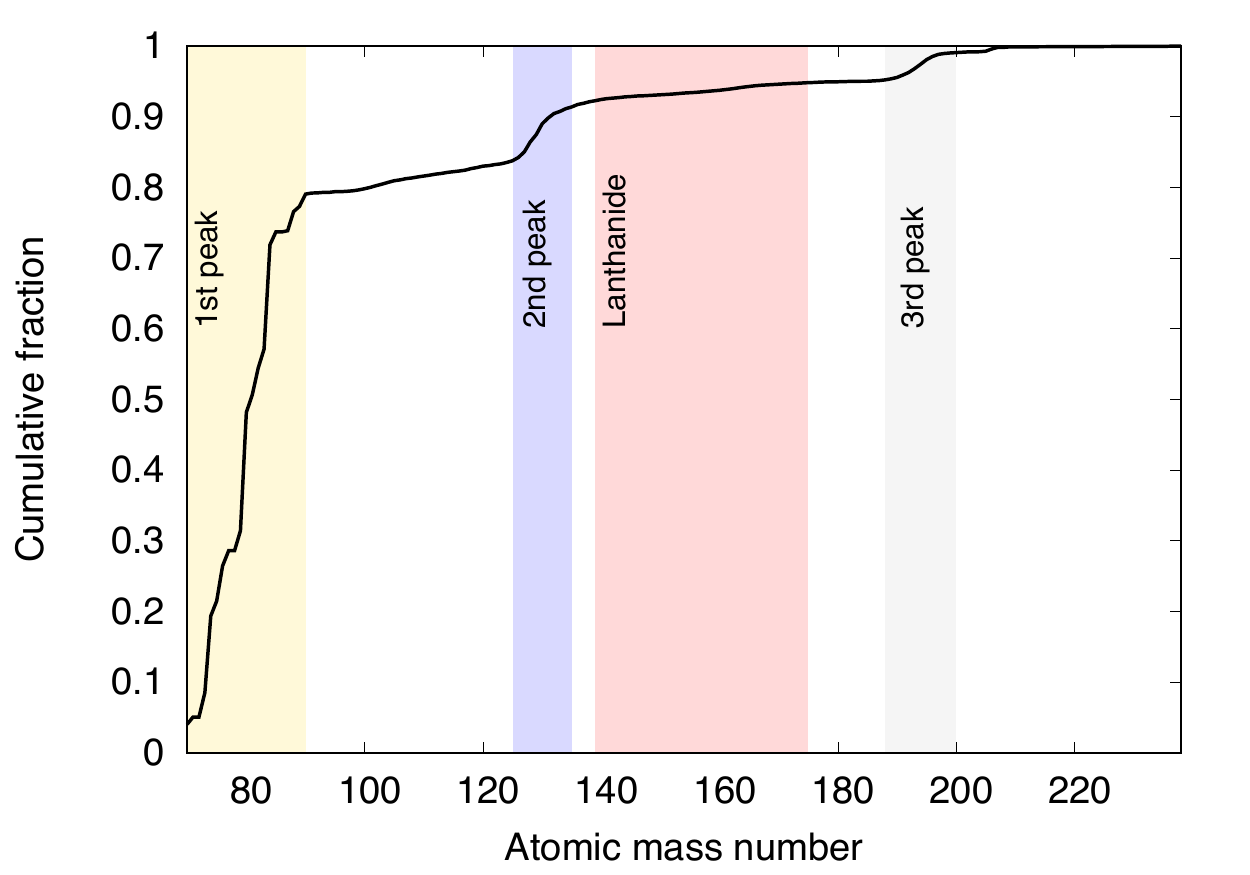}}
\caption{Top: the solar abundance pattern of r-process elements \cite{Goriely:1999}. Bottom:
their cumulative distribution.
}
\label{fig:abundance}
\end{figure}

\begin{table}
\caption{New and archival ultraviolet, optical, and near-infrared
  photometry of \ot. For each observation we give the observation
  date, time since GW170817, telescope, instrument, filter, and AB
  magnitude.   Upper limits are all at 5-$\sigma$ confidence. We did not use photometry from several telescopes where the results reported in circulars were uncertain \cite{LVCC21895,LVCC21560,LVCC21584,LVCC21608}.
  \label{tab:phot}}
{\small
\begin{tabular}{c c c c c c c}
\hline
\multicolumn{1}{c}{Observation Date} & \multicolumn{1}{c}{$\Delta t$} & \multicolumn{1}{c}{Telescope} & \multicolumn{1}{c}{Instrument} & \multicolumn{1}{c}{Filter} & \multicolumn{1}{c}{AB Magnitude} & \multicolumn{1}{c}{Reference}\\
\multicolumn{1}{c}{(UTC)} & \multicolumn{1}{c}{(d)} &  &  &  &  & \\
\hline
2017-08-17 23:31 &  \phantom{0}0.45 &  Swope & directCCD & $i$ &  $17.33 \pm 0.10$ &  \cite{LVCC21567,Coulter2017,Drout2017}\\
2017-08-18 00:01 &  \phantom{0}0.47 &  PROMPT5 & CCD & $R$ &  $17.52 \pm 0.20$ &  \cite{LVCC21531,Valenti2017}\\
2017-08-18 00:04 &  \phantom{0}0.47 &  CTIO & DECam & $i$ &  $17.50 \pm 0.10$ &  \cite{LVCC21541,SoaresSantos2017}\\
2017-08-18 00:04 &  \phantom{0}0.47 &  CTIO & DECam & $z$ &  $17.50 \pm 0.10$ &  \cite{LVCC21541,SoaresSantos2017}\\
2017-08-18 00:10 &  \phantom{0}0.48 &  ESO-VISTA & VIRCAM & $J$ &  $18.42 \pm 0.10$ &  \cite{LVCC21544}\\
2017-08-18 00:15 &  \phantom{0}0.48 &  Magellan-Clay & MEGACAM & $g$ &  $17.20 \pm 0.10$ &  \cite{LVCC21551,Drout2017}\\
2017-08-18 01:30 &  \phantom{0}0.53 &  Gemini-S & FLAMINGOS-2 & $K_s$ &  $18.42 \pm 0.04$ &  this paper\\
2017-08-18 03:44 &  \phantom{0}0.63 &  \textit{Swift} & UVOT & UVM2 &  $21.32 \pm 0.22$ &  \cite{Evans17}\\
2017-08-18 03:54 &  \phantom{0}0.63 &  \textit{Swift} & UVOT & UVW1 &  $19.62 \pm 0.11$ &  \cite{Evans17}\\
2017-08-18 04:01 &  \phantom{0}0.64 &  \textit{Swift} & UVOT & $U$ &  $18.23 \pm 0.08$ &  \cite{Evans17}\\
2017-08-18 04:07 &  \phantom{0}0.64 &  \textit{Swift} & UVOT & UVW2 &  $21.48 \pm 0.25$ &  \cite{Evans17}\\
2017-08-18 05:33 &  \phantom{0}0.70 &  Pan-STARRS & GPC1 & $y$ &  $17.28 \pm 0.13$ &  \cite{LVCC21553,Smartt2017}\\
2017-08-18 05:33 &  \phantom{0}0.70 &  Pan-STARRS & GPC1 & $z$ &  $17.31 \pm 0.09$ &  \cite{LVCC21553,Smartt2017}\\
2017-08-18 05:34 &  \phantom{0}0.70 &  Pan-STARRS & GPC1 & $i$ &  $17.23 \pm 0.10$ &  \cite{LVCC21553,Smartt2017}\\
2017-08-18 05:35 &  \phantom{0}0.70 &  Pan-STARRS & GPC1 & $y$ &  $17.47 \pm 0.16$ &  \cite{LVCC21553,Smartt2017}\\
2017-08-18 05:36 &  \phantom{0}0.71 &  Pan-STARRS & GPC1 & $z$ &  $17.21 \pm 0.08$ &  \cite{LVCC21553,Smartt2017}\\
2017-08-18 05:37 &  \phantom{0}0.71 &  Pan-STARRS & GPC1 & $i$ &  $17.26 \pm 0.08$ &  \cite{LVCC21553,Smartt2017}\\
2017-08-18 05:40 &  \phantom{0}0.71 &  Subaru & HSC & $z$ &  $17.30 \pm 0.10$ &  \cite{LVCC21561}\\
2017-08-18 09:42 &  \phantom{0}0.88 &  SSO & CCD & $g$ &  $17.50 \pm 0.20$ &  \cite{LVCC21566}\\
2017-08-18 13:39 &  \phantom{0}1.04 &  \textit{Swift} & UVOT & UVW1 &  $20.36 \pm 0.21$ &  \cite{Evans17}\\
2017-08-18 13:43 &  \phantom{0}1.04 &  \textit{Swift} & UVOT & $U$ &  $19.05 \pm 0.16$ &  \cite{Evans17}\\
2017-08-18 13:48 &  \phantom{0}1.05 &  \textit{Swift} & UVOT & UVW2 &  $>21.19$ &  \cite{Evans17}\\
2017-08-18 14:11 &  \phantom{0}1.06 &  AST3-2 & CCD & $g$ &  $18.15 \pm 0.10$ &  \cite{LVCC21883,Andreoni2017}\\
2017-08-18 14:15 &  \phantom{0}1.07 &  \textit{Swift} & UVOT & UVM2 &  $>21.25$ &  \cite{Evans17}\\
2017-08-18 17:17 &  \phantom{0}1.19 &  KMTNet & 18KCCD & $V$ &  $17.83 \pm 0.10$ &  \cite{LVCC21632,Troja2017}\\
2017-08-18 17:17 &  \phantom{0}1.19 &  KMTNet & 18KCCD & $I$ &  $17.75 \pm 0.10$ &  \cite{LVCC21632,Troja2017}\\
2017-08-18 17:17 &  \phantom{0}1.19 &  KMTNet & 18KCCD & $R$ &  $17.62 \pm 0.10$ &  \cite{LVCC21632,Troja2017}\\
2017-08-18 17:17 &  \phantom{0}1.19 &  MASTER-II & NA & $R$ &  $17.42 \pm 0.20$ &  \cite{LVCC21687,Lipunov2017}\\
2017-08-18 17:34 &  \phantom{0}1.20 &  MASTER-II & NA & $B$ &  $18.42 \pm 0.10$ &  \cite{LVCC21687,Lipunov2017}\\
2017-08-18 20:42 &  \phantom{0}1.33 &  NOT & NOTCam & $J$ &  $17.13 \pm 0.11$ &  \cite{LVCC21591}\\
2017-08-18 20:42 &  \phantom{0}1.33 &  NOT & NOTCam & $K_s$ &  $17.70 \pm 0.25$ &  \cite{LVCC21591}\\
2017-08-18 23:45 &  \phantom{0}1.46 &  KMTNet & 18KCCD & $R$ &  $17.82 \pm 0.10$ &  \cite{LVCC21632,Troja2017}\\
2017-08-18 23:46 &  \phantom{0}1.46 &  KMTNet & 18KCCD & $I$ &  $17.85 \pm 0.10$ &  \cite{LVCC21632,Troja2017}\\
2017-08-18 23:46 &  \phantom{0}1.46 &  KMTNet & 18KCCD & $V$ &  $18.33 \pm 0.10$ &  \cite{LVCC21632,Troja2017}\\
2017-08-18 23:59 &  \phantom{0}1.47 &  PROMPT5 & CCD & $R$ &  $18.22 \pm 0.06$ &  \cite{LVCC21579,Valenti2017}\\
2017-08-19 00:44 &  \phantom{0}1.50 &  \textit{Swift} & UVOT & $U$ &  $>19.50$ &  \cite{Evans17}\\
2017-08-19 00:50 &  \phantom{0}1.51 &  \textit{Swift} & UVOT & UVW2 &  $>21.53$ &  \cite{Evans17}\\
2017-08-19 01:15 &  \phantom{0}1.52 &  \textit{Swift} & UVOT & UVM2 &  $>21.71$ &  \cite{Evans17}\\
\hline
\end{tabular}

}
\end{table}

\begin{table}
\contcaption{New and archival ultraviolet, optical, and near-infrared photometry of \ot\ (continued).}
            {\small
              \begin{tabular}{c c c c c c c}
\hline
\multicolumn{1}{c}{Observation Date} & \multicolumn{1}{c}{$\Delta t$} & \multicolumn{1}{c}{Telescope} & \multicolumn{1}{c}{Instrument} & \multicolumn{1}{c}{Filter} & \multicolumn{1}{c}{AB Magnitude} & \multicolumn{1}{c}{Reference}\\
\multicolumn{1}{c}{(UTC)} & \multicolumn{1}{c}{(d)} &  &  &  &  & \\
\hline
2017-08-19 01:23 &  \phantom{0}1.53 &  \textit{Swift} & UVOT & UVW1 &  $>20.86$ &  \cite{Evans17}\\
2017-08-19 01:55 &  \phantom{0}1.55 &  Swope & directCCD & $B$ &  $18.70 \pm 0.10$ &  \cite{LVCC21583,Coulter2017,Drout2017}\\
2017-08-19 01:55 &  \phantom{0}1.55 &  Swope & directCCD & $V$ &  $18.14 \pm 0.10$ &  \cite{LVCC21583,Coulter2017,Drout2017}\\
2017-08-19 01:55 &  \phantom{0}1.55 &  Swope & directCCD & $i$ &  $17.43 \pm 0.10$ &  \cite{LVCC21583,Coulter2017,Drout2017}\\
2017-08-19 05:39 &  \phantom{0}1.71 &  Subaru & HSC & $z$ &  $17.80 \pm 0.10$ &  \cite{LVCC21595}\\
2017-08-19 05:46 &  \phantom{0}1.71 &  Pan-STARRS & GPC1 & $i$ &  $17.91 \pm 0.05$ &  \cite{LVCC21590,Smartt2017}\\
2017-08-19 05:46 &  \phantom{0}1.71 &  Pan-STARRS & GPC1 & $y$ &  $17.59 \pm 0.07$ &  \cite{LVCC21590,Smartt2017}\\
2017-08-19 05:46 &  \phantom{0}1.71 &  Pan-STARRS & GPC1 & $z$ &  $17.80 \pm 0.05$ &  \cite{LVCC21590,Smartt2017}\\
2017-08-19 10:59 &  \phantom{0}1.93 &  Zadko & AndorIKON-L & $r$ &  $18.46 \pm 0.17$ &  \cite{LVCC21744,Andreoni2017}\\
2017-08-19 14:09 &  \phantom{0}2.06 &  \textit{Swift} & UVOT & $U$ &  $>21.02$ &  \cite{Evans17}\\
2017-08-19 14:17 &  \phantom{0}2.07 &  \textit{Swift} & UVOT & UVW2 &  $>21.55$ &  \cite{Evans17}\\
2017-08-19 14:25 &  \phantom{0}2.07 &  \textit{Swift} & UVOT & UVM2 &  $>21.59$ &  \cite{Evans17}\\
2017-08-19 14:33 &  \phantom{0}2.08 &  \textit{Swift} & UVOT & UVW1 &  $>21.57$ &  \cite{Evans17}\\
2017-08-19 17:17 &  \phantom{0}2.19 &  KMTNet & 18KCCD & $R$ &  $18.52 \pm 0.10$ &  \cite{LVCC21632,Troja2017}\\
2017-08-19 17:17 &  \phantom{0}2.19 &  KMTNet & 18KCCD & $V$ &  $19.43 \pm 0.10$ &  \cite{LVCC21632,Troja2017}\\
2017-08-19 17:17 &  \phantom{0}2.19 &  KMTNet & 18KCCD & $I$ &  $18.35 \pm 0.10$ &  \cite{LVCC21632,Troja2017}\\
2017-08-19 17:53 &  \phantom{0}2.22 &  MASTER-II & NA & $R$ &  $18.42 \pm 0.30$ &  \cite{LVCC21687,Lipunov2017}\\
2017-08-19 23:23 &  \phantom{0}2.45 &  CTIO & DECam & $i$ &  $17.80 \pm 0.10$ &  \cite{LVCC21580,SoaresSantos2017,Cowperthwaite2017}\\
2017-08-19 23:23 &  \phantom{0}2.45 &  CTIO & DECam & $z$ &  $17.60 \pm 0.10$ &  \cite{LVCC21580,SoaresSantos2017,Cowperthwaite2017}\\
2017-08-19 23:31 &  \phantom{0}2.45 &  KMTNet & 18KCCD & $R$ &  $18.62 \pm 0.10$ &  \cite{LVCC21632,Troja2017}\\
2017-08-19 23:31 &  \phantom{0}2.45 &  KMTNet & 18KCCD & $I$ &  $18.65 \pm 0.10$ &  \cite{LVCC21632,Troja2017}\\
2017-08-19 23:31 &  \phantom{0}2.45 &  KMTNet & 18KCCD & $V$ &  $19.83 \pm 0.10$ &  \cite{LVCC21632,Troja2017}\\
2017-08-19 23:50 &  \phantom{0}2.46 &  PROMPT5 & CCD & $r$ &  $18.96 \pm 0.10$ &  \cite{LVCC21606,Valenti2017}\\
2017-08-20 00:19 &  \phantom{0}2.49 &  Gemini-S & FLAMINGOS-2 & $K_s$ &  $17.60 \pm 0.04$ &  this paper\\
2017-08-20 00:27 &  \phantom{0}2.49 &  Gemini-S & FLAMINGOS-2 & $J$ &  $17.76 \pm 0.02$ &  this paper\\
2017-08-20 05:46 &  \phantom{0}2.71 &  Pan-STARRS & GPC1 & $y$ &  $18.08 \pm 0.07$ &  \cite{LVCC21617,Smartt2017}\\
2017-08-20 05:46 &  \phantom{0}2.71 &  Pan-STARRS & GPC1 & $i$ &  $18.47 \pm 0.08$ &  \cite{LVCC21617,Smartt2017}\\
2017-08-20 05:46 &  \phantom{0}2.71 &  Pan-STARRS & GPC1 & $z$ &  $18.31 \pm 0.06$ &  \cite{LVCC21617,Smartt2017}\\
2017-08-20 11:14 &  \phantom{0}2.94 &  Zadko & AndorIKON-L & $r$ &  $19.18 \pm 0.12$ &  \cite{LVCC21744,Andreoni2017}\\
2017-08-20 11:47 &  \phantom{0}2.96 &  \textit{Swift} & UVOT & UVW1 &  $>22.18$ &  \cite{Evans17}\\
2017-08-20 13:08 &  \phantom{0}3.02 &  \textit{Swift} & UVOT & $U$ &  $>21.87$ &  \cite{Evans17}\\
2017-08-20 13:11 &  \phantom{0}3.02 &  \textit{Swift} & UVOT & $B$ &  $>19.82$ &  \cite{Evans17}\\
2017-08-20 13:17 &  \phantom{0}3.03 &  \textit{Swift} & UVOT & UVW2 &  $>22.76$ &  \cite{Evans17}\\
2017-08-20 13:25 &  \phantom{0}3.03 &  \textit{Swift} & UVOT & $V$ &  $>19.04$ &  \cite{Evans17}\\
2017-08-20 15:11 &  \phantom{0}3.10 &  \textit{Swift} & UVOT & UVM2 &  $>23.29$ &  \cite{Evans17}\\
2017-08-21 00:35 &  \phantom{0}3.50 &  Gemini-S & FLAMINGOS-2 & $J$ &  $17.93 \pm 0.06$ &  this paper\\
2017-08-21 00:38 &  \phantom{0}3.50 &  Gemini-S & FLAMINGOS-2 & $H$ &  $17.72 \pm 0.04$ &  this paper\\
2017-08-21 00:40 &  \phantom{0}3.50 &  Gemini-S & FLAMINGOS-2 & $K_s$ &  $17.61 \pm 0.06$ &  this paper\\
\hline
\end{tabular}

}
\end{table}
\begin{table}
\contcaption{New and archival ultraviolet, optical, and near-infrared photometry of \ot\ (continued). }
{\small
\begin{tabular}{c c c c c c c}
\hline
\multicolumn{1}{c}{Observation Date} & \multicolumn{1}{c}{$\Delta t$} & \multicolumn{1}{c}{Telescope} & \multicolumn{1}{c}{Instrument} & \multicolumn{1}{c}{Filter} & \multicolumn{1}{c}{AB Magnitude} & \multicolumn{1}{c}{Reference}\\
\multicolumn{1}{c}{(UTC)} & \multicolumn{1}{c}{(d)} &  &  &  &  & \\
\hline
2017-08-21 05:31 &  \phantom{0}3.70 &  Pan-STARRS & GPC1 & $y$ &  $>18.00$ &  \cite{LVCC21633,Smartt2017}\\
2017-08-21 05:46 &  \phantom{0}3.71 &  Pan-STARRS & GPC1 & $z$ &  $18.27 \pm 0.33$ &  \cite{LVCC21633,Smartt2017}\\
2017-08-21 05:46 &  \phantom{0}3.71 &  Pan-STARRS & GPC1 & $i$ &  $>18.50$ &  \cite{LVCC21633,Smartt2017}\\
2017-08-21 09:30 &  \phantom{0}3.87 &  Zadko & AndorIKON-L & $r$ &  $19.86 \pm 0.21$ &  \cite{LVCC21744,Andreoni2017}\\
2017-08-22 01:00 &  \phantom{0}4.51 &  Gemini-S & FLAMINGOS-2 & $K_s$ &  $17.72 \pm 0.09$ &  this paper\\
2017-08-22 01:03 &  \phantom{0}4.52 &  Gemini-S & FLAMINGOS-2 & $J$ &  $18.15 \pm 0.06$ &  this paper\\
2017-08-22 01:06 &  \phantom{0}4.52 &  Gemini-S & FLAMINGOS-2 & $H$ &  $18.02 \pm 0.07$ &  this paper\\
2017-08-22 09:43 &  \phantom{0}4.88 &  \textit{Swift} & UVOT & UVW1 &  $>22.14$ &  \cite{Evans17}\\
2017-08-22 11:43 &  \phantom{0}4.96 &  Zadko & AndorIKON-L & $r$ &  $20.20 \pm 0.23$ &  \cite{LVCC21744,Andreoni2017}\\
2017-08-22 20:19 &  \phantom{0}5.32 &  \textit{HST} & WFC3/UVIS & F336W &  $24.97 \pm 0.11$ &  this paper\\
2017-08-22 20:28 &  \phantom{0}5.32 &  \textit{HST} & WFC3/UVIS & F336W &  $25.05 \pm 0.11$ &  this paper\\
2017-08-22 21:01 &  \phantom{0}5.35 &  \textit{HST} & WFC3/UVIS & F336W &  $25.18 \pm 0.11$ &  this paper\\
2017-08-22 23:23 &  \phantom{0}5.45 &  ESO-VST & OMEGACAM & $i$ &  $21.31 \pm 0.10$ &  \cite{LVCC21703}\\
2017-08-22 23:23 &  \phantom{0}5.45 &  ESO-VST & OMEGACAM & $r$ &  $22.43 \pm 0.10$ &  \cite{LVCC21703}\\
2017-08-22 23:23 &  \phantom{0}5.45 &  ESO-VST & OMEGACAM & $g$ &  $23.27 \pm 0.10$ &  \cite{LVCC21703}\\
2017-08-23 08:04 &  \phantom{0}5.81 &  \textit{Swift} & UVOT & $U$ &  $>20.74$ &  \cite{Evans17}\\
2017-08-23 08:05 &  \phantom{0}5.81 &  \textit{Swift} & UVOT & $B$ &  $>19.94$ &  \cite{Evans17}\\
2017-08-23 08:07 &  \phantom{0}5.81 &  \textit{Swift} & UVOT & UVW2 &  $>23.03$ &  \cite{Evans17}\\
2017-08-23 08:09 &  \phantom{0}5.81 &  \textit{Swift} & UVOT & $V$ &  $>18.91$ &  \cite{Evans17}\\
2017-08-23 08:11 &  \phantom{0}5.81 &  \textit{Swift} & UVOT & UVM2 &  $>22.66$ &  \cite{Evans17}\\
2017-08-23 11:48 &  \phantom{0}5.96 &  Zadko & AndorIKON-L & $r$ &  $>20.60$ &  \cite{LVCC21744,Andreoni2017}\\
2017-08-23 17:22 &  \phantom{0}6.20 &  IRSF & SIRIUS & $H$ &  $18.60 \pm 0.18$ &  this paper\\
2017-08-23 17:22 &  \phantom{0}6.20 &  IRSF & SIRIUS & $K_s$ &  $18.01 \pm 0.10$ &  this paper\\
2017-08-23 17:22 &  \phantom{0}6.20 &  IRSF & SIRIUS & $J$ &  $18.65 \pm 0.19$ &  this paper\\
2017-08-23 23:35 &  \phantom{0}6.45 &  VLT & VISIR & $J8.9$ &  $>8.26$ &  this paper\\
2017-08-24 11:55 &  \phantom{0}6.97 &  Zadko & AndorIKON-L & $r$ &  $>20.60$ &  \cite{LVCC21744,Andreoni2017}\\
2017-08-24 16:51 &  \phantom{0}7.17 &  IRSF & SIRIUS & $J$ &  $18.95 \pm 0.32$ &  this paper\\
2017-08-24 16:51 &  \phantom{0}7.17 &  IRSF & SIRIUS & $H$ &  $18.53 \pm 0.17$ &  this paper\\
2017-08-24 16:51 &  \phantom{0}7.17 &  IRSF & SIRIUS & $K_s$ &  $18.02 \pm 0.12$ &  this paper\\
2017-08-24 23:20 &  \phantom{0}7.44 &  CTIO1.3m & ANDICAM & $K$ &  $18.06 \pm 0.17$ &  this paper\\
2017-08-25 11:52 &  \phantom{0}7.97 &  Zadko & AndorIKON-L & $r$ &  $>20.60$ &  \cite{LVCC21744,Andreoni2017}\\
2017-08-25 16:42 &  \phantom{0}8.17 &  \textit{Swift} & UVOT & UVW1 &  $>22.51$ &  \cite{Evans17}\\
2017-08-25 23:29 &  \phantom{0}8.45 &  CTIO1.3m & ANDICAM & $K$ &  $18.44 \pm 0.18$ &  this paper\\
2017-08-26 16:57 &  \phantom{0}9.18 &  IRSF & SIRIUS & $H$ &  $18.83 \pm 0.23$ &  this paper\\
2017-08-26 16:57 &  \phantom{0}9.18 &  IRSF & SIRIUS & $K_s$ &  $18.25 \pm 0.21$ &  this paper\\
2017-08-26 16:57 &  \phantom{0}9.18 &  IRSF & SIRIUS & $J$ &  $>18.87$ &  this paper\\
2017-08-26 22:56 &  \phantom{0}9.43 &  Gemini-S & FLAMINGOS-2 & $K_s$ &  $18.50 \pm 0.08$ &  this paper\\
2017-08-26 23:01 &  \phantom{0}9.43 &  Gemini-S & FLAMINGOS-2 & $J$ &  $20.57 \pm 0.20$ &  this paper\\
\hline
\end{tabular}

}
\end{table}

\begin{table}
\contcaption{New and archival ultraviolet, optical, and near-infrared photometry of \ot\ (continued).}
{\small 
\begin{tabular}{c c c c c c c}
\hline
\multicolumn{1}{c}{Observation Date} & \multicolumn{1}{c}{$\Delta t$} & \multicolumn{1}{c}{Telescope} & \multicolumn{1}{c}{Instrument} & \multicolumn{1}{c}{Filter} & \multicolumn{1}{c}{AB Magnitude} & \multicolumn{1}{c}{Reference}\\
\multicolumn{1}{c}{(UTC)} & \multicolumn{1}{c}{(d)} &  &  &  &  & \\
\hline
2017-08-26 23:05 &  \phantom{0}9.43 &  Gemini-S & FLAMINGOS-2 & $H$ &  $19.68 \pm 0.08$ &  this paper\\
2017-08-26 23:21 &  \phantom{0}9.44 &  CTIO1.3m & ANDICAM & $K$ &  $18.43 \pm 0.17$ &  this paper\\
2017-08-27 02:15 &  \phantom{0}9.57 &  APO & NICFPS & $K_s$ &  $>17.99$ &  this paper\\
2017-08-27 02:49 &  \phantom{0}9.59 &  Palomar5m & WHIRC & $K_s$ &  $>17.64$ &  this paper\\
2017-08-27 23:07 &  10.43 &  Gemini-S & FLAMINGOS-2 & $K_s$ &  $18.77 \pm 0.07$ &  this paper\\
2017-08-27 23:10 &  10.44 &  Gemini-S & FLAMINGOS-2 & $H$ &  $19.63 \pm 0.08$ &  this paper\\
2017-08-27 23:16 &  10.44 &  Gemini-S & FLAMINGOS-2 & $J$ &  $21.33 \pm 0.30$ &  this paper\\
2017-08-27 23:18 &  10.44 &  CTIO1.3m & ANDICAM & $K$ &  $18.91 \pm 0.19$ &  this paper\\
2017-08-28 16:40 &  11.17 &  \textit{Swift} & UVOT & $B$ &  $>20.48$ &  \cite{Evans17}\\
2017-08-28 16:44 &  11.17 &  \textit{Swift} & UVOT & UVW2 &  $>24.32$ &  \cite{Evans17}\\
2017-08-28 16:47 &  11.17 &  \textit{Swift} & UVOT & $V$ &  $>20.07$ &  \cite{Evans17}\\
2017-08-28 16:50 &  11.17 &  \textit{Swift} & UVOT & UVM2 &  $>23.29$ &  \cite{Evans17}\\
2017-08-28 16:52 &  11.17 &  IRSF & SIRIUS & $J$ &  $>18.37$ &  this paper\\
2017-08-28 16:52 &  11.17 &  IRSF & SIRIUS & $K_s$ &  $>18.48$ &  this paper\\
2017-08-28 16:52 &  11.17 &  IRSF & SIRIUS & $H$ &  $>18.43$ &  this paper\\
2017-08-28 17:21 &  11.20 &  \textit{Swift} & UVOT & $U$ &  $>21.44$ &  \cite{Evans17}\\
2017-08-28 23:17 &  11.44 &  CTIO1.3m & ANDICAM & $K$ &  $>19.11$ &  this paper\\
2017-08-28 23:35 &  11.45 &  Gemini-S & FLAMINGOS-2 & $K_s$ &  $19.41 \pm 0.09$ &  this paper\\
2017-08-28 23:40 &  11.46 &  Gemini-S & FLAMINGOS-2 & $H$ &  $>20.63$ &  this paper\\
2017-08-28 23:45 &  11.46 &  Gemini-S & FLAMINGOS-2 & $J$ &  $>21.07$ &  this paper\\
2017-08-29 00:36 &  11.50 &  \textit{HST} & WFC3/UVIS & F225W &  $>26.04$ &  this paper\\
2017-08-29 00:36 &  11.50 &  \textit{HST} & WFC3/UVIS & F336W &  $>26.37$ &  this paper\\
2017-08-29 00:36 &  11.50 &  \textit{HST} & WFC3/UVIS & F275W &  $>26.13$ &  this paper\\
2017-08-29 10:44 &  11.92 &  \textit{Swift} & UVOT & UVW1 &  $>22.26$ &  \cite{Evans17}\\
2017-08-29 23:10 &  12.44 &  Gemini-S & FLAMINGOS-2 & $K_s$ &  $19.44 \pm 0.08$ &  this paper\\
2017-08-29 23:23 &  12.45 &  Gemini-S & FLAMINGOS-2 & $H$ &  $20.99 \pm 0.21$ &  this paper\\
2017-08-29 23:41 &  12.46 &  Gemini-S & FLAMINGOS-2 & $J$ &  $>21.55$ &  this paper\\
2017-08-30 23:01 &  13.43 &  Gemini-S & FLAMINGOS-2 & $K_s$ &  $19.84 \pm 0.09$ &  this paper\\
2017-08-30 23:29 &  13.45 &  Gemini-S & FLAMINGOS-2 & $H$ &  $21.48 \pm 0.30$ &  this paper\\
2017-08-30 23:43 &  13.46 &  Gemini-S & FLAMINGOS-2 & $J$ &  $>21.94$ &  this paper\\
2017-08-31 23:03 &  14.43 &  Gemini-S & FLAMINGOS-2 & $K_s$ &  $20.06 \pm 0.10$ &  this paper\\
2017-08-31 23:18 &  14.44 &  VLT & VISIR & $J8.9$ &  $>7.74$ &  this paper\\
2017-08-31 23:50 &  14.47 &  Gemini-S & FLAMINGOS-2 & $H$ &  $21.63 \pm 0.36$ &  this paper\\
2017-09-01 00:18 &  14.48 &  Gemini-S & GMOS & $i$ &  $>23.20$ &  this paper\\
2017-09-01 23:18 &  15.44 &  VLT & VISIR & $J8.9$ &  $>7.57$ &  this paper\\
2017-09-01 23:24 &  15.45 &  Gemini-S & FLAMINGOS-2 & $K_s$ &  $20.43 \pm 0.13$ &  this paper\\
2017-09-02 00:46 &  15.50 &  Gemini-S & GMOS & $i$ &  $>23.40$ &  this paper\\
2017-09-02 23:22 &  16.45 &  Gemini-S & FLAMINGOS-2 & $K_s$ &  $20.95 \pm 0.18$ &  this paper\\
\hline
\end{tabular}

}
\end{table}

\begin{table}
\contcaption{New and archival ultraviolet, optical, and near-infrared photometry of \ot\ (continued).}
{\small 
\begin{tabular}{c c c c c c c}
\hline
\multicolumn{1}{c}{Observation Date} & \multicolumn{1}{c}{$\Delta t$} & \multicolumn{1}{c}{Telescope} & \multicolumn{1}{c}{Instrument} & \multicolumn{1}{c}{Filter} & \multicolumn{1}{c}{AB Magnitude} & \multicolumn{1}{c}{Reference}\\
\multicolumn{1}{c}{(UTC)} & \multicolumn{1}{c}{(d)} &  &  &  &  & \\
\hline
2017-09-03 00:03 &  16.47 &  Gemini-S & GMOS & $r$ &  $>21.18$ &  this paper\\
2017-09-03 23:36 &  17.46 &  Gemini-S & FLAMINGOS-2 & $K_s$ &  $>19.92$ &  this paper\\
2017-09-04 00:16 &  17.48 &  Gemini-S & GMOS & $r$ &  $>21.98$ &  this paper\\
2017-09-04 23:28 &  18.45 &  Gemini-S & FLAMINGOS-2 & $K_s$ &  $21.04 \pm 0.09$ &  this paper\\
2017-09-05 00:03 &  18.47 &  Gemini-S & GMOS & $i$ &  $>21.90$ &  this paper\\
2017-09-05 23:48 &  19.46 &  Gemini-S & FLAMINGOS-2 & $K_s$ &  $21.23 \pm 0.37$ &  this paper\\
2017-09-06 23:30 &  20.45 &  Gemini-S & FLAMINGOS-2 & $H$ &  $>21.22$ &  this paper\\
2017-09-06 23:33 &  20.45 &  VLT & VISIR & $J8.9$ &  $>7.42$ &  this paper\\
2017-09-07 23:39 &  21.46 &  Gemini-S & FLAMINGOS-2 & $K_s$ &  $>21.48$ &  this paper\\
2017-09-11 23:39 &  25.46 &  Gemini-S & FLAMINGOS-2 & $J$ &  $>20.21$ &  this paper\\
2017-09-14 23:14 &  28.44 &  Gemini-S & FLAMINGOS-2 & $K_s$ &  $>19.96$ &  this paper\\
2017-09-15 23:19 &  29.44 &  Gemini-S & FLAMINGOS-2 & $K_s$ &  $>20.60$ &  this paper\\
\hline
\end{tabular}

}
\end{table}

\begin{table}
\caption{Optical and near-IR spectroscopic observations of \ot.  For
  each observation we give the observation date, the time since
  GW170817, the telescope, instrument, exposure time, approximate
  wavelength range, and spectral resolving power.
  \label{tab:spec}}
{\small
\begin{tabular}{c c c c c c c}
\hline
\hline
\multicolumn{1}{c}{Observation Date} & \multicolumn{1}{c}{$\Delta t$} & \multicolumn{1}{c}{Telescope} & \multicolumn{1}{c}{Instrument} & \multicolumn{1}{c}{Exposure} & \multicolumn{1}{c}{Wavelength Range} & \multicolumn{1}{c}{$\lambda/\Delta\lambda$}\\
\multicolumn{1}{c}{(UTC)} & \multicolumn{1}{c}{(days)} &  &  & \multicolumn{1}{c}{(s)}  & (\AA) & \\
\hline
2017-08-20 01:08 &  \phantom{0}2.52 &  Gemini-S & GMOS        & \phantom{0}$2 \times 300$  & 6000--9000             & 1900 \\
2017-08-21 00:15 &  \phantom{0}3.48 &  Gemini-S & GMOS        & \phantom{0}$4 \times 360$  & 3800--9200             & 1700 \\
2017-08-22 00:21 &  \phantom{0}4.49 &  Gemini-S & FLAMINGOS-2 & \phantom{0}$8 \times 150$  & 12980--25070           & 600 \\
2017-08-22 00:47 &  \phantom{0}4.50 &  Gemini-S & FLAMINGOS-2 & \phantom{0}$6 \times 150$  & \phantom{0}9840--18020 & 600 \\
2017-08-25 05:45 &  \phantom{0}7.71 &  Keck I   & LRIS        &     $300 + 2\times 600$                          &   2000--10300                     &  1000   \\
2017-08-29 00:23 &  11.49           &  Gemini-S & FLAMINGOS-2 & $16 \times 150$            & 12980--25070           & 600 \\
\hline
\end{tabular}

}
\end{table}

\begin{sidewaystable*}
\caption{Census of the Local Universe (CLU) galaxies within the
  localization volume of GW170817 \cite{LVCC21535}.  For each galaxy
  we give the coordinates (J2000), distance, far-ultraviolet (FUV) magnitude from \textit{GALEX}, mid-infrared magnitudes from \textit{WISE}, estimated star-formation rate (SFR) based on the FUV magnitude, estimated stellar mass based on the mid-infrared magnitudes, and probability within the GW170817 localization volume.}
{\small
\begin{tabular}{cccccccccc}
\hline
\hline
Galaxy         & RA	       & DEC	       & $D$   & FUV    & WISE1  & WISE4  & $\log_{10}({\rm SFR})$ (FUV)	         & $\log_{10}(M_\star)$         & Prob.    \\ 
	       & (deg)     & (deg)     & (Mpc) &(AB)    &(Vega)  &(Vega)  &($M_{\odot}$yr$^{-1}$)& ($M_{\odot}$) &  \\ 
\hline

NGC 4970 & 196.8906 & $-$24.0086 & 46.50 & $\ldots$ & \phn9.78 & 8.01 & $\ldots$ & 10.42 & 0.68 \\
NGC 4830 & 194.3663 & $-$19.6913 & 47.90 & 18.88 & \phn9.87 & 8.34 & $-$1.32 & 10.41 & 0.75 \\
NGC 4993 & 197.4487 & $-$23.3839 & 41.66 & 20.13 & \phn9.94 & 7.47 & $-$1.35 & 10.26 & 0.41 \\
NGC 4968 & 196.7749 & $-$23.6770 & 42.24 & 18.93 & \phn9.99 & 2.41 & \phantom{$-$}0.63 & 10.25 & 0.51 \\
IC 4197 & 197.0180 & $-$23.7969 & 43.24 & $\ldots$ & 10.07 & 7.45 & $\ldots$ & 10.24 & 0.48 \\
IC 4180 & 196.7354 & $-$23.9171 & 42.46 & 18.05 & 10.19 & 5.62 & $-$0.58 & 10.17 & 0.60 \\
ESO 508$-$ G 033 & 199.0969 & $-$26.5614 & 45.59 & 17.02 & 10.90 & 3.70 & \phantom{$-$}0.21 & \phn9.95 & 0.83 \\
MCG $-$02$-$33$-$036 & 193.1066 & $-$15.5172 & 53.87 & 19.81 & 11.50 & 7.47 & $-$1.11 & \phn9.86 & 0.88 \\
ESO 508$-$ G 010 & 196.9080 & $-$23.5790 & 43.04 & $\ldots$ & 11.88 & 5.97 & $\ldots$ & \phn9.51 & 0.44 \\
MCG $-$03$-$33$-$023 & 194.2521 & $-$17.3202 & 56.79 & $\ldots$ & 12.93 & 6.37 & $\ldots$ & \phn9.33 & 0.90 \\
ESO 575$-$ G 053 & 196.2705 & $-$22.3839 & 36.37 & 16.46 & 11.98 & 6.47 & $-$0.68 & \phn9.33 & 0.23 \\
2MASX J12525109$-$1529300 & 193.2130 & $-$15.4916 & 52.26 & $\ldots$ & 12.80 & 8.55 & $\ldots$ & \phn9.31 & 0.84 \\
2MASX J12505229$-$1454238 & 192.7180 & $-$14.9066 & 52.96 & 17.92 & 12.89 & 7.10 & $-$0.79 & \phn9.29 & 0.89 \\
2MASX J12573271$-$1942006 & 194.3863 & $-$19.7002 & 52.39 & 20.65 & 12.96 & 9.18 & $-$1.73 & \phn9.25 & 0.87 \\
ESO 576$-$ G 003 & 197.6488 & $-$21.7482 & 42.04 & $\ldots$ & 12.65 & 6.59 & $\ldots$ & \phn9.18 & 0.89 \\
UGCA 331 & 197.6914 & $-$23.8657 & 40.82 & $\ldots$ & 12.61 & 7.33 & $\ldots$ & \phn9.17 & 0.47 \\
IC 3825 & 192.6544 & $-$14.4828 & 51.04 & 18.24 & 13.10 & 7.33 & $-$0.93 & \phn9.17 & 0.87 \\
ESO 575$-$ G 055 & 196.6663 & $-$22.4561 & 44.49 & 17.11 & 13.04 & 7.29 & $-$0.79 & \phn9.07 & 0.32 \\
ESO 508$-$ G 003 & 196.6000 & $-$24.1641 & 40.52 & 16.93 & 12.86 & 6.89 & $-$0.77 & \phn9.06 & 0.73 \\
ESO 508$-$ G 019 & 197.4663 & $-$24.2391 & 41.79 & $\ldots$ & 13.13 & 6.66 & $\ldots$ & \phn8.98 & 0.47 \\
ESO 575$-$ G 029 & 193.9986 & $-$19.2691 & 45.21 & $\ldots$ & 13.36 & 7.96 & $\ldots$ & \phn8.96 & 0.78 \\
2MASX J13073768$-$2356181 & 196.9071 & $-$23.9384 & 49.73 & $\ldots$ & 13.66 & 8.44 & $\ldots$ & \phn8.92 & 0.75 \\
2MFGC 10461 & 197.1774 & $-$23.7756 & 41.39 & $\ldots$ & 13.32 & 8.61 & $\ldots$ & \phn8.90 & 0.40 \\
2MFGC 10484 & 197.4617 & $-$24.2419 & 42.31 & $\ldots$ & 13.46 & $\ldots$ & $\ldots$ & \phn8.86 & 0.48 \\
2MASX J13061939$-$2258491 & 196.5805 & $-$22.9804 & 41.51 & 17.52 & 13.51 & 7.30 & $-$0.96 & \phn8.83 & 0.35 \\
UGCA 327 & 196.9370 & $-$22.8579 & 37.29 & $\ldots$ & 13.33 & 7.02 & $\ldots$ & \phn8.81 & 0.24 \\
GALEXASC J125520.46$-$170546.9 & 193.8364 & $-$17.0966 & 56.69 & 18.39 & 14.58 & 8.31 & $-$1.06 & \phn8.67 & 0.89 \\
WINGS J125412.84$-$153523.6 & 193.5534 & $-$15.5899 & 50.96 & $\ldots$ & 14.43 & 8.33 & $\ldots$ & \phn8.64 & 0.82 \\
ESO 508$-$ G 004 & 196.7177 & $-$22.8405 & 41.37 & 16.59 & 14.05 & 8.54 & $-$0.78 & \phn8.61 & 0.26 \\
ESO 508$-$ G 014 & 197.1342 & $-$23.3469 & 46.61 & 18.44 & 14.32 & 8.83 & $-$1.31 & \phn8.60 & 0.50 \\
\hline

\label{tab:clumatch}
\end{tabular}
}
\end{sidewaystable*}

\begin{sidewaystable*}
\contcaption{Census of the Local Universe (CLU) galaxies within the localization volume of GW170817 \cite{LVCC21535}, continued.}
{\small
\begin{tabular}{cccccccccc}
\hline
\hline
Galaxy         & RA	       & DEC	       & $D$   & FUV    & WISE1  & WISE4  & $\log_{10}({\rm SFR})$ (FUV)	         & $\log_{10}(M_\star)$         & Prob.    \\ 
	       & (deg)     & (deg)     & (Mpc) &(AB)    &(Vega)  &(Vega)  &($M_{\odot}$yr$^{-1}$)& ($M_{\odot}$) &  \\ 
\hline

6dF J1254495$-$160308 & 193.7063 & $-$16.0523 & 48.02 & $\ldots$ & 14.60 & 8.60 & $\ldots$ & \phn8.52 & 0.64 \\
GALEXASC J125811.97$-$210246.3 & 194.5501 & $-$21.0461 & 43.89 & 17.69 & 14.51 & 8.86 & $-$1.13 & \phn8.47 & 0.90 \\
GALEXASC J130525.30$-$233008.8 & 196.3546 & $-$23.5025 & 45.90 & 17.44 & 14.73 & 7.16 & $-$0.83 & \phn8.43 & 0.76 \\
GALEXASC J125259.36$-$152150.9 & 193.2474 & $-$15.3639 & 49.87 & 18.44 & 14.96 & 8.63 & $-$1.23 & \phn8.41 & 0.78 \\
GALEXASC J125301.39$-$151007.7 & 193.2552 & $-$15.1693 & 53.40 & 20.94 & 15.25 & 8.93 & $-$1.67 & \phn8.35 & 0.90 \\
PGC45429 & 196.7822 & $-$24.1104 & 41.50 & $\ldots$ & 14.82 & 9.12 & $\ldots$ & \phn8.30 & 0.65 \\
6dF J1305235$-$233121 & 196.3478 & $-$23.5224 & 41.71 & 17.31 & 14.90 & 8.41 & $-$1.02 & \phn8.28 & 0.68 \\
UGCA 325 & 196.7796 & $-$24.1119 & 42.70 & $\ldots$ & 15.05 & $\ldots$ & $\ldots$ & \phn8.24 & 0.67 \\
GALEXASC J130415.26$-$225251.3 & 196.0633 & $-$22.8814 & 41.23 & 18.41 & 15.05 & 8.72 & $-$1.40 & \phn8.20 & 0.57 \\
6dF J1309177$-$242256 & 197.3241 & $-$24.3821 & 40.36 & $\ldots$ & 15.23 & 7.34 & $\ldots$ & \phn8.12 & 0.49 \\
UGCA 328 & 197.3298 & $-$24.3866 & 41.03 & $\ldots$ & 15.27 & 8.31 & $\ldots$ & \phn8.11 & 0.50 \\
GALEXASC J125157.02$-$160617.8 & 192.9872 & $-$16.1047 & 50.30 & 21.04 & 15.85 & 8.20 & $-$1.49 & \phn8.06 & 0.76 \\
ESO 508$-$ G 035 & 199.4497 & $-$26.9025 & 37.77 & 17.95 & 15.33 & 9.19 & $-$1.37 & \phn8.02 & 0.85 \\
PGC45611 & 197.3286 & $-$24.3846 & 39.81 & 15.87 & 15.55 & 8.29 & $-$0.54 & \phn7.98 & 0.49 \\
{ABELL 1664\_11:[PSE2006] 2506} & 196.8922 & $-$23.8153 & 42.80 & 21.49 & 15.83 & 9.09 & $-$1.96 & \phn7.93 & 0.50 \\
GALEXASC J131426.62$-$271242.6 & 198.6106 & $-$27.2120 & 29.32 & 18.39 & 15.28 & 8.77 & $-$1.69 & \phn7.82 & 0.89 \\
{ABELL 1631:[CZ2003]B0295[024]} & 192.8695 & $-$15.8723 & 53.56 & $\ldots$ & 16.70 & 8.46 & $\ldots$ & \phn7.77 & 0.87 \\
WINGS J125701.38$-$172325.2 & 194.2558 & $-$17.3903 & 26.13 & $\ldots$ & 16.83 & 9.31 & $\ldots$ & \phn7.10 & 0.27 \\
HIPASS J1255$-$15 & 193.8983 & $-$15.0175 & 27.32 & $\ldots$ & $\ldots$ & $\ldots$ & $\ldots$ & \phn0.00 & 0.83 \\

\hline
\end{tabular}
}
\end{sidewaystable*}

\end{document}